\documentclass[a4paper]{article} %pre=prf
\usepackage{a4wide}
\usepackage{color}
\usepackage{amsbsy,amssymb,amsgen,amsfonts,graphics}
\usepackage{amsmath,amsfonts,color,latexsym,gensymb}
\usepackage[english]{babel}
\usepackage{inputenc}
\usepackage[pdftex,bookmarks=true,colorlinks=true,linkcolor=blue,citecolor=blue,
filecolor=blue,urlcolor=blue]{hyperref}
\usepackage[T1]{fontenc}
\usepackage{graphicx}
\usepackage{sectsty}
\allsectionsfont{\sffamily}
\usepackage[font={small}]{caption}
\usepackage{mathpazo}
\usepackage{tikz}
\usetikzlibrary{shapes,arrows,positioning,calc,through,backgrounds}
\tikzstyle{every picture}+=[remember picture]
\usetikzlibrary{patterns}
\definecolor{lightgray}{rgb}{0.75,0.75,0.75}
\definecolor{green}{rgb}{0.0, 0.75, 0.0}
\definecolor{cyan}{rgb}{0.0, 0.75, 0.75}
\definecolor{magenta}{rgb}{0.75, 0.0, 0.75}
\definecolor{grey}{rgb}{0.45, 0.45, 0.45}
\def\solidthick{\protect\rule[2pt]{10.pt}{1pt}}
\def\solidshort{\protect\rule[2pt]{3.pt}{1pt}}
\def\dashed{\solidshort$\,$\solidshort$\,$\solidshort}

\usepackage{relsize}
\usepackage[percent]{overpic}
\newcommand{\typo} [2] {{\color{black}{#2}}}
\usepackage[square,sort,comma,numbers]{natbib}
\begin{document}

\title{
\Large{\textbf{\textsf{
On the influence of forced homogeneous-isotropic turbulence on the
settling and clustering of finite-size particles
}}}
}
\author{\large Agathe Chouippe\footnote{\tt agathe.chouippe@kit.edu} \
 and Markus Uhlmann\footnote{\tt markus.uhlmann@kit.edu}
 \\[0.5ex]
 \small{Institute for Hydromechanics, Karlsruhe Institute of Technology (KIT), 76131 Karlsruhe, Germany}\\[0.5ex]
\footnotesize
\today, Manuscript accepted for publication in \textit{Acta Mechanica}
 }
\date{}
\maketitle
%
% ------------------------------------------------------------------------ %
%
%% -----------------------------------------------------------------%
\begin{abstract}
  %%-----------------------------------------------------------------%
  We investigate the motion of heavy particles with a diameter of
  several multiples of the Kolmogorov length scale in the presence of 
  forced turbulence and gravity, resorting to interface-resolved
  direct numerical simulation based on an immersed boundary
  method. The values of the particles' relative density (1.5) and of
  the Galileo number (180) are such that strong wake-induced
  particle clustering would occur in the absence of turbulence
  \citep{uhlmann:14a}.
  The forced turbulence in the two present cases (with Taylor-scale
  Reynolds number 95 and 140) would lead to mild levels of clustering
  in the absence of gravity \citep{uhlmann:16a}. 
  Here we detect a tendency to cluster with an intensity (quantified
  via the standard deviation of the distribution of Vorono\"i cell
  volumes) which is intermediate between these two limiting cases,
  meaning that forced background turbulence decreases the level of
  clustering otherwise observed under ambient settling. However, the
  clustering strength does not monotonously decay with the relative
  turbulence intensity. Various mechanisms by which coherent
  structures can affect particle motion are discussed. It is
  argued that the reduced interaction time due to particle settling
  through the surrounding eddy (crossing trajectories) has the effect
  of shifting upwards the range of eddies with a time scale matching
  the characteristic time scale of the particle. In the present cases
  this shift might bring the particles into resonance with the
  energetic eddies of the turbulent spectrum. 
  Concerning the average particle settling velocity we find very small
  deviations (of the order of one percent) from the value obtained for
  an isolated particle in ambient fluid when defining the relative
  velocity as an apparent slip velocity (i.e.\ as the difference
  between the averages computed separately for the velocities of each
  phase). This is consistent with simple estimates of the non-linear
  drag effect. However, the relative velocity based upon the fluid
  velocity seen by each particle (computed via local averaging over a
  particle-attached sphere) has on average a smaller magnitude (by
  5-7\%) than the ambient single-particle value.\\[1.0ex]
  %
  %%-----------------------------------------------------------------%
%  {\textbf{keywords:}
%      multiphase and particle-laden flows, turbulence simulation, finite-size particles, settling
%    } 
\end{abstract}
%% -----------------------------------------------------------------%

% %
% %% ************************************************************************* 
%--------------------------------------------------------------------%
\section{Introduction}
%--------------------------------------------------------------------%
Many fluidic systems in geophysics and engineering involve
suspended mobile particles of some kind.
Prominent examples 
are droplets in atmospheric
clouds and plankton in ocean currents,
while a large number of technical applications can be found
practically across all engineering disciplines. 

Among the many relevant aspects of fluid-mediated particle motion,
the settling velocity of heavy particles in a turbulent flow
is of
particular interest. 
While this quantity is clearly important in various fields of
research, it has received
much
attention in the meteorological community
due to its relevance in precipitation forecasting
\citep{shaw:03,grabowski:13}. 
The available data, which will be reviewed in more detail below,
shows that as a result of the interaction with a turbulent flow
field the mean particle settling velocity can be significantly 
enhanced or reduced when compared to the reference value of a single 
particle settling in ambient fluid. 

In many cases (e.g.\ small water droplets in atmospheric clouds) the
particle size is much smaller than the smallest (i.e.\ Kolmogorov)
flow scales and the flow at the particle scale can be considered as
Stokesian. This scenario has been relatively well explored in the
past, in particular through
Eulerian-Lagrangian point-particle simulations. 
Conversely, when the particle diameter exceeds the Kolmogorov length
scale and the Reynolds number of the flow around the particles is not
negligible, it becomes necessary to consider the fluid-solid
interaction problem at the level of each ``finite-size'' particle in
all details.
This more complex situation is the subject of the present work. 
Systems where finite-size effects become important arise for instance
in meteorology when focusing on larger hydrometeors (e.g.\
hailstones), or
{in many chemical engineering processes.} 
Across the large parameter space of the fluid-particle system a rich
set of features emerges due to the action of turbulence, buoyancy,
the collective action of particles, and the interplay between
these different actors.

Let us first consider fluid-particle systems in the absence of gravity
or buoyancy. 
For this configuration it is now well established that the particle
motion can be considerably affected by the action of turbulent 
coherent structures, leading to an inhomogeneous sampling of the flow
field.
In a multi-particle system, the disperse phase will then typically be
non-randomly distributed in space, a phenomenon which
is commonly referred to as ``preferential concentration''
\citep{squires:91}.

A number of mechanisms has been proposed to account for the
phenomenon 
of particle clustering in turbulence.  
According to the centrifugal mechanism dense particles will tend to
be expelled from vortical flow regions and accumulate preferentially
in strain-dominated regions.
Through asymptotic expansion \citet{maxey:87} has shown how this
argument applies to the simplest point particle model in the limit of
small Stokes number (i.e.\ the particle-to-fluid ratio of time
scales). 
For this class of particles, the Stokes number is generally considered 
to be the significant parameter with respect to the tendency to form
clusters \citep{hogan:01,balachandar:10,monchaux:12}.
However, \citet{yoshimoto:07} have shown that the definition of the
fluid time scale should probably take into account the entire range of
available flow scales.
Experimental support for this multi-scale view of particle clustering
has recently been provided by the study of \citet{sumbekova:16} who
have found no significant dependency of the cluster statistics upon
the conventional Stokes number.
Also based upon the analysis of \citet{maxey:87} 
a scenario dubbed ``sweep-stick'' mechanism 
has been proposed by \citet{goto:08} and further elaborated by
\citet{coleman:09}. 
Since the leading order velocity difference between point particles
(in the small Stokes number limit) and the carrier flow is
proportional to the fluid acceleration, particles can be expected to
``stick'' to zero-acceleration points, which is what has indeed
been observed in DNS \citep{coleman:09}. 
Concerning finite-size particles,
the data from particle-resolved DNS of \citet{uhlmann:16a}
suggests that the sweep-stick mechanism also applies to particles
whose diameter exceeds the Kolmogorov scale, leading to small but
significant levels of clustering in their work. 
Based upon a Lagrangian description of the relative motion between
particle pairs \citet{zaichik:03}, \citet{chun:05} and \citet{zaichik:07}
have derived models for the particles' radial distribution
function. This type of approach is non-local, since the history of the
fluid velocity field seen by the particles enters the description
\citep{bragg:15a,gustavsson:16}.
Note that the relevance of this non-local clustering mechanism for
finite-size particles has not been verified to date. 

Let us now turn to the case of an ensemble of particles settling in
the absence of a priori turbulence. 
In this configuration a new mechanism for generating
a non-trivial spatial particle distribution comes into play through
the effect of wakes. Roughly speaking, at a sufficiently large
particle Reynolds number the reduced drag acting upon a
vertically trailing particle induces 
``drafting-kissing-tumbling'' motion in particle pairs
\citep{fortes:87,wu:98}, which -- at least transiently -- brings a 
particle pair closer together. 
Essentially the same wake-attraction effect is also active in
many-particle configurations, where it can lead to the formation of
massive columnar 
clusters which reach a sustained statistically stationary state in
particle-resolved direct numerical simulations
\citep{kajishima:02,kajishima:04b}.  
\citet{uhlmann:14a} and \citet{doychev:14} have shown that 
wake-induced clustering appears above a critical Galileo number, 
a parameter which can be understood as a particle Reynolds number with
a velocity scale constructed from buoyancy. 
The critical value in these studies approximately coincides with the
value at which a change in the wake structure (from axi-symmetric to
double-threaded oblique) is observed for a single settling particle. 
As a consequence of wake-induced clustering the average particle
settling velocity is significantly enhanced in the many-particle case
as compared to a single isolated particle. 
The formation of clusters as well as an increase of the settling velocity have also been evidenced experimentally by \citet{huisman:16}.
It should be noted that despite the absence of a priori turbulence,
the particle-induced flow field exhibits quite considerable levels of
turbulence intensity even in relatively dilute systems if the Galileo
number is significant
(e.g.\ the standard deviation of the vertical fluid
velocity component exceeds one fourth of the particle settling
velocity in the simulations of \citet[cf.\ their figure~6$a$]{uhlmann:14a}).  
However, the ``pseudo-turbulence'' induced by the settling particles
is distinct from classical turbulence, as it is characterized by a
$(-3)$ power-law spectrum in terms of wavenumber
\citep{lance:91,risso:11,doychev:14}.  

Finally, let us return to the actual case of interest where heavy
particles are settling in a homogeneous turbulent ``background'' flow.
In this configuration we intuitively expect that for sufficiently high
turbulence intensity the background flow will tend to suppress
wake-induced cluster formation.
On the other hand, the existence of gravity is expected to have a
significant impact on the various mechanisms of preferential 
concentration discussed above.
\citet{wang:93} have shown that turbulence leads to a significant
enhancement of the average particle settling velocity in the framework
of the point-particle model with Stokes drag and one-way coupling, a
result which has subsequently been confirmed by various authors
\citep{bec:14,good:14,ireland:16b}. It has been observed by
\citet{wang:93} and others that particles under these conditions
preferentially sample the downward sweeping part of the vortical
structures
(an effect which is sometimes also referred to as
``fast-tracking'' \citep{nielsen:93}).

\citet{aliseda:02} have performed experiments with water droplets in
decaying grid-turbulence in a wind-tunnel, where the particle diameter
was much smaller than the Kolmogorov scale. They observed particle
clustering and substantial enhancement of the average settling
velocity, the latter increasing with the solid volume fraction. This
is, therefore, clearly a collective effect, as opposed to the
non-collective effect of \citet{wang:93} and others who have resorted
to one-way coupled descriptions. \citet{aliseda:02} argue that the
particle clusters (presumably formed through the centrifugal effect)
settle at an increased speed due to the fact that the Reynolds
number of the cluster is much larger than the particle Reynolds
number. 
This interpretation has received support through the recent work of
\citet{monchaux:17} who have performed two-way coupled point-particle
simulations with Stokes drag,
resulting in a similar dependency of settling enhancement upon the
solid volume fraction as in the experiments of \citet{aliseda:02}.

Several experiments have observed a reduction in the average settling
speed due to turbulence.
\citet{yang:03} have measured the motion of tungsten and glass
particles with diameters in the range of $0.4\ldots4$ times the
Kolmogorov scale in a water tank with oscillating grids, albeit at
very low Reynolds number (the Taylor micro-scale Reynolds number was
in the range of $20\ldots30$). They mostly observed an enhancement of
the average settling velocity, except for their largest particles
settling at the highest turbulence intensity for which a reduction by 
several percent was determined. Unfortunately these authors did not
analyze particle clustering or preferential concentration. 
\citet{good:14} have performed measurements of sub-Kolmogorov size
water droplets in homogeneous-isotropic air turbulence generated by a
loudspeaker arrangement in a spherical enclosure. 
In this system the average settling velocity is reduced for
the smallest relative turbulence intensity (i.e.\ for the largest and
fastest settling particles). 

It has long been recognized that a non-linear drag law can lead to a 
reduction of the average particle settling velocity in unsteady
flows with zero mean \citep{tunstall:68,nielsen:93,wang:93}. 
In the framework of the one-way coupled point-particle approach,
\citet{good:14} have shown that imposing a non-linear drag law (instead
of the usual linear Stokes drag) can indeed lead to a reduction in
settling speed, in accordance with what is measured in their companion
experiments for the largest particles (which are still of
sub-Kolmogorov size). 
\citet{homann:13} have performed particle-resolved DNS of a single
sphere which steadily translates through homogeneous-isotropic
turbulence. They observe a significant increase in the average drag
coefficient, the increase scaling with the square of the relative
turbulence intensity (i.e.\ normalized with the translation
velocity). They also show that the observed scaling is consistent with
the non-linear drag mechanism, at least for low to moderate values of
this latter control parameter. 
\citet{rosa:16} have likewise investigated the effect of a non-linear drag law upon the settling behavior of one-way coupled point particles. Contrary to \citet{good:14}, however, they did not find any reduction in the mean partice settling velocity, despite careful consideration of the numerical set-up (e.g.\ the turbulence forcing scheme). One possible cause for the observed discrepancy is the duration of the sampling interval, which was significantly longer in the study of \citet{rosa:16}.
\citet{chouippe:15a} have performed the first particle-resolved
simulations of settling under forced homogeneous turbulence
conditions. In their moderate-size systems a slight reduction of the
average settling velocity has been observed for particles with a
diameter equal to approximately 7 Kolmogorov lengths and Galileo
number measuring 120. While preferential concentration was not
detected, it was found that the non-linear drag model of
\citet{homann:13} indeed provides a reasonable prediction for the
reduced settling speed. 

\citet{fornari:16a} and \citet{fornari:16b} have performed
particle-resolved DNS of forced homogeneous-isotropic turbulence with
several hundred nearly neutrally-buoyant particles (solid-to-fluid
density ratio less than $1.04$) with a diameter equivalent to
approximately 12 Kolmogorov length scales. 
While particle clustering was not detected, they observed a reduction
in the average settling velocity of up to
{55\%} percent at the smallest
value of the Galileo number
{(measuring 19), for which the relative turbulence intensity
  was approximately 5 times as large as the average settling velocity}.  
\citet{fornari:16a} perform a force analysis which shows that unsteady
forces are significant in their
{parameter range.} 
The analysis in the study of \citet{fornari:16b} suggests that the
horizontal velocity perturbations are mainly responsible for an 
average drag increase
{in this low-excess-density and
  high-turbulence-intensity case}, hence causing the reduced settling
speed. 

Finally let us mention the only available experimental data (to our
knowledge) on clustering of finite-size heavy particles in
approximately homogeneous-isotropic turbulence.
\citet{fiabane:12} have measured the positions of glass particles in 
water, while turbulence was forced through an array of impellers in an
icosahedral enclosure. For particle diameters equivalent to 2 to 5
times the Kolmogorov length and a Galileo number of approximately 13
the authors detected a strong tendency to cluster. Unfortunately,
however, the average settling velocity was not measured in that study. 

In the present study we are considering heavy particles settling under
gravity while a turbulent background flow is maintained via 
homogeneous-isotropic, large-scale, random forcing. We have performed
interface-resolved DNS for two cases with much higher turbulent
Reynolds number than previously investigated in \citet{chouippe:15a},
providing the largest-scale data-sets of this kind to date.
The present study specifically builds upon the earlier work of
\citet{uhlmann:14a} for ambient settling
and of \citet{uhlmann:16a} for particle-turbulence interaction without
gravity 
by maintaining most parameters at similar values, while
essentially varying the ratio between the turbulence intensity and the
terminal particle velocity.
In this fashion it is possible to analyze the particular influence of
the relative turbulence intensity upon the particle settling behavior
and upon their tendency to cluster.

{The present manuscript is organized as follows.
  In section~2 we describe the numerical methodology and the
  set-up of the simulations.
  Section~3 is devoted to the presentation of the
  fluid phase statistics, 
  while we analyze the settling velocity and the spatial distribution
  of the dispersed phase in detail in section~4.
  The different mechanisms for cluster formation and modification of
  the particle settling velocity are discussed in section~5 before
  closing with a summary and outlook.}

\section{Methodology}
% ######################################################################### %
We simulate the motion of multiple particles in forced homogeneous isotropic turbulence with the aid of the Immersed Boundary Method of \citet{Uhlmann_JCP2005} coupled to the turbulence forcing of \citet{Eswaran_CF1988}. This configuration has been previously investigated at lower Reynolds number in \citet{chouippe:15a} and in the absence of gravity in \citet{uhlmann:16a} while using the same methodology. 
The flow is governed by the incompressible Navier-Stokes equations (eq. \ref{eq:momentum_NS} and \ref{eq:continuity_NS}) where $\mathbf{u}$ denotes the fluid velocity, $\rho_f$ and $\nu$ its density and kinematic viscosity (which are both supposed to be constant). The forcing term $\mathcal{F}$ in the momentum equation accounts for the turbulence forcing $\mathbf{f}^{(t)}$ and the influence of the particles on the flow $\mathbf{f}^{(ibm)}$.
\begin{eqnarray}
   \frac{\partial \mathbf{u}}{\partial t}+ (\mathbf{u}\cdot\nabla)\mathbf{u}
   +\frac{1}{\rho_f} \nabla p
   & = &
   \nu \nabla^2\mathbf{u} + \boldsymbol{\mathcal{F}} \label{eq:momentum_NS}\\
   %
   % Continuity
   \nabla \cdot \mathbf{u} 
   & = & 
   0 \label{eq:continuity_NS}\\
   {\mathcal{F}}
   & = &
   \mathbf{f}^{(t)} + \mathbf{f}^{(ibm)}
\end{eqnarray}
The solver employs a fractional step method with an implicit treatment of the viscous term (Crank-Nicolson) and a three step Runge Kutta method for the nonlinear terms. The spatial discretization relies on a second order finite differences method on a staggered mesh with a uniform discretization in the three directions.
The forcing term $\mathbf{f}^{(ibm)}$ imposes the fluid velocity to be equal to the local particle velocity at the particle surface according to the Immersed Boundary Method of \citet{Uhlmann_JCP2005}.
The motion of the particles is computed by the time integration of the Newton equation for the linear and angular motion.
We focus here on very dilute regimes and treat collisions with a simple repulsive force mechanism \citep{glowinski:99}.
The background flow is forced by the large scale turbulence forcing of \citet{Eswaran_CF1988}. It consists in adding a forcing term $\mathbf{f}^{(t)}$ in the momentum equation that randomly forces the largest scales of the flow, and more specifically every wavenumber smaller than a given wavenumber $\kappa_f$. In the absence of buoyancy effects this would lead to the development of a homogeneous isotropic turbulent flow.
For more details on the forcing method and its implementation in physical space
in an IBM framework the reader is referred to \citep{chouippe:15a}.
\begin{table}[t]
   \small
   \centering
   \renewcommand{\arraystretch}{1.25}
   \setlength{\tabcolsep}{0.5ex}
   % ----------------------------------------------------------------------
   %\begin{tabular}{rccccccccccc}
   \begin{tabular}{rccccccccccc}
      \hline\noalign{\smallskip}
      case & $Ga$ & $D/\eta^{SP}$ & $D/\eta^{TP}$ 
           & $Re_\lambda^{SP}$ & $Re_\lambda^{TP}$
           & $\Phi_s$ 
           & $\rho_p/\rho_f$ 
           & $u_{rms}^{SP}/u_{g}$ & $u_{rms}^{TP}/u_{g}$ 
     & initial positions
     & source\\
      \noalign{\smallskip}\hline\noalign{\smallskip}
      {\bf G178-R95} & $178$ & $6.8$ & \typo{$1.5$}{$13.1$}
           & $94.76$ & $35.12$
           & $0.005$ 
           & $1.5$ 
           & $0.19$ & $0.22$
     & final state of G178
     & present\\
      {\bf G180-R140} & $180$ & $8.5$ & \typo{$2.6$}{$13.5$} 
           & $142.2$ & $64.37$
           & $0.005$ 
           & $1.5$ 
           & $0.29$ & $0.31$ 
     & random
     & present\\[1ex]
      G178 & $178$ & $-$ & \typo{$0.69$}{$6.1$} 
           & $0$ & $170.2$
           & $0.005$ 
           & $1.5$ 
           & $0$ & $0.23$ 
     & random
     & \citep{uhlmann:14a} \\
      G0-R120 & $0$ & $5.5$ & $5.5$ 
           & $119.0$ & $116.8$
           & $0.005$ 
           & $1.5$ 
           & $\infty$ & $\infty$ 
     & random
     & \citep{uhlmann:16a}\\
      R95 & $-$ & $-$ & $-$ 
           & $94.76$ & $-$
           & $0$ 
           & $-$ 
           & $-$ & $-$ 
     & $-$
     & present\\
     R140 %}
           & $-$ & $-$ & $-$ 
           & $142.2$ & $-$
           & $0$ 
           & $-$ 
           & $-$ & $-$ 
     & $-$
     & \citep{chouippe:15a} \\
     R120 
           & $-$ & $-$ & $-$ 
           & $119$ & $-$
           & $0$ 
           & $-$ 
           & $-$ & $-$ 
     & $-$
     & \citep{uhlmann:16a}\\
      \noalign{\smallskip}\hline
   \end{tabular}
   % ---
   \caption{Main physical parameters of the cases investigated in the present study: Galileo number $Ga$, length scale ratio $D/\eta$, Taylor-scale Reynolds number $Re_\lambda=\lambda u_{rms}/\nu$, global solid volume fraction $\Phi_s$, density ratio $\rho_p/\rho_f$, and resulting velocity ratio $u_{rms}/u_g$.   For each quantity  resulting from the flow characteristics ($\eta$, $Re_\lambda$,$u_{rms}$), the distinction is made between the value in single phase configuration and particle laden configuration with the respective SP and TP subscripts. Two types of initializations have been used for the particle positions: they are either randomly distributed in the domain or initialized with the final set of position of the case G178 which features columnar accumulation.}
   \label{table:physical_parameters}
\end{table}
\mbox{}\\

The main physical parameters of the dispersed and continuous phases are given in the table \ref{table:physical_parameters}. 
The characteristics of the background flow are given in terms of the Reynolds number $Re_\lambda$, Kolmogorov length scale $\eta$ and velocity scale of the turbulence $u_{rms}$.
We use the following notation to distinguish between particle and fluid velocity. $\mathbf{u}=(u_x,u_y,u_z)^T$ refers to the fluid velocity and $\mathbf{v^{(i)}}=(v^{(i)}_x,v^{(i)}_y,v^{(i)}_z)^T$ to the velocity of ith the particle.
We introduce the operator $\langle . \rangle_{\Omega_f}$ corresponding to space-averaging over regions occupied by the fluid, and use the decomposition
\begin{eqnarray}
   \mathbf{u}(\mathbf{x},t) 
   & = & 
   \langle \mathbf{u} \rangle_{\Omega_f} (\mathbf{x},t) 
   + \mathbf{u}^{'}(\mathbf{x},t) 
\end{eqnarray}
The dissipation $\varepsilon$ is then defined by
\begin{eqnarray}
   \varepsilon(t)
   & = &
   2\nu\langle S_{ij}^{'}S_{ij}^{'}\rangle_{\Omega_f},
   \;\;\;\;\;\; \text{with } \;
   S_{ij}=\frac{1}{2}\left( \frac{\partial u_{i}}{\partial x_j}
   +  \frac{\partial u_{j}}{\partial x_i} \right),
\end{eqnarray}
and the kinetic energy of the fluctuation $k$ as well as the characteristic velocity scale $u_{rms}$ according to
\begin{eqnarray}
   k(t) & = & \frac{1}{2}\langle \mathbf{u}^{'}\cdot\mathbf{u}^{'}\rangle_{\Omega_f} = \frac{3}{2} u_{rms}^2
\end{eqnarray}
From those quantities we define the Kolmogorov length scale $\eta=(\nu^3/\varepsilon)^{1/4}$, the Taylor microscale $\lambda=(15 \nu u_{rms}^2/\varepsilon)^{1/2}$, the corresponding Reynolds number $Re_\lambda=\lambda u_{rms}/\nu$, the large-eddy length scale $L=k^{3/2}/\varepsilon$, the large-eddy turnover time $\tau_e=u_{rms}^2/\varepsilon$, the Kolmogorov time scale $\tau_\eta=(\nu/\varepsilon)^{1/2}$. 
We will make the distinction between scales related to the corresponding single phase simulation (which is obtained with exactly the same forcing conditions, but no particles) and scales of the particle laden configurations with the subscripts SP and TP respectively.
In a single phase configuration the system would be parametrized by the Reynolds number $Re_\lambda$ while a multi-particle system requires additional four non-dimensional parameters. We propose to use the density ratio $\rho_p/\rho_f$ of the particle with respect to the fluid, the length scale ratio $D/\eta$ ($D$ being the diameter of the particles)
the solid volume fraction $\phi_s=N_p\mathcal{V}_p/\mathcal{L}_x\mathcal{L}_y\mathcal{L}_z$ (where $N_p$ is the number of particles, $\mathcal{V}_p=\pi D^3/6$ their volume, and $\mathcal{L}_i$ the size of the domain in the direction $i$), and the Galileo number defined by $Ga=u_gD/\nu$ with $u_g=(\lvert \rho_p/\rho_f-1\rvert \lvert \mathbf{g} \rvert D)^{1/2}$ the gravitational velocity scale.

The Galileo number can be thought of as a Reynolds number built with $u_g$.
In the case of one single spherical particle falling in a still fluid, its wake
is known to follow different regimes according to the Galileo number: for the
lowest Ga the particle fall straight and its wake is steady axisymmetric. Then
when Ga increases the wake loses its axisymmetry to become planar symmetric and the particle follows an oblique path. This regime is known as the steady oblique regime. A further increase of the Galileo number will make the regime become unsteady with the oblique oscillating regime and followed then by the chaotic regime \citep{Jenny_JFM2004}.
For a density ratio of $\rho_p/\rho_f$=1.5 and a Galileo number of 180 one particle would follow the steady oblique regime.
In the case of multi-particle settling, for a solid volume fraction $\phi_s=0.005$ this Galileo number is characterized by the formation of columnar clusters and an enhancement of the settling velocity \citep{uhlmann:14a,huisman:16}.

In the current paper we consider three types of configurations: 
they can be either purely affected by the settling, or purely by the turbulence, or affected by the interaction of both settling and turbulence.
The background turbulence is forced for the cases G178-R95, G180-R140 and G0-R120 with a Galileo number of 178 and 180 for the cases G178-R95, G180-R140, and no influence of gravity ($Ga=0$) for G0-R120.
In order to estimate the influence of the particle induced turbulence we also
included the case G178 corresponding to multiple particle settling in an
initially ambient flow. This case has been previously characterized in the former work of \citet{uhlmann:14a}, and is characterized by the formation of columnar clusters, large scale columnar fluid structures and an enhancement of the settling velocity. 
The parameters of the cases G178 and G178-R95 differ therefore only by the nature of the background turbulence: while cases G178 features only particle-induced turbulence (also called pseudo-turbulence \citep{lance:91,Riboux_JFM2010,risso:11}), the background flow of G178-R95 results in the interaction of the forced and the particle-induced turbulence.
Cases G178-R95 and G180-R140 differ by the scales of the resulting forced turbulence that would be obtained in a single-phase configuration and consequently by the turbulence intensity $I$ which we define as the ratio between the characteristic velocity $u_{rms}$ and settling velocity of one single particle falling in a still fluid. 
\begin{table}[t] % Physical parameters
   \small
   \centering
   \renewcommand{\arraystretch}{1.25}
   \setlength{\tabcolsep}{1ex}
   % ----------------------------------------------------------------------
   \begin{tabular}{rccc}
      \hline\noalign{\smallskip}
      Name & $\kappa_f/\kappa_1$ & $T_L\nu/\mathcal{L}_x^2$ 
           & $\epsilon^{*}\mathcal{L}_x^4/\nu^3$\\
      \noalign{\smallskip}\hline\noalign{\smallskip}
      G178-R95 and R95 & $3.91$ & $1.00 \times 10^{-4}$ 
           & $5.14 \times 10^{8}$\\
     G180-R140 
     and R140 
           & $2.5$ & $5.94 \times 10^{-5}$ 
           & $3.52 \times 10^{9}$\\
     G0-R120 
     and R120 
           & $3.61$ & $5.37 \times 10^{-5}$ 
           & $1.31 \times 10^9$\\
      \noalign{\smallskip}\hline
   \end{tabular}
   % ---
   \caption{Imposed parameters used for the forcing of the background turbulence with the method of \cite{Eswaran_CF1988} with the notation of \cite{chouippe:15a}: forcing cutoff wavenumber $\kappa_f$ normalized by the smallest discrete wavenumber $\kappa_1$ in the horizontal direction, characteristic time of the random forcing $T_L$, dissipation-rate parameter $\varepsilon^{*}$}
     \label{table:forcing_parameters}
\end{table}

The turbulence forcing is parametrized by the largest wavenumber forced, $\kappa_f$, by a timescale $T_L$ and a coefficient $\varepsilon^{*}$ which controls the amplitude of this forcing \citep{Eswaran_CF1988,chouippe:15a}. The coefficients used for the different cases are listed in a dimensionless form in table \ref{table:forcing_parameters}.

\begin{table}[t] 
   \small
   \centering
   \renewcommand{\arraystretch}{1.25}
   \setlength{\tabcolsep}{0.55ex}
   % ----------------------------------------------------------------------
   \begin{tabular}{rcccccccccc}
      \hline\noalign{\smallskip}
      Name & $N_p$ & $N_x \times N_y \times N_z$ 
           & $\mathcal{L}_z/\mathcal{L}_x$ & $L^{SP}/\mathcal{L}_x$
           & $\eta^{SP}/\Delta x$ & $\mathcal{L}_x/D$ & $D/\Delta x$ & $T_{obs}/\tau_e^{SP}$
           & $T_{obs}/\tau_p$ & $T_{obs}/\tau_g$ \\
      \noalign{\smallskip}\hline\noalign{\smallskip}
      {\bf G178-R95} & $11867$ & $2048 \times 2048 \times 4096$ 
           & $2$ & $0.39$
           & $3.57$ & $85$ & $24$ & $8.7$
           & $53.28$ & $837$\\
      {\bf G180-R140} & $11868$ & $2048 \times 2048 \times 4096$ 
           & $2$ & $0.56$
           & $2.82$ & $85$ & $24$ & $14.9$
           & $90.64$ & $1350$\\[1ex]
      G178  & $11867$ & $2048 \times 2048 \times 4096$ 
           & $2$ & $-$
           & $-$ & $85$ & $24$ & $-$
           & $98.98$ & $1472$\\
      G0-R120 & $20026$ & $2048 \times 2048 \times 2048$
           & $1$ & $0.43$
           & $2.89$ & $128$ & $16$ & $25.0$
           & $460.64$ & $-$\\
      R95 & $0$ & $1024 \times 1024 \times 1024$ 
           & $1$ & $0.39$
           & $3.57$ & $-$ & $-$ & $28.5$
           & $-$ & $-$\\
     R140 
           & $0$ & $512 \times 512 \times 512$ 
           & $1$ & $0.56$
           & $0.704$ & $-$ & $-$ & $80.5$
           & $-$ & $-$\\
     R120 
           & $0$ & $512 \times 512 \times 512$
           & $1$ & $0.43$
           & $0.73$ & $-$ & $-$ & $24.3$
           & $-$ & $-$\\
      
      \noalign{\smallskip}\hline
   \end{tabular}
   % ---
   \caption{Main numerical parameters used for the present simulations: Number of particles $N_p$, extent $\mathcal{L}_i$ of the domain in the direction $i$ and corresponding number of Eulerian grid points $N_i$, 
elongation of the domain in the vertical direction $\mathcal{L}_z/\mathcal{L}_x$, ratio between the single phase large-eddy length scale and box size $L^{SP}/\mathcal{L}_x$ and between the single phase Kolmogorov length scale and grid width $\eta^{SP}/\Delta x$, observation time $T_{obs}$ normalized by the single phase large-eddy time scale $\tau_e^{SP}$, particle time scale $\tau_p=D^2\rho_p/(18\nu\rho_f)$ and gravitational time scale $\tau_g=D/u_g$. 
For all cases the length of the domain in both horizontal directions are set to be equal ($\mathcal{L}_x=\mathcal{L}_y$).
}
     \label{table:numerical_parameters}
\end{table}

The numerical parameters are listed in table
\ref{table:numerical_parameters}. 
We use periodic boundary conditions in the three directions and the 
cases with non-zero Galileo numbers have a computational domain elongated in the vertical direction in order to reduce the effect of periodicity in the presence of vertically-elongated structures; this point will be further discussed in section \ref{sec:General_Stats}. 
For those cases we used a discretization at the particle scale $D/\Delta x=24$ in accordance to the former work of \citet{Uhlmann_IJMF2014} and \citet{uhlmann:14a}, while the case without gravity uses $D/\Delta x=16$ \citep{uhlmann:16a}.

\begin{figure}
   \begin{minipage}{0.29\linewidth}
      \centerline{(a)}
      \includegraphics[width=1.0\linewidth]
         {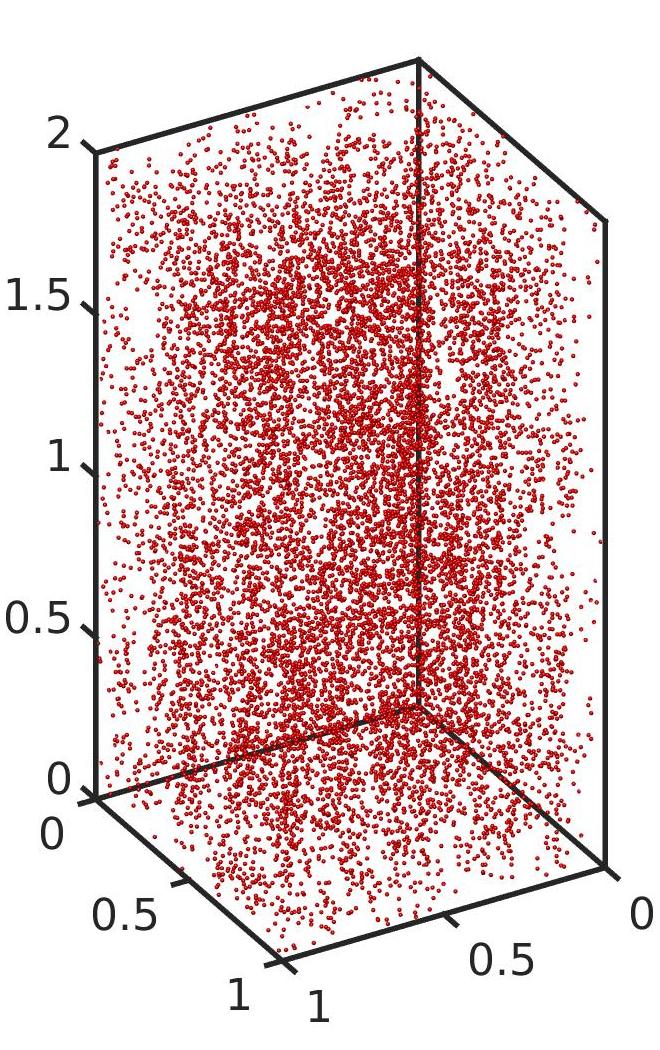}\\
   \end{minipage}
   \begin{minipage}{0.2\linewidth}
      \centerline{(b)}
      \includegraphics[width=1.0\linewidth]
         {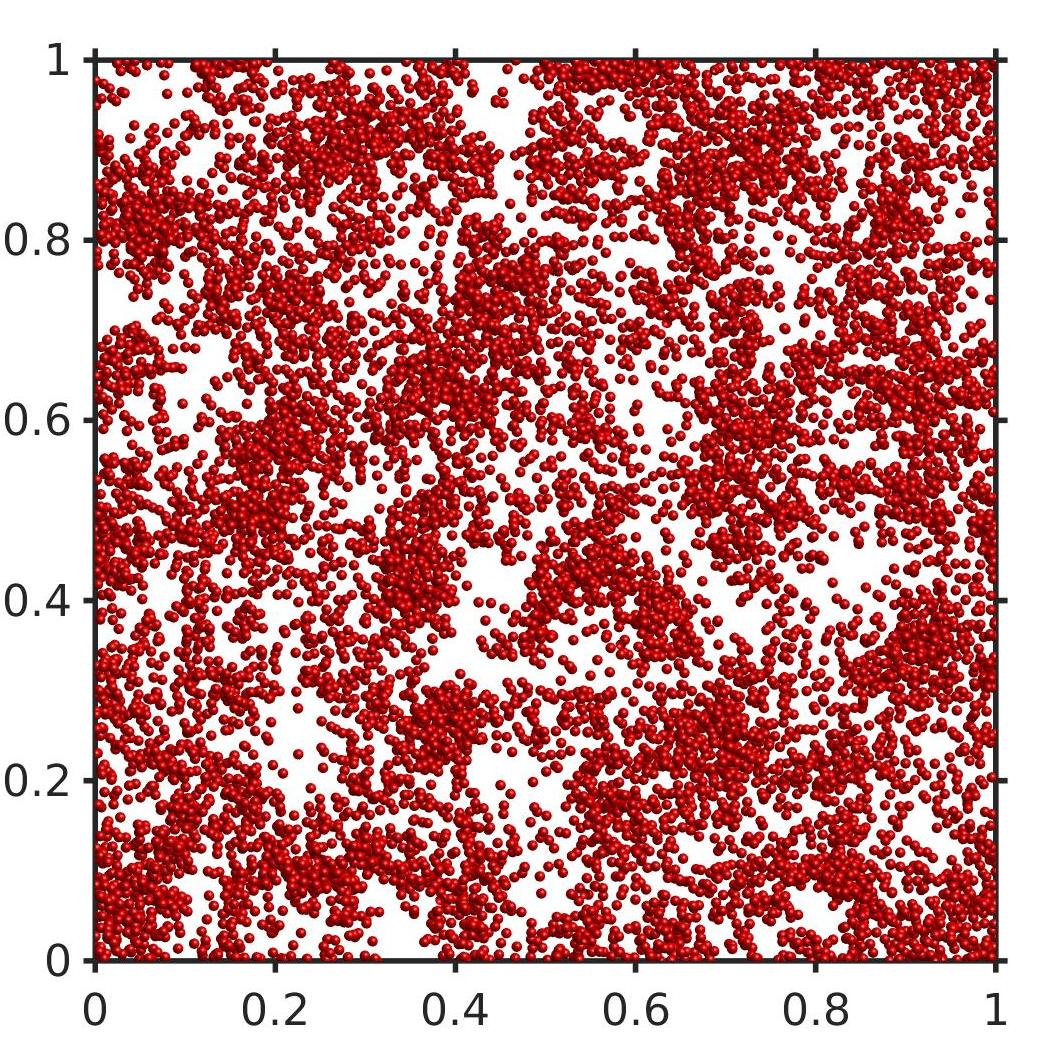}\\
      \centerline{(c)}
         \includegraphics[width=1.0\linewidth]
         {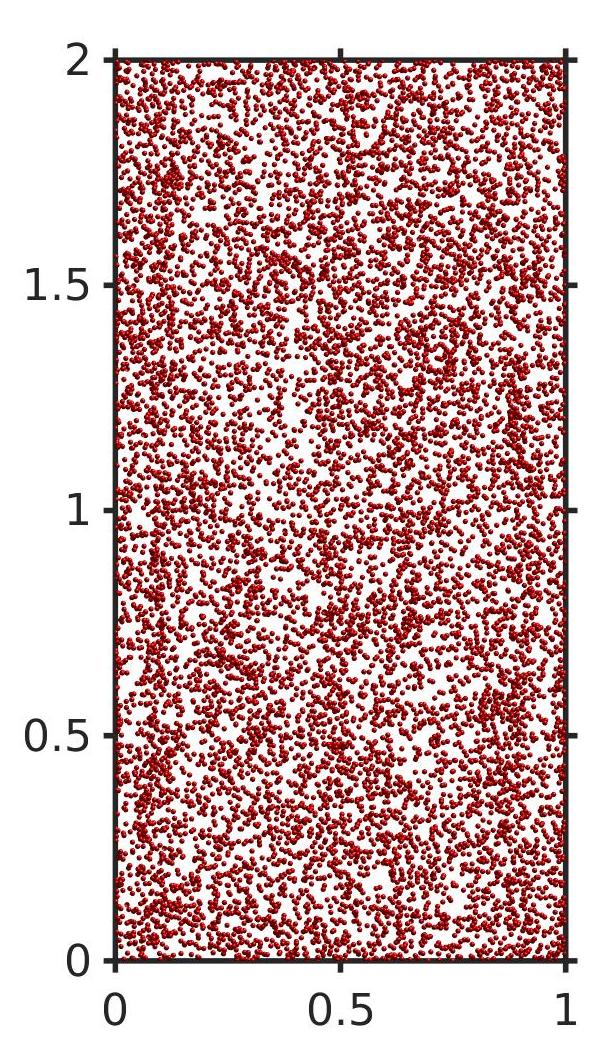}\\
   \end{minipage}
   % -----------------
   \begin{minipage}{0.29\linewidth}
      \centerline{(d)}
      \includegraphics[width=1.0\linewidth]
         {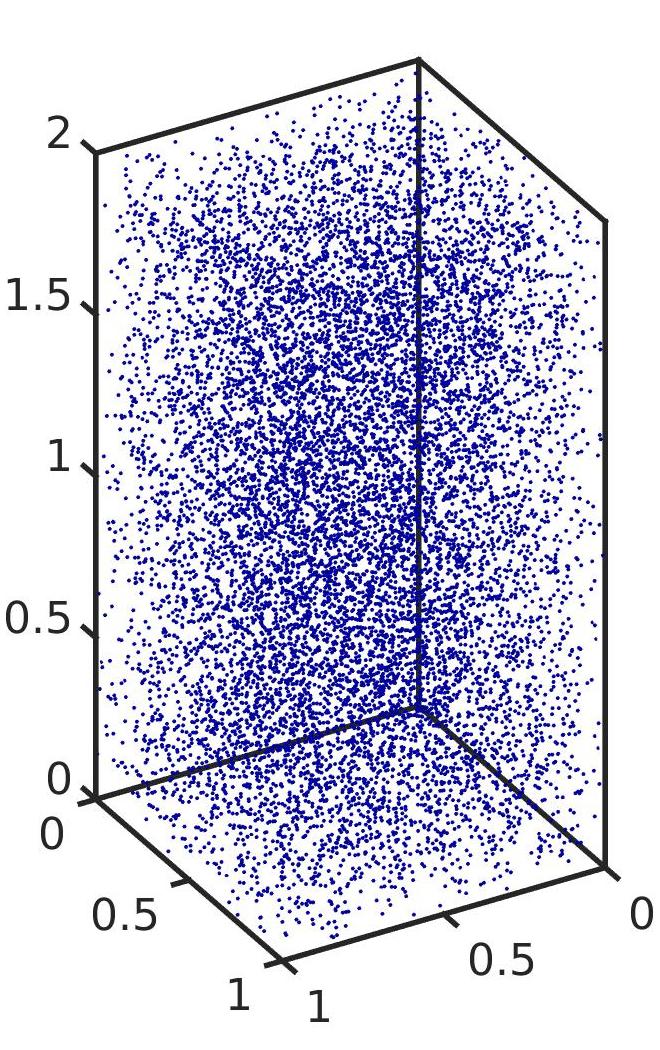}\\
   \end{minipage}
   \begin{minipage}{0.2\linewidth}
      \centerline{(e)}
      \includegraphics[width=1.0\linewidth]
         {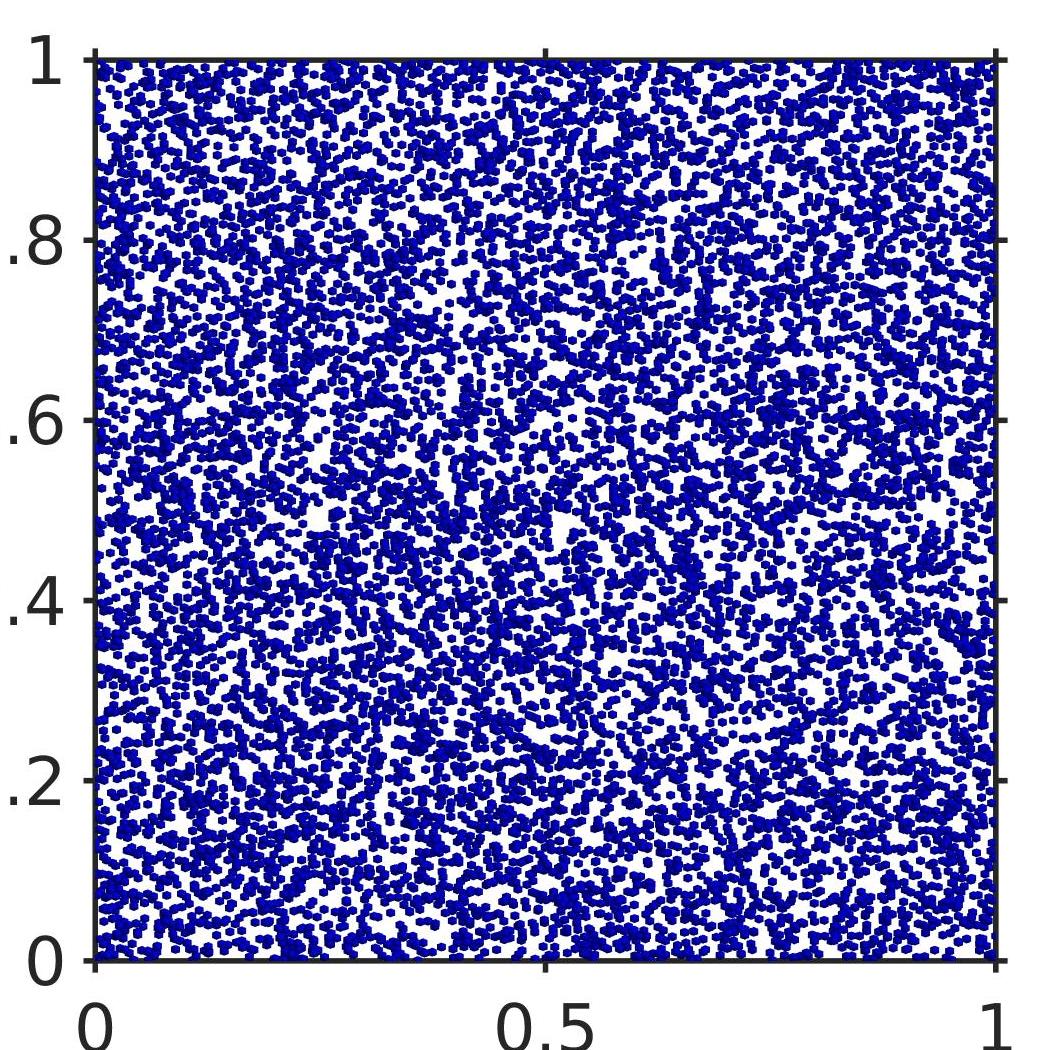}\\
      \centerline{(f)}
         \includegraphics[width=1.0\linewidth]
         {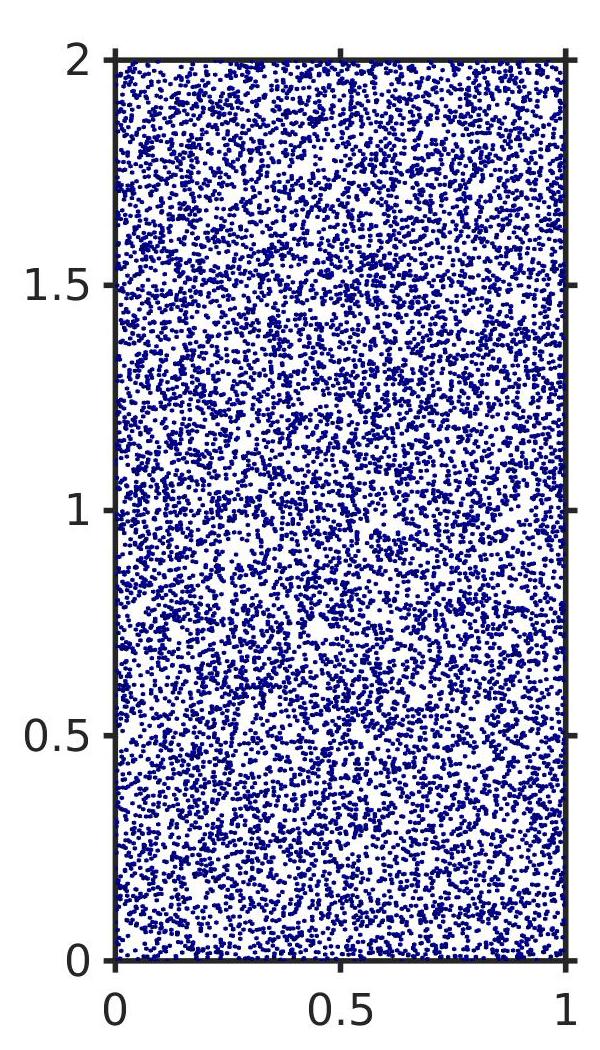}\\
   \end{minipage}
%---
\caption{Visualization of particle positions for both types of initial state used in the current study: (a-c) columns (corresponding to the final state of the case G178) used in G178-R95, (d-f) Randomly distributed particles used in G180-R140. }
\label{fig:initial_positions}
\end{figure}

The particles of the cases G178, G180-R140, and G0-R120 are initially randomly distributed. 
We tested the resistance of the clusters formed in G178 to turbulence by using the final state of G178 as an initial state for G178-R95. 
Figure \ref{fig:initial_positions} gives a visual impression of both types of initial positions (columnar vs random).
In G178-R95 the system starts with developed columns of particles and the associated particle-induced turbulence, then the turbulence is activated. In a single phase case, turbulence is known to become homogeneous isotropic after a transient phase of typically 3 large-eddy time scales \citep{Eswaran_CF1988,chouippe:15a} and we ensured our computations to have an observation time larger than $8\tau_e$.
In the case G180-R140 we proceeded conversely: turbulence is forced until statistically steady state is reached, then we add the particle and keep them fixed with respect to the computational grid while adding a constant positive vertical velocity $w_{sh}$ to the initial flow. During this phase the space resolution is low $D/\Delta x=12$, then we increase the resolution to $D/\Delta x=24$ by linear interpolation and release the particles.
For both cases G178-R95 and G180-R140 the turbulence forcing is
adapted to account for this shift in the vertical component of the
velocity \citep{chouippe:15a}.
The cases with $2048^2\times4096$ grid points required roughly 10 million core-hours each on 8000 processor cores at LRZ Munich, Germany.
Apart from differing in the initial condition, the cases G178-R95 and G180-R140 also differ by the Reynolds number of the forced turbulence, and, as a consequence, the resulting turbulence intensity $u_{rms}/V_T$  differs from 0.14 to 0.22, and $D/\eta$ is slightly different (6.8 vs 8.5 cf. tables \ref{table:physical_parameters} and \ref{table:velocity_scaled}).
%
% ------------------------------------------------------------------------- %

\section{
  Description of the fluid phase
  % }
}
\label{sec:General_Stats}
% ######################################################################### %
We discuss in this section the behaviour of the fluid phase, focusing
first on the evolution of the energy budget and the modification of
the scales of the background flow, and then moving on to the
description of the coherent flow structures. 
% ------------------------------------------------------------------------- %
\subsection{Influence of the particles on the background flow}
\begin{figure}
   \begin{minipage}{2ex}
      \rotatebox{90}{\centerline{Budget}}
   \end{minipage}
   \begin{minipage}{0.45\linewidth}
      \centerline{(a)}
      \includegraphics[width=0.95\linewidth]
         {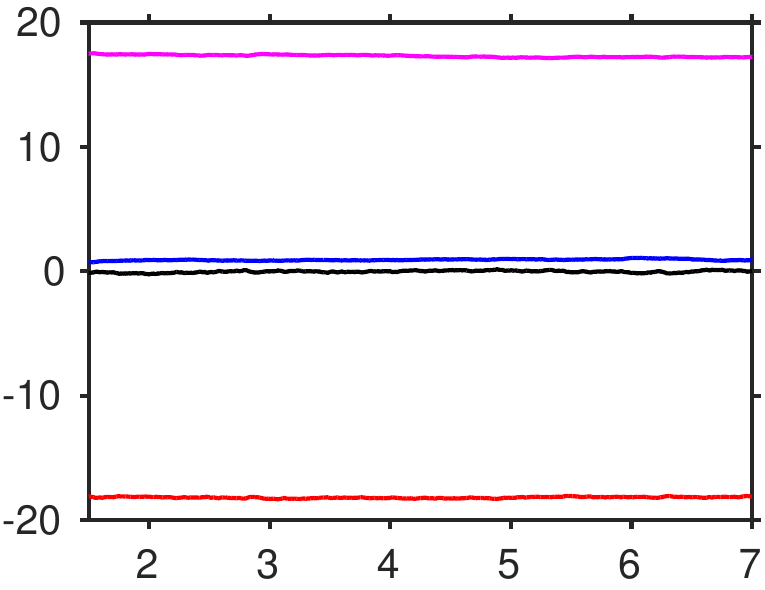}\\
      \centerline{$t/\tau_e$}
   \end{minipage}
   \begin{minipage}{2ex}
      \rotatebox{90}{\centerline{Budget}}
   \end{minipage}
   \begin{minipage}{0.45\linewidth}
      \centerline{(b)}
      \includegraphics[width=0.95\linewidth]
         {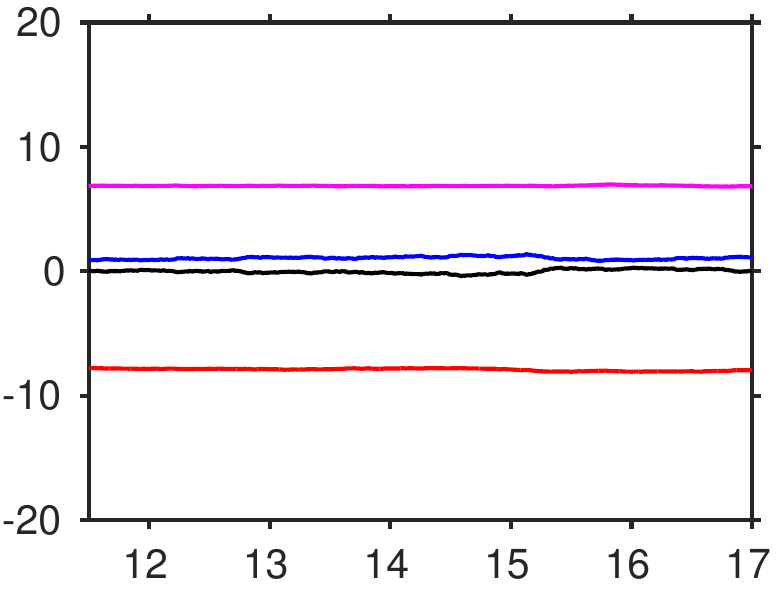}\\
      \centerline{$t/\tau_e$}
   \end{minipage}
   %
%---
\caption{Time evolution of the different terms in the budget of the volume averaged kinetic energy: $0=-\mbox{d}\left< E_k \right > _{\Omega}/\mbox{d}t-\varepsilon_\Omega+\Psi^{(t)}+\Psi^{(p)}$ scaled by the dissipation of the corresponding single phase simulation for case G178-R95 (a) and G180-R140 (b). 
 Linestyle: $\solidthick$ $(-\mbox{d}\left<E_k\right>_{\Omega}/\mbox{d}t)$, $\color{red}{\solidthick}$ $(-\varepsilon_\Omega)$, $\color{blue}{\solidthick}$ $\Psi^{(t)}$, $\color{magenta}{\solidthick}$ $\Psi^{(p)}$}
\label{fig:energy_balance}
\end{figure}

% ------------------------------------------------------------------------- %
We propose to focus first on the energy budget in order to discuss how particles affect the global scales of the flow. For this we introduce the instantaneous kinetic energy $E_k(t)=\mathbf{u}\cdot\mathbf{u}/2$. Its transport equation can be derived from the momentum equation \ref{eq:momentum_NS} and then integrated over the entire domain $\Omega$ (occupied by both fluid and particles) with the space averaging operator $\langle . \rangle_{\Omega}$ \citep{chouippe:15a}.
This leads to: 
\begin{eqnarray}
   \frac{\mathrm{d}\left<E_k\right>_\Omega}{\mathrm{d}t} 
      &=&
      -\varepsilon_\Omega
      +\psi^{(t)}+\psi^{(p)} \label{eq:budget}
\end{eqnarray}
where $\varepsilon_\Omega(t)=2\nu\langle S_{ij}^{'}S_{ij}^{'}\rangle_{\Omega}$ is the instantaneous box-averaged dissipation, 
$ \psi^{(t)}(t)= \left< \mathbf{u} \cdot \mathbf{f}^{(t)}\right>$ the work done by the turbulence forcing and $\psi^{(p)}$ the fluid-particle coupling term (eq. \ref{eq:fluid_particle_coupling}).
\begin{eqnarray}
  \psi^{(p)}(t)&=&\left<\mathbf{u}\cdot\mathbf{f}^{(ibm)}\right>_\Omega
        -\left<\mathbf{u}\right>_\Omega\cdot\left<\mathbf{f}^{(ibm)}\right>_\Omega
        \label{eq:fluid_particle_coupling}
\end{eqnarray}
% ------------------------------------------------------------------------- %
Figure \ref{fig:energy_balance} gives the time evolution of the
different terms in the budget of the volume averaged kinetic energy,
scaled by the equivalent single phase dissipation $\varepsilon^{SP}$.
Our previous works \citep{chouippe:15a,uhlmann:16a}
showed that in the absence of gravity, and for $D/\eta$ and $\Phi_s$ of similar range as the values explored in the current paper, the budget is barely affected by the presence of the particles and the system can be consequently seen as macroscopically one-way coupled.
For a given non-zero Galileo number we observe here that the energy budget, and more specifically the dissipation, gets more affected while decreasing $D/\eta$.
This can be emphasized by developing the fluid-particle forcing term into three contributions \citep{chouippe:15a}:
\begin{eqnarray}
   \psi^{(p)}(t)
   & = &
   \phi_s \left( \frac{\rho_p}{\rho_f} -1 \right) \mathbf{u}_{rel,\Omega} \cdot \mathbf{g}
         + \psi_{accel}^{(p)}(t) + \psi_{coll}^{(p)}(t)
         \,,
   \label{eq:phi_p_decomposition}
\end{eqnarray}
with the apparent slip velocity $\mathbf{u}_{rel,\Omega}=\langle
\mathbf{v}^{(i)} \rangle_p- \langle \mathbf{u} \rangle_\Omega$ (where the
operator $\langle . \rangle_p$ refers to the average over the particles), $\psi_{accel}^{(p)}$ the contribution due to particle acceleration and $\psi_{coll}^{(p)}$ the contribution due to inter-particle collisions.
The reader is referred to the appendix of \citet{chouippe:15a} for
more details on the derivation of eq. \ref{eq:phi_p_decomposition}. 
In the dilute regime, the term $\psi_{accel}^{(p)}(t) + \psi_{coll}^{(p)}(t)$ is negligible compared to $\psi_{pot}^{(p)}(t)=\phi_s \left( \frac{\rho_p}{\rho_f} -1 \right) \mathbf{u}_{rel,\Omega} \cdot \mathbf{g}$, which can be normalized according to
\begin{eqnarray}
   \frac{\psi_{pot}^{(p)}}{\varepsilon^{SP}}
   & = &
   \phi_s Ga^3 \left(\frac{D}{\eta^{SP}}\right)^{-4}\frac{w_{rel,\Omega}}{u_g}
\end{eqnarray}
Interestingly, if we propose to estimate, as a first level of
approximation, the relative velocity $w_{rel} \approx V_T^{SN}$ with
$V_T^{SN}$ the terminal velocity of one single particle estimated
according to the drag law of Schiller and Naumann \cite{Schiller_1935}
it turns to give $\psi^{(p)}/\psi_{pot}^{SN} \sim 0.94 (0.92)$ for G178-R95 (G180-R140).
As $\phi_s$, $Ga$, $V_T^{SN}/u_g$ have the same order of magnitude for both cases, this shows the importance of $D/\eta^{SP}$ on the enhancement of the dissipation level.
The estimate that the average settling velocity is equal to the terminal settling velocity of a single particle in ambient fluid ($w_{rel}\approx V_T^{SN}$) is reasonable; below we will find that for relative turbulence intensities around 0.2 deviations of the order of one percent are observed, while in the absence of background turbulence the difference amounts to 12 percent (cf.\ section \ref{sec:settling_velocity}).

%
% ---------------------------------------------------------------------------
Once averaged over the domain filled by the fluid we get a ratio for the two-phase dissipation $\varepsilon_{\Omega_f}^{TP}/\varepsilon^{SP}$ of the order of 18 for G178-R95 and 6.4 for G180-R140.
We observe that the turbulence forcing power input is barely affected by the presence of the particles meaning that the main discrepancy is carried by $\psi^{(p)}$.
We introduce a surrogate dissipation $\tilde{\varepsilon}^{SP}=\varepsilon_{\Omega}^{TP}-\psi^{(p)}$, and observe that it equals $0.88\varepsilon^{SP}$  for G178-R95 and $1.02\varepsilon^{SP}$ for G180-R140.
As a first level of approximation the increase of the dissipation to be observed on figure \ref{eq:budget} would follow $\varepsilon^{SP} \sim \varepsilon_\Omega^{TP}-\psi_{pot}^{SN}$ which attributes this increase to the release of potential energy of the particles during the settling, in accordance with the former observations of \citet{chouippe:15a} and \citet{Hwang_JFM2006}.

\begin{figure}
   \centering
   \begin{minipage}{2ex}
      \rotatebox{90}{\centerline{$\mathcal{R}_{ww}$}}
   \end{minipage}
   \begin{minipage}{0.45\linewidth}
      \centerline{(a)}
      \includegraphics[width=0.95\linewidth]
         {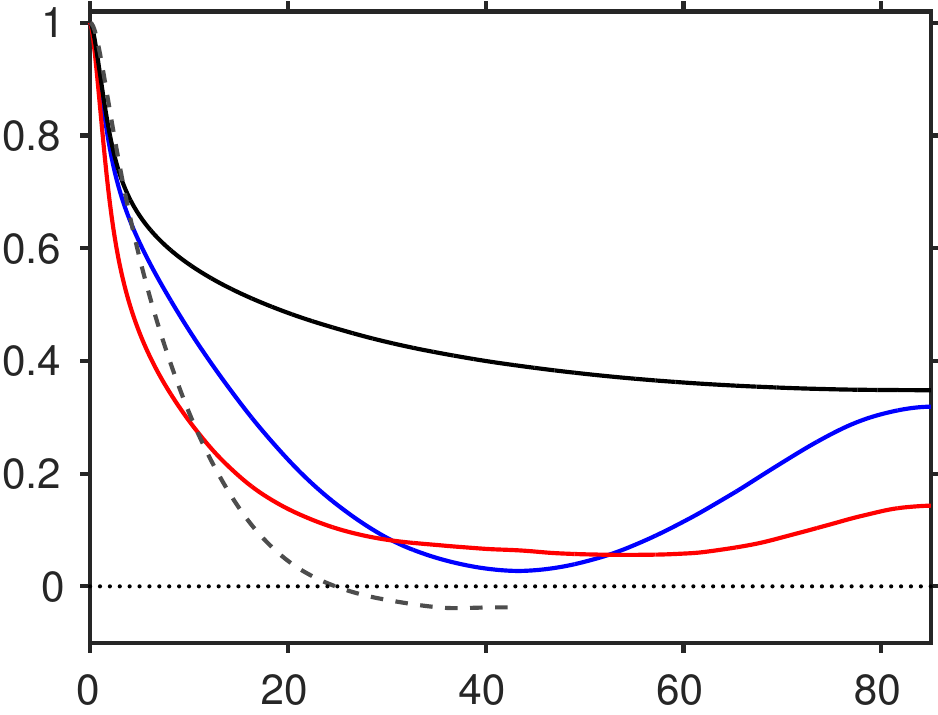}\\
      \centerline{$r_z/D$}
   \end{minipage}
   %
%---
\caption{(a) Longitudinal two-point correlation function in the z-coordinate direction, $\mathcal{R}_{ww}(r_z)$. 
Linestyle:  
$\color{red}{\solidthick}$ G178-R95, 
$\color{blue}{\solidthick}$ G180-R140, 
${\solidthick}$ G178, 
$\dashed$ R95 }
\label{fig:two_point_corr}
\end{figure}

%
% ---------------------------------------------------------------------------
Figure \ref{fig:two_point_corr} gives the evolution of the autocorrelation of
the fluid velocity fluctuations in the z-coordinate
direction.
In
\citet{chouippe:15a}
we have seen that the single phase
configuration with elongated domains in the vertical direction follows its cubic
equivalent up to distances equal 
to the maximal separation in the direction x ($\mathcal{L}_x/2$) 
where we observe almost no correlation, then the
autocorrelation follows a monotonic increase until the maximal separation. This
behavior is to be interpreted as a signature of the turbulence forcing that
injects energy for every two modes in the z-direction, and is conserved in the
turbulent particle-laden cases: $\mathcal{R}_{ww}$ first decreases until
$\mathcal{L}_x/2$ where the correlation is weak (this point is consistent with
the single phase case R95), and then increases monotonically until the maximal separation where we observe that the flow is not fully decorrelated.
The successive decrease and increase can hence be attributed to the forcing
scheme, but the amount of correlation at the maximal separation is also a
consequence of the wakes influence. Indeed in the pure multiparticle case G178 we observe that the flow does not fully decorrelate \citep{uhlmann:14a}.
The level of correlation at $\mathcal{L}_x/2$ also monotonically increases with $u_{ver,rms}^{TP}/u_g$, which is counter-intuitive as one would expect turbulence to decrease the influence of the wakes. 
This trend can still be seen as a signature of particles, and more specifically to the formation of clusters, which will be discussed in the next section, since we observe that the level of autocorrelation actually increases with the clustering level.
\citet{uhlmann:14a} have suggested to increase the domain size in
the vertical direction to decrease this correlation, but they also
estimated that even an elongation of $\mathcal{L}_z/\mathcal{L}_x=8$
would be insufficient to resolve this issue.  

% ---------------------------------------------------------------------------
\begin{table}[bt]
   \small
   \centering
   \renewcommand{\arraystretch}{1.25}
   \setlength{\tabcolsep}{1ex}
   % ----------------------------------------------------------------------
   \begin{tabular}{rccccccc}
      \hline\noalign{\smallskip}
      Name 
           & $u_{rms}^{SP}/V_T$ 
           & $u_{rms}^{TP}/V_T$
           & $u_{rms,hor}^{TP}/V_T$
           & $u_{rms,ver}^{TP}/V_T$
           & $V_T^{turb}/V_T$
           & $v_{rms,hor}/V_T$
           & $v_{rms,ver}/V_T$ \\
      \noalign{\smallskip}\hline\noalign{\smallskip}
      {\bf G178-R95} 
           & $0.14$ 
           & $0.16$
           & $0.14$
           & $0.20$
           & $1.011$
           & $0.19$
           & $0.18$ \\
      {\bf G180-R140} 
           & $0.22$ 
           & $0.23$
           & $0.22$
           & $0.25$
           & $0.994$
           & $0.30$
           & $0.22$ \\[1ex]
     G178  
           & $-$ 
           & $0.17$
           & $0.09$
           & $0.27$
           & $1.12$
           & $0.08$
           & $0.26$ \\
      \noalign{\smallskip}\hline
   \end{tabular}
   % ---
   \caption{Scales of the fluid velocity $\mathbf{u}$ and the particle velocity $\mathbf{v}$, scaled by the settling velocity of one particle falling in quiescent fluid $V_T$. The velocity $V_T^{turb}$ corresponds to the terminal velocity for multiparticle settling in forced turbulence
     $\langle \lvert w_{rel} \rvert \rangle_t$. 
 }
   \label{table:velocity_scaled}
\end{table}

We now turn to the influence of the forced turbulence and the wakes on
the scales of the flow velocity. Two contributions are here in
competition as noted by previous authors
\citep{Hwang_JFM2006, Tanaka_PRL2008, chouippe:15a}: on the one hand the increase of the dissipation due to the gradients at the fluid-particle interface should tend to decrease the level of turbulent kinetic energy, but on the other hand the development of the wakes should induce the formation of small scale structures which increases velocity fluctuations.
Table \ref{table:velocity_scaled} gives the values for the ratios between the fluctuating velocities and settling velocity for both fluid and particle phases. 
We first observe that particles at the current Galileo numbers tend to globally enhance the fluctuating velocity as $u_{rms}^{TP}$ is larger than $u_{rms}^{SP}$ for both G178-R95 and G180-R140 cases. We choose to split the amount of fluctuations into its horizontal and vertical fluctuations, respectively $u_{rms,hor}^{TP}=[0.5((u_x^{'})^2+(u_y^{'})^2)]^{1/2}$ and $u_{rms,ver}^{TP}=((u_z^{'})^2)^{1/2}$.
We observe that the horizontal contribution is of the same order as its single phase equivalent, meaning that the wakes mostly affect the vertical component of velocity fluctuations.
Besides the turbulent cases feature lower anisotropy levels than in the pure multi-particle settling, with a ratio $u_{rms,ver}^{TP}/u_{rms,hor}^{TP}\approx 1.61$ for G178-R95 and 1.19 for G180-R140, while $u_{rms,ver}^{TP}/u_{rms,hor}^{TP} \approx 9.90$ for the original Ga178 configuration.

Despite this clear anisotropy of the flow we use the standard definitions of the Kolmogorov length scale, Taylor micro scale, fluctuating velocity and associated Reynolds number. 
The increase of the dissipation and the formation of small scale structures in
the wakes leads on one side to an increase of the Kolmogorov length scale $\eta^{TP}$, but on the other side hardly affects the level of turbulent kinetic energy. 
This point can be partially attributed to the fact that the level of fluctuations induced by the particles $u_{rms}^{TP}$ of the pure settling case G178 is actually of the same order as the level of fluctuations $u_{rms}^{SP}$ used to force turbulence in the case G178-R95. The amount of particle-induced turbulence is therefore probably not strong enough to induce by itself larger modifications of the fluctuations amplitude.
We also introduced two Taylor micro scale deduced from the transverse autocorrelation functions, namely $\lambda_{hor}$ and $\lambda_{ver}$ respectively related to $\mathcal{R}_{uu}(r_z)$ and $\mathcal{R}_{ww}(r_x)$. 
Based on this we introduce two separate Reynolds numbers to account for the anisotropy, namely:
\begin{eqnarray}
   Re_{\lambda,hor}^{TP} & = & \frac{\lambda_{hor} \; u_{rms,hor}^{TP}}{\nu} 
   \label{eq:Re_hor}   \\
   Re_{\lambda,ver}^{TP} & = & \frac{\lambda_{ver} \; u_{rms,ver}^{TP}}{\nu}
   \;.   \label{eq:Re_ver}
\end{eqnarray}
The resulting Reynolds numbers are listed in table \ref{tab:Re_lambda_hor_ver}.\\
% ------------------------------------------------------------------------- %
\begin{table}[h]
   \small
   \centering
   \renewcommand{\arraystretch}{1.25}
   \setlength{\tabcolsep}{1ex}
   % -----
   \begin{tabular}{rcc}
      \hline\noalign{\smallskip}
      Name & $Re_{\lambda,hor}^{TP}$ & $Re_{\lambda,ver}^{TP}$ \\
      \noalign{\smallskip}\hline\noalign{\smallskip}
      G178-R95 & $90.00$ & $63.03$ \\
      G180-R140 & $168.52$ & $107.50$ \\
      G178 & $31.64$ & $108.55$ \\
      \noalign{\smallskip}\hline
   \end{tabular}
   % -----
   \caption{Two-Phase Reynolds numbers computed according to equations \ref{eq:Re_hor} and \ref{eq:Re_ver} for which the Taylor microscale has been deduced from the transverse two-point correlations. The distinction is also made for the characteristic fluctuating velocities in the horizontal and vertical directions.}
   \label{tab:Re_lambda_hor_ver}
\end{table}
% ------------------------------------------------------------------------- %

We observe that the Taylor microscale in the horizontal direction are closer to the single phase value while it is closer in the vertical direction to the pure multiparticle settling configuration, resulting in an increase of the Taylor microscale in the horizontal direction and a decrease in the vertical direction. We observed that the presence of the particles barely affects the amplitude of the horizontal fluctuations and tend to increase this amplitude in the vertical direction. This explains that the presence of the particles tends to increase the Reynolds numbers in the horizontal direction and decreases them in the vertical direction.
%
% ------------------------------------------------------------------------- %
\subsection{Large and small scale structures of the flow}
\begin{figure}
   \begin{minipage}{2ex}
      \rotatebox{90}{\centerline{p.d.f.}}
   \end{minipage}
   \begin{minipage}{0.45\linewidth}
      \includegraphics[width=0.95\linewidth]
         {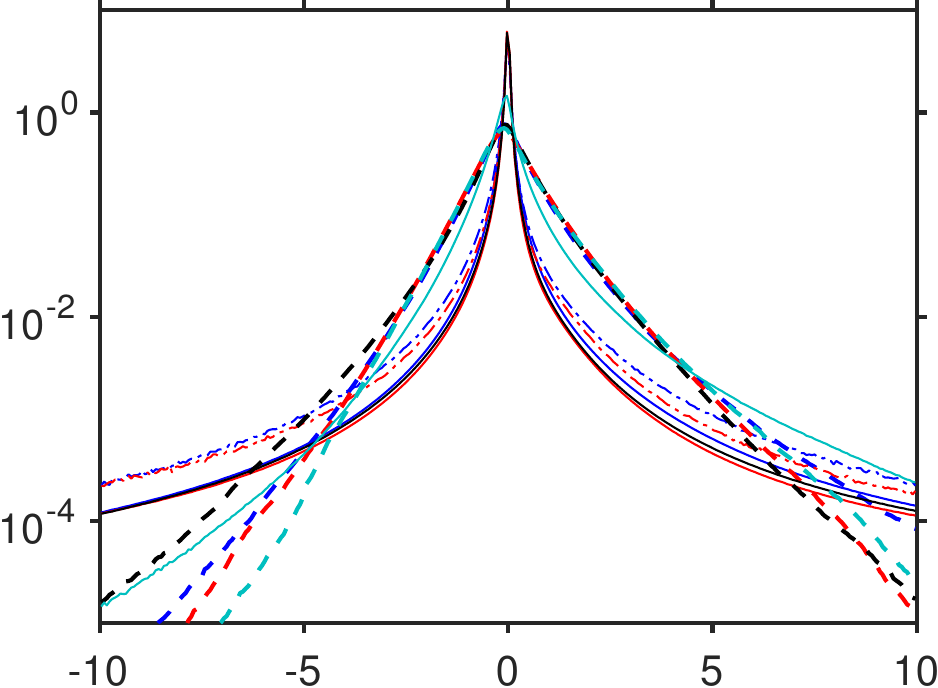}\\
      \centerline{$Q/\sigma(Q)$}
   \end{minipage}
   \begin{minipage}{0.05\linewidth}
       $\;$
   \end{minipage}
   \begin{minipage}{0.4\linewidth}
   % ----------------------------------------------------------------------
   \begin{tabular}{rcc}
      \hline\noalign{\smallskip}
      case & $\sigma(Q)/\omega_{rms}$ & $\sigma(Q^{filt})/\omega_{rms}$ \\
      \noalign{\smallskip}\hline\noalign{\smallskip}
      {\bf G178-R95} 
         & $2.96 \times 10^{1}$ & $6.59 \times 10^{-1}$ \\
     {\bf G180-R140}
          & $6.22 \times 10^{1}$ & $2.66 $ \\[1ex]
      G178
          & $3.07 \times 10^{1}$ & $6.23 \times 10^{-1}$ \\
      R95
          & $1.49 \times 10^{1}$ & $1.95 \times 10^{-1}$ \\
      \noalign{\smallskip}\hline\noalign{\smallskip}
   \end{tabular}
   % ---
   \end{minipage}
%---
   \caption{
     P.d.f.\ of the second invariant of the velocity gradient
  tensor, $Q$, \citep{hunt:88} for the unfiltered field (solid lines)
  and the filtered field (dashed lines). The field has been filtered
  with a box-filter of width $\Delta_{filt}=89\Delta x$ for the
  particle-laden cases  
(corresponding to $\Delta_{filt}=3.7D$ for the settling cases and
$\Delta_{filt}=5.6D$ for G0-R120) 
and $\Delta_{filt}=45\Delta x$ for the single phase case 
such as to keep the same ratio $\Delta_{filt}/\eta^{SP}$ as G178-R95. 
The dashed-dotted lines correspond to the values of $Q$ sampled on the
spheres $\mathcal{S}_{(i)}$ centered on the particles.
 Linestyle:   
$\color{blue}{\solidthick}$ G180-R140, 
$\color{red}{\solidthick}$ G178-R95, 
${\solidthick}$ G178, 
$\color{cyan}{\solidthick}$ R95.
The table lists the standard deviation of $Q$ computed for the
unfiltered field ($Q$) and the filtered field ($Q^{filt}$).
} 
\label{fig:qhunt_pdf_filtered}
\end{figure}

\begin{figure}
   \centering
   \begin{minipage}{0.46\linewidth}
      \centerline{(a)}
      \includegraphics[width=1.0\linewidth]
         {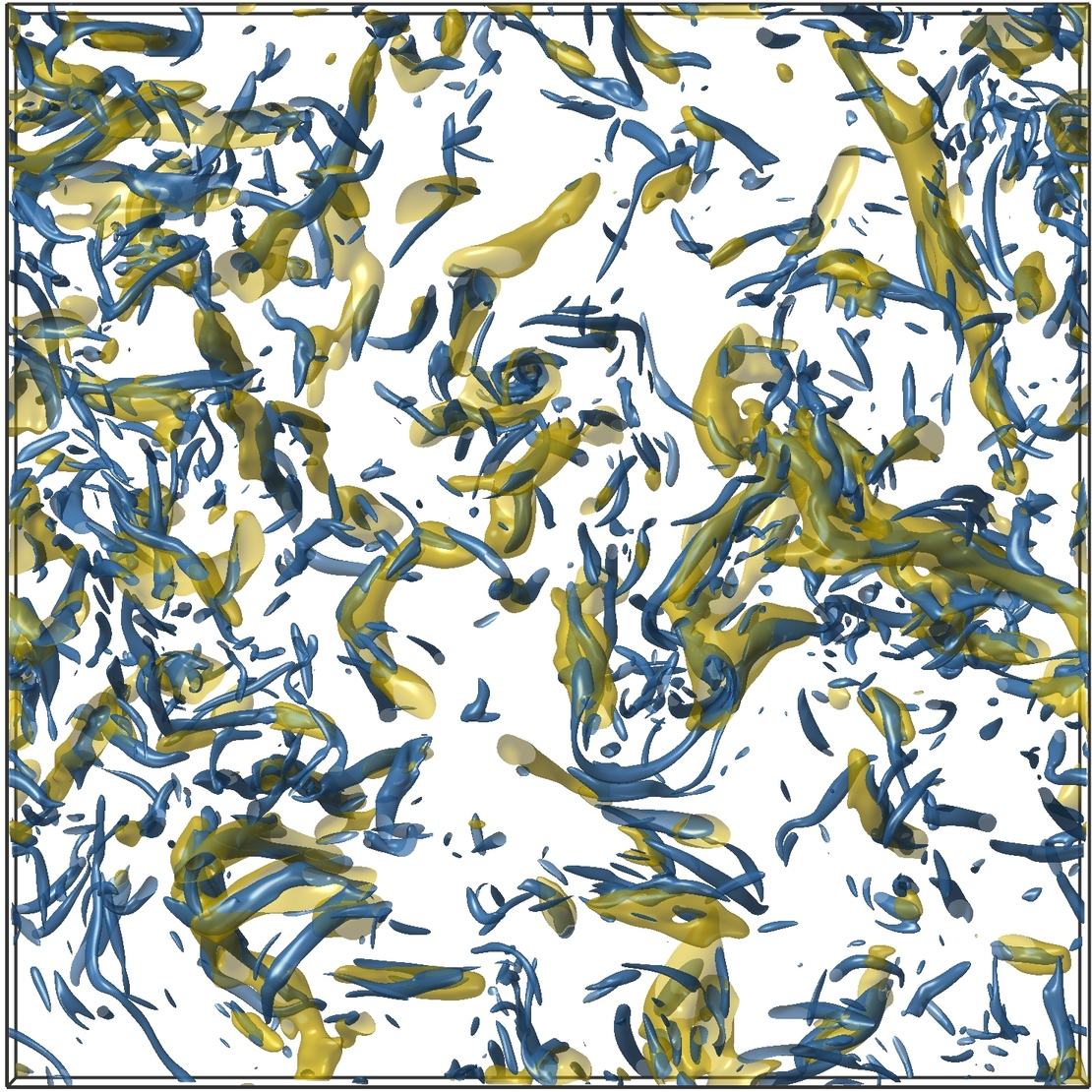}
      \centerline{(b)}
      \includegraphics[width=1.0\linewidth]
         {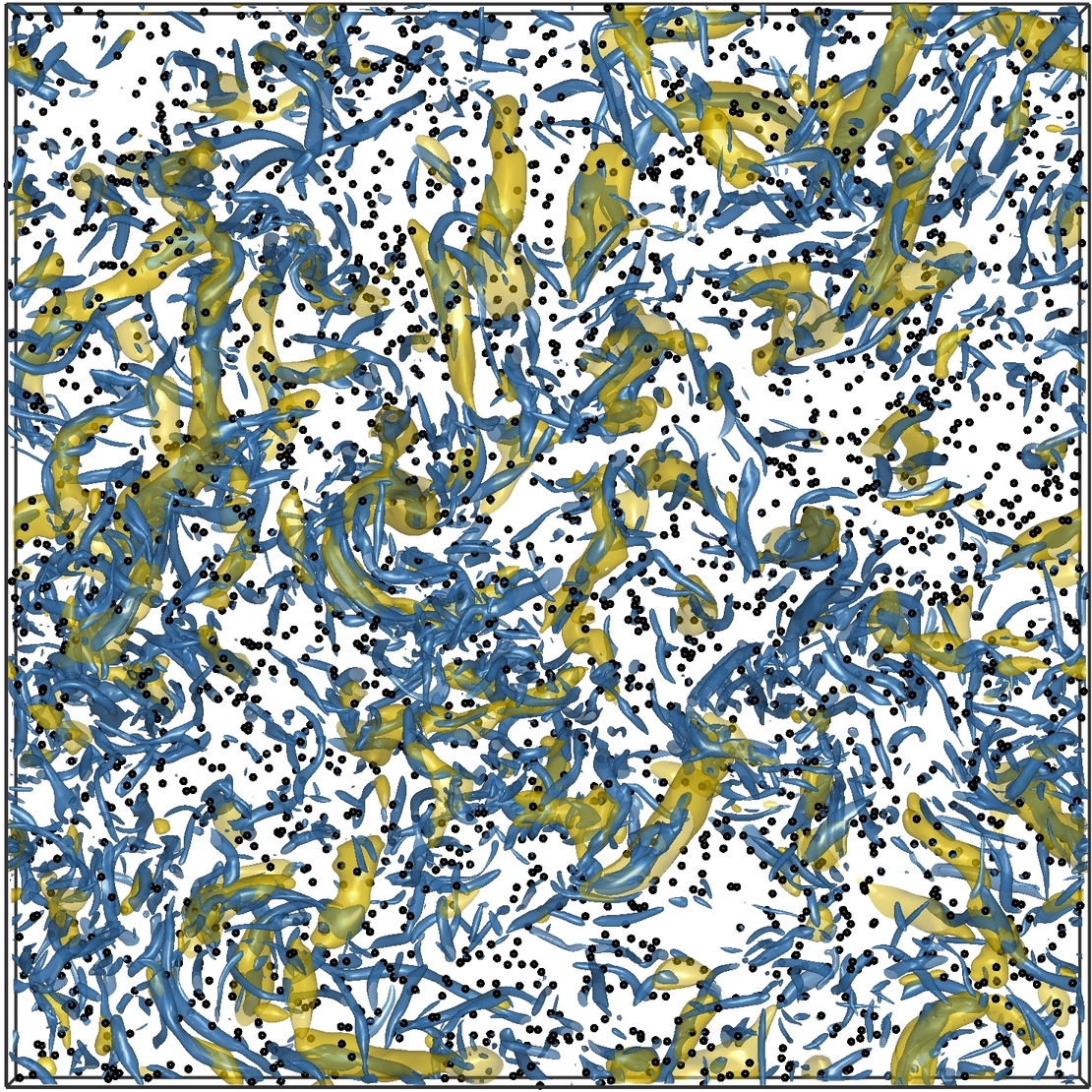}
   \end{minipage}
   \begin{minipage}{0.49\linewidth}
      \centerline{(c)}
      \includegraphics[width=1.0\linewidth]
         {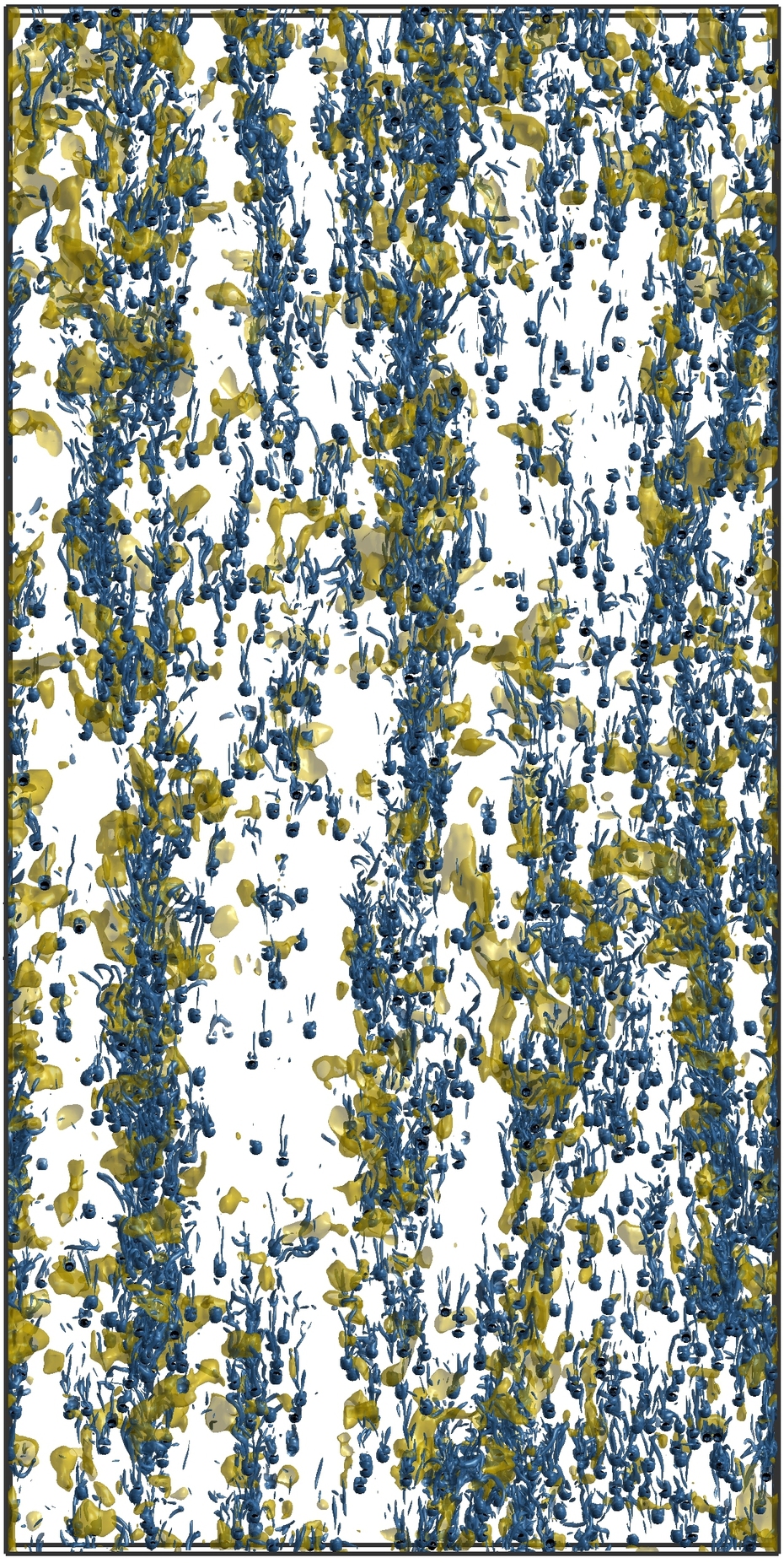}
   \end{minipage}
   %
%---
   \caption{Isocontour of $Q$ %criterion
     for the unfiltered field (blue) and the filtered field (yellow)
     for the single phase case R95 (a), the case G0-R120 (b) and the
     case G178 (d). The filter width is equal to
     $\Delta_{filt}=89\Delta x$ for both cases G0-R114 and G178, and
     to $\Delta_{filt}=45\Delta x$ for R95. For the unfiltered field
     the isocontour corresponds to $Q=1.5 \sigma(Q)$ in the absence of
     gravity and $Q=0.5 \sigma(Q)$ for M178, while the filtered field
     corresponds to $Q_{filt}=1.5 \sigma(Q_{filt})$. For the three
     figures the visualizations represent one eighth of the domain in
     the
     depth (into the page) 
     and the total domain in the other two directions.}
\label{fig:visus_coherent_structures_wakes1}
\end{figure}

\begin{figure}
   \centering
   \begin{minipage}{0.49\linewidth}
      \centerline{(a)}
      \includegraphics[width=1.0\linewidth]
         {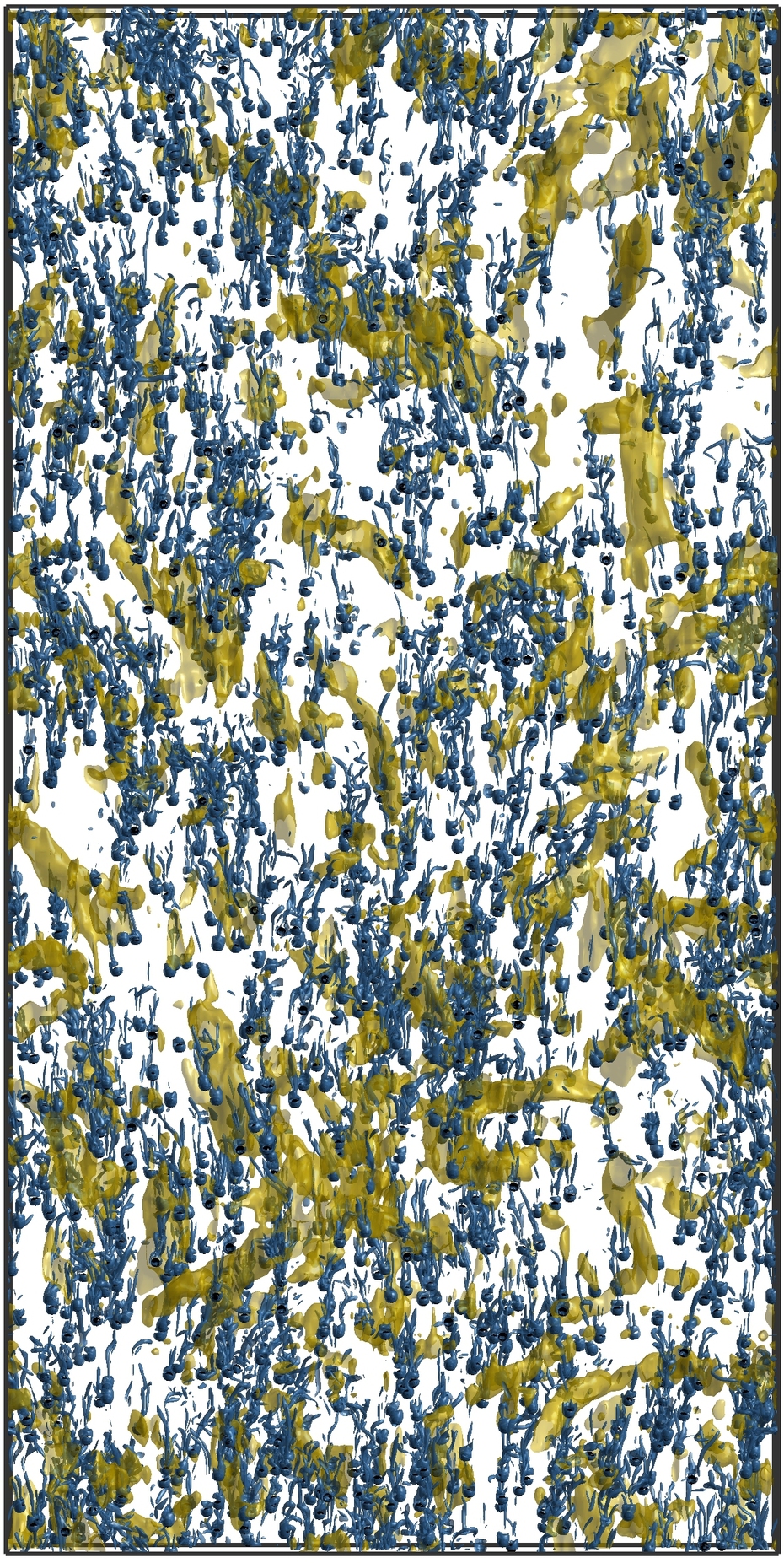}
   \end{minipage}
   \begin{minipage}{0.49\linewidth}
      \centerline{(b)}
      \includegraphics[width=1.0\linewidth]
         {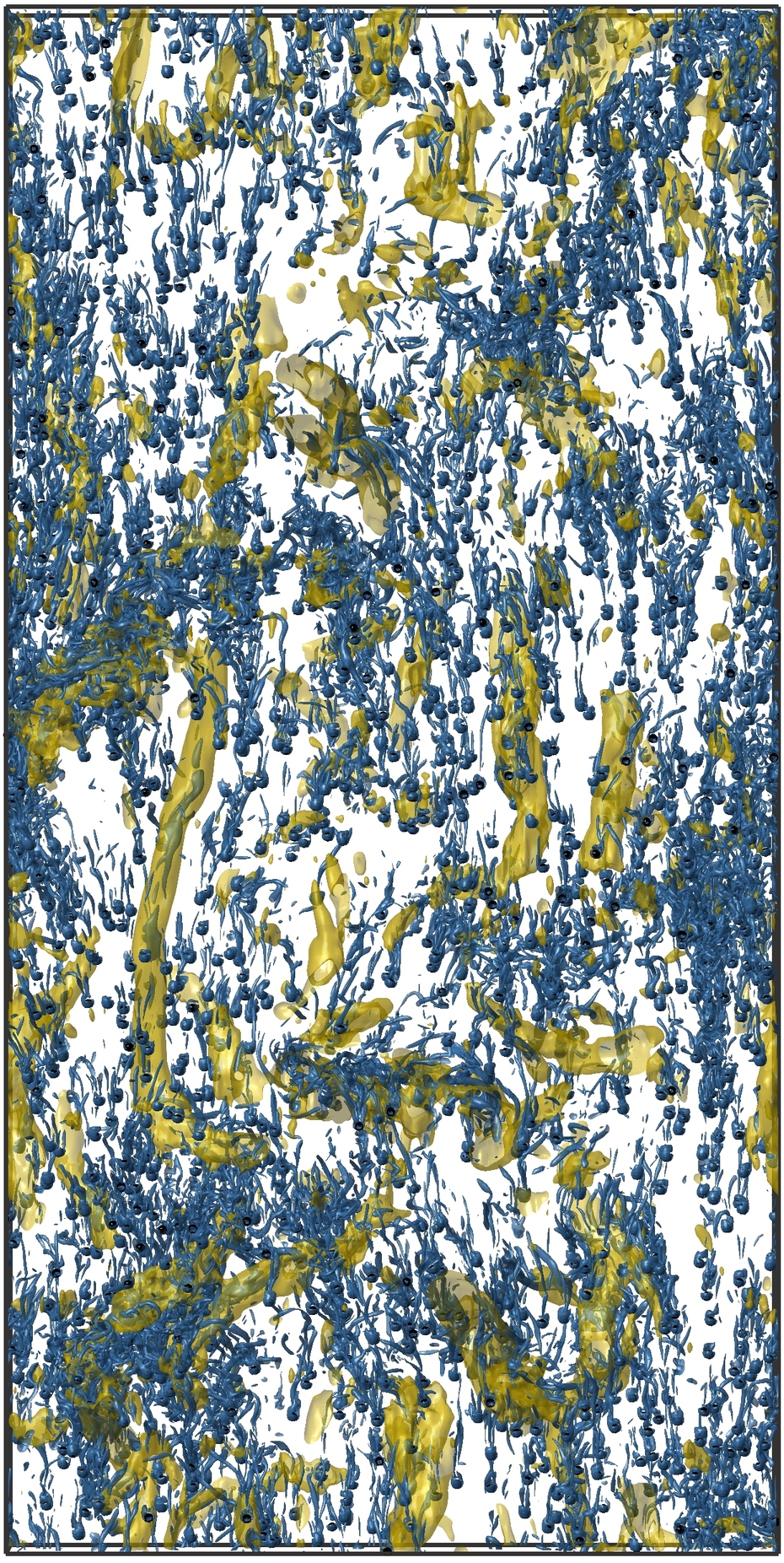}
   \end{minipage}
   %
%---
\caption{Isocontour of the Q criterion for the unfiltered field (blue)
  and the filtered field (yellow) for the case M178-R95 (a) and
  G180-R140 (b). The filter width is equal to $\Delta_{filt}=89\Delta
  x$ for both cases. For the unfiltered field the isocontour
  corresponds to $Q=0.5 \sigma(Q)$ while the filtered field
  corresponds to $Q_{filt}=1.5 \sigma(Q_{filt})$. For both figures the
  visualizations represent one
  eighth of the domain in the
  depth (into the page) 
  direction and the total domain in the other two directions.}
\label{fig:visus_coherent_structures_wakes2}
\end{figure}

% ------------------------------------------------------------------------- %
We now consider the turbulent structures, and propose for this to base our analysis on the Q criterion proposed by \citet{hunt:88}, and defined as $Q=u_{i,i}u_{i,i}/2-u_{i,j}u_{j,i}/2$.
Figure \ref{fig:qhunt_pdf_filtered} gives the probability distribution of $Q$;
the figure also states the values of the corresponding standard deviations scaled by $\omega_{rms}$ (defined as $\omega_{rms}=\sqrt{\varepsilon/\nu}$). 
It shows that turbulence tends to increase the amplitude of the peak of $Q$ compared to the mutliparticle settling case G178. The presence of the particles increases the standard deviation of $Q$ due to the enhancement of the velocity gradients at the surface of the particles.
We represent the coherent structures by isocontours of $Q$ taken at a
given threshold of the standard deviation (blue contours on figures
\ref{fig:visus_coherent_structures_wakes1} and
\ref{fig:visus_coherent_structures_wakes2}).
We
chose as threshold 
1.5$\sigma(Q)$ in the absence of gravity and 0.5$\sigma(Q)$ for the settling cases.
The adjustment of the threshold for the settling cases was motivated
by the choice to stay at a roughly equal probability to meet the
corresponding Q among the cases 
(i.e. the depicted iso-surfaces roughly enclose the same total volumes
in figures \ref{fig:visus_coherent_structures_wakes1} and
\ref{fig:visus_coherent_structures_wakes2}). 
The visualizations feature 
the well known elongated worm-like vortices for both single phase and particle-laden configurations (figures \ref{fig:visus_coherent_structures_wakes1} and \ref{fig:visus_coherent_structures_wakes2}).
\citet{uhlmann:16a} have shown the presence of wakes (with volumes below the particle volume) in the vicinity of the particles for $Ga=0$. In the settling cases, the small-scale structures are on contrary mainly dominated by the wakes arising from the mean settling, and which lead to the increase of the standard deviation of $Q$.
This does not mean that worm-like structures are completely destroyed by the wakes however, but that they are not the dominant small scale vortical structures at 0.5$\sigma(Q)$ (those worms are for example still present in the visualization of G180-R140 which features the largest turbulence intensity). 
We recall that, in the absence of forced turbulence, the large-scale
structure of the flow is characterized by large columns of wakes, that
are here clearly visible
(fig. \ref{fig:visus_coherent_structures_wakes1}). 
For the forced cases those large patterns are less evident though, and we filter the velocity field with a filter width $\Delta_{filt}$ with the aim of extracting more clearly those large eddies. We took a filter width $\Delta_{filt}=3.7D$ for G0-R120, and $\Delta_{filt}=5.6D$ for the other cases (we used the same width in multiple of the Kolmogorov length in the single-phase case R95).
As expected, filtering smoothes out the velocity gradients, decreases the standard deviation of $Q$ and lets its pdf tend towards Gaussian (cf. figure \ref{fig:qhunt_pdf_filtered}).
We used this criterion to visualize the large-scale structures of the flow in the same manner as for the unfiltered field, this time with the threshold 1.5$\sigma(Q_{filt})$ for all cases (where $Q_{filt}$ is the box-filtered $Q$ field). No adjustment of the threshold was here necessary for the settling cases to ensure that the corresponding probabilities of $Q_{filt}$ stay roughly equal, and that this thresholds provided similar probability to its unfiltered equivalent (yellow contours on figures \ref{fig:visus_coherent_structures_wakes1} and \ref{fig:visus_coherent_structures_wakes2}).
The large scale structure obtained with the filtering procedure features similar characteristics as those obtained by representation of the low pressure regions of the unfiltered field.
In the absence of gravity both single phase and particle-laden cases
feature elongated vortices of larger size in the vicinity of the
smaller worms that concentrate inside and around those large
eddies. Note that this ``vortex-within-vortex'' scenario has been
documented in previous single-phase studies (cf.\ e.g.\ the excellent
visualization of \citet{buerger:12a}). 
In the pure sedimentation case, we observe the presence of large
structures that most probably result from the formation of vortical
structures in the mixing zones induced by the wake columns, in
accordance with the observation of \citet{uhlmann:14a}. 
As expected, forced turbulence reduces the signature of those vertical columns that become barely visible on the visualizations (fig \ref{fig:visus_coherent_structures_wakes2}).
It also seems that large-scale structures of similar size as the single phase equivalent are still present, with an orientation mostly directed in the vertical and horizontal direction.
This behaviour is consistent with the description of the
``banana-shape'' structures proposed by
\citet{ferrante:03}, 
but a more systematic analysis should be performed to measure the impact of the settling on the large eddies of the flow.

Cases G178-R95 and G180-R140 differ mainly by the interaction between large vortices and the wakes, and consequently the concentration of small-scale vortices with respect to the larger ones. 
On the one hand the wakes mostly disrupt and shorten the large-scale vortical structures for the case G178-R95, while on the other hand they also accumulate in the vicinity of the large eddies for the case G180-R140.  As a result, relatively large regions with a more important density of wakes develop for G180-R140, while for G178-R95 those regions have a smaller extent and do not seem to sample preferential regions of the filtered field.
This trend is consistent with the level of concentration that will be described in the next section.
G178-R95 and G180-R140 do not only differ by their turbulence intensity, but also by the length scales of the flow with respect to the particle size.  The Kolmogorov length is slightly smaller for Ga180-R140 ($\eta^{SP}=0.12D$) than for Ga178-R95 ($\eta^{SP}=0.15D$) while large-eddy length are larger for Ga180-R140 ($L^{SP}=47.8D$) than for Ga178-R95 ($L^{SP}=32.8D$).
As an effect, one could expect the particles and their wakes to be more affected by the large scales of the forced turbulence for the case G180-R140 than G178-R95, by getting expelled outside of those large scale vortices according to the model initially proposed for point-particles. 
This could explain the observation on figure \ref{fig:visus_coherent_structures_wakes2} that, contrarily to the case G178-R95,  some particles seem to be located in the vicinity of the large structures in the case G180-R140 and probably rotating around them according to the orientation of their wakes.
A more systematic investigation of those events
(including an analysis of time-resolved data)
is however necessary to conclude on their statistical evidence as well as their actual impact on potential particle accumulation.

\section{The dispersed phase}
\label{sec:results_part}
% ######################################################################### %
We now
the behaviour of the dispersed phase and propose to focus first on the possible modification of the settling velocity of the particles and considering then the ability of the particles to form large or small scale clusters.
% ======================================================================== %
\subsection{Settling velocity \label{sec:settling_velocity}}
\begin{figure}
   \begin{minipage}{3ex}
      \rotatebox{90}{\centerline{$\lvert w_{rel}(t)\lvert/u_g$, $\lvert w_{rel}^{\mathcal{S}}(t)\lvert/u_g$}}
   \end{minipage}
   \begin{minipage}{0.45\linewidth}
      \centerline{(a)}
      \includegraphics[width=0.95\linewidth]
         {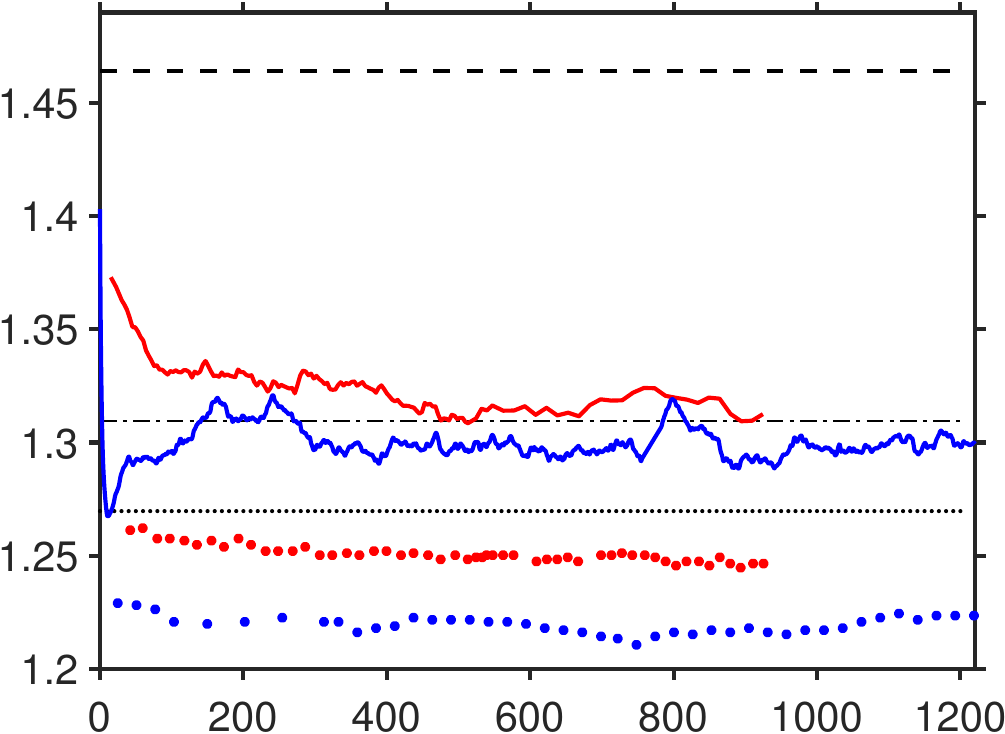}\\
      \centerline{$t/\tau_g$}
   \end{minipage}
   \begin{minipage}{3ex}
      \rotatebox{90}{\centerline{$\langle \lvert w_{rel}(t)\rvert \rangle_t/V_T$, $\langle \lvert w_{rel}^{\mathcal{S}}(t)\rvert \rangle_t/V_T$}}
   \end{minipage}
   \begin{minipage}{0.45\linewidth}
      \centerline{(b)}
      \includegraphics[width=0.95\linewidth]
         {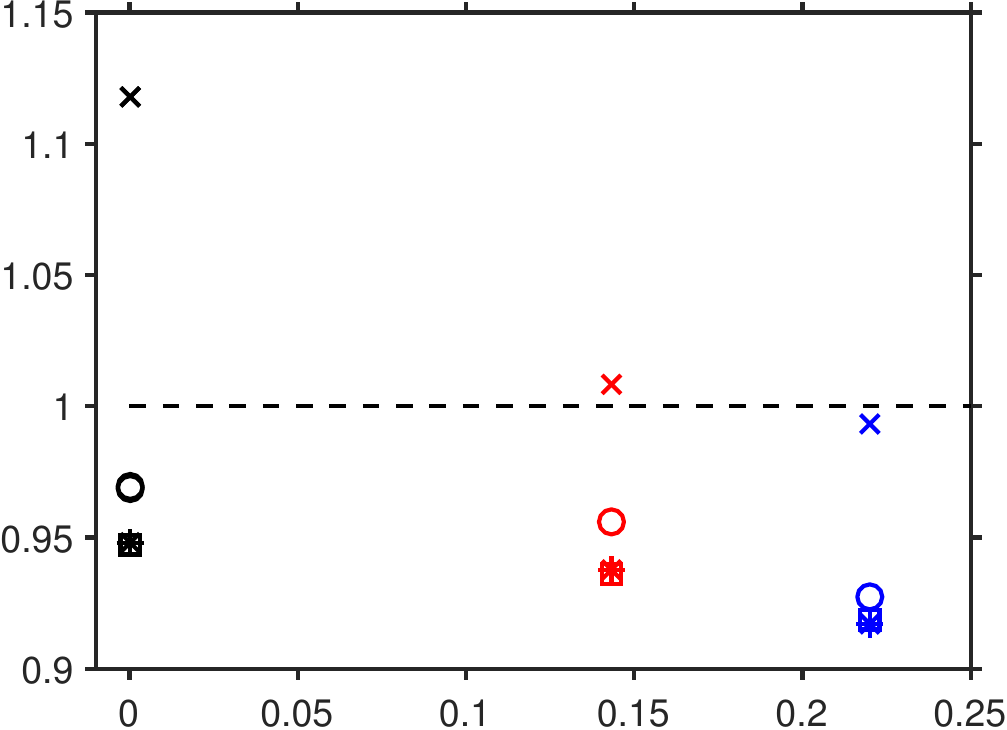}\\
      \centerline{$I$}
   \end{minipage}
%---
\caption{(a) Time evolution of the settling velocity (continuous and
  dashed lines: $w_{rel}=\left< u_{rel,z}^{(i)}\right>_p$,
  dots: $w_{rel}^{\mathcal{S}}=\langle
  {u}_{rel,z}^{\mathcal{S}_{(i)}}\rangle_p$ (cf.\
  equ.~\ref{eq:sphere_averaged_wrel}),  
$\solidshort \cdot \solidshort$ one single particle in ambient flow,
${\solidshort\,\solidshort\,\solidshort}$ mean settling velocity for
case G178, ${\cdot \cdot \cdot}$ mean settling velocity based on
the velocity seen by the particles ($w_{rel}^{\mathcal{S}}$) for
case G178 (note that for G178 the two definitions correspond to the average both in time and over the set of particles). 
(b)
Time average of the mean settling velocity as a function of the turbulence intensity $I=u_{rms}^{SP}/V_T$. The crosses correspond to the apparent relative velocity $w_{rel}$ and circles to the relative velocity based upon the sphere averaging procedure $w_{rel}^{\mathcal{S}}$, square to $w_{rel,f1}^{\mathcal{S}}$ and stars to $w_{rel,f2}^{\mathcal{S}}$.
 Linestyle:  
$\color{red}{\solidthick}$ G178-R95, 
$\color{blue}{\solidthick}$ G180-R140, 
${\solidthick}$ G178} 
\label{fig:settling_velocities}
\end{figure}

% ------------------------------------------------------------------------- %
The interaction with the background turbulence will also lead to a modification of the settling velocity of the particles. To estimate this settling velocity we need to define a relative velocity between the particles and the fluid. 
To do so we first use an apparent velocity $\mathbf{u}_{rel}^{(i)}=
(u_{rel,x}^{(i)},u_{rel,y}^{(i)},u_{rel,z}^{(i)})^T$ corresponding to the difference between the ith particle velocity and the mean flow, viz.
\begin{equation}
   \mathbf{u}_{rel}^{(i)}(t)= \mathbf{v}^{(i)}(t) - \langle \mathbf{u}
   \rangle_{\Omega_f}(t)
   \,.
\end{equation}
We introduce then the average settling velocity $w_{rel}$ corresponding to the average over all particles of the vertical component of this apparent slip velocity, namely
\begin{eqnarray}
  w_{rel}(t) & = & \left< u_{rel,z}^{(i)}\right>_p(t)
                   \,.
\end{eqnarray}
%}
Figure \ref{fig:settling_velocities} gives the time evolution of this velocity, scaled by the gravitational velocity scale $u_g$. It shows that turbulence tends to decrease the settling velocity for both cases G178-R95 and G180-R140 with respect to the multiple particle settling configuration G178, in accordance with the observations of \citet{fornari:16a} and \citet{chouippe:15a}. 
It shows that the resulting settling velocity tends towards the value
of one single particle falling in quiescent flow, with a difference of
few percents.
Case G178-R95 features an increase of 1.1$\%$ while G180-R140 shows a decrease of 0.6$\%$. 
Nonlinear drag effects are known to induce a reduction of the settling velocity
\citep{tunstall:68,nielsen:93,wang:93}, and \citet{homann:13} proposed a model
to estimate the amplitude of this reduction depending on the turbulence
intensity. This model relies on adding a Gaussian perturbation that possesses
the same standard deviation as the forced turbulence to the velocity seen by the particles,
and to estimate how the induced modification of the drag would affect the mean settling velocity. We applied this model to the configurations explored here and
found that turbulence should induce a reduction of the settling velocity of 0.7$\%$ (3.1$\%$) for case G178-R95 (G180-R140).
This model is not meant to reproduce collective effects (e.g. wake attractions, modification of the flow structures), but partially explains the reductions observed here.

In order to account for the local properties of the velocity field in
the vicinity of the particles, we introduce for each particle with
index ``$i$'' a velocity $\mathbf{u}^{\mathcal{S}_{(i)}}$ defined as the average velocity over a sphere $\mathcal{S}_{(i)}$ of radius $R_S$ and centered at the particle's center.
We then define the corresponding relative velocity $\mathbf{u}_{rel}^{\mathcal{S}_{(i)}}$, according to
\begin{eqnarray}
   \mathbf{u}_{rel}^{\mathcal{S}_{(i)}}(t) &=&
      \mathbf{v}^{(i)}(t) - \mathbf{u}^{\mathcal{S}_{(i)}}(t).
      \label{eq:sphere_averaged_wrel}
\end{eqnarray}
This definition has been introduced in \cite{Kidanemariam_NJP2013} and
tested for different shell radii $R_S$. Using a radius $R_S=1.5D$ has
been shown to be a good compromise between not being too much affected
by the particle wakes and still being representative of the local
properties of the surrounding velocity field. 
We introduce a subsequent settling velocity $w_{ref}^{\mathcal{S}}$
defined as the vertical component of the particle average relative
velocity based upon sphere-averaging, viz.\
$w_{ref}^{\mathcal{S}}=\langle
{u}_{rel,z}^{\mathcal{S}_{(i)}}\rangle_p$. 
The time evolution of this settling velocity is also represented on figure \ref{fig:settling_velocities}.
Similar to the pure multiparticle settling case, it shows that the
Reynolds number based upon the sphere-averaged velocity
(eq. \ref{eq:sphere_averaged_wrel}) is always smaller than its global
equivalent, meaning that the particle velocity tends to adapt to its
surrounding. 
Here as well a monotonous trend is observed with respect to the influence of the relative intensity of the turbulence, with a smaller falling velocity obtained for the largest turbulence intensity.
It appears indeed that the turbulence intensity tends to monotonously affect the mean apparent relative velocity $w_{rel}$ as well as the mean vertical component representative of the velocity difference between the particles and their vicinity $w_{rel}^{\mathcal{S}}$ (cf. figure \ref{fig:settling_velocities}).

\begin{figure}
   \begin{minipage}{2ex}
      \rotatebox{90}{\centerline{p.d.f.}}
   \end{minipage}
   \begin{minipage}{0.45\linewidth}
      \centerline{(a)}
      \includegraphics[width=0.95\linewidth]
         {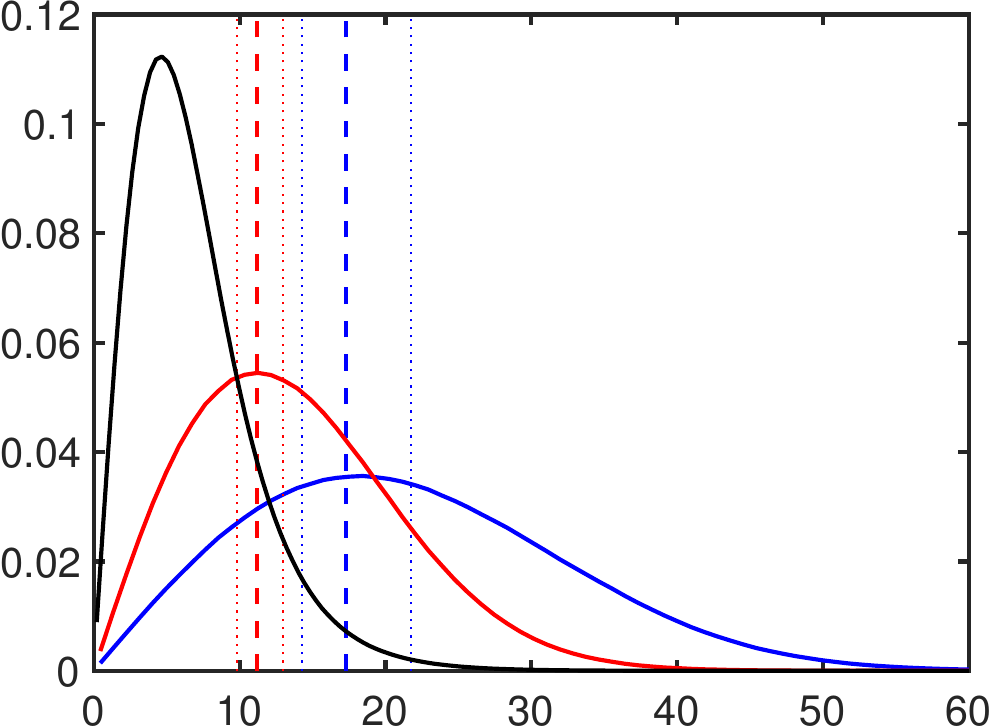}\\
      \centerline{$\alpha_p$}
   \end{minipage}
   \begin{minipage}{2ex}
      \rotatebox{90}{\centerline{p.d.f.}}
   \end{minipage}
   \begin{minipage}{0.45\linewidth}
      \centerline{(b)}
      \includegraphics[width=0.95\linewidth]
         {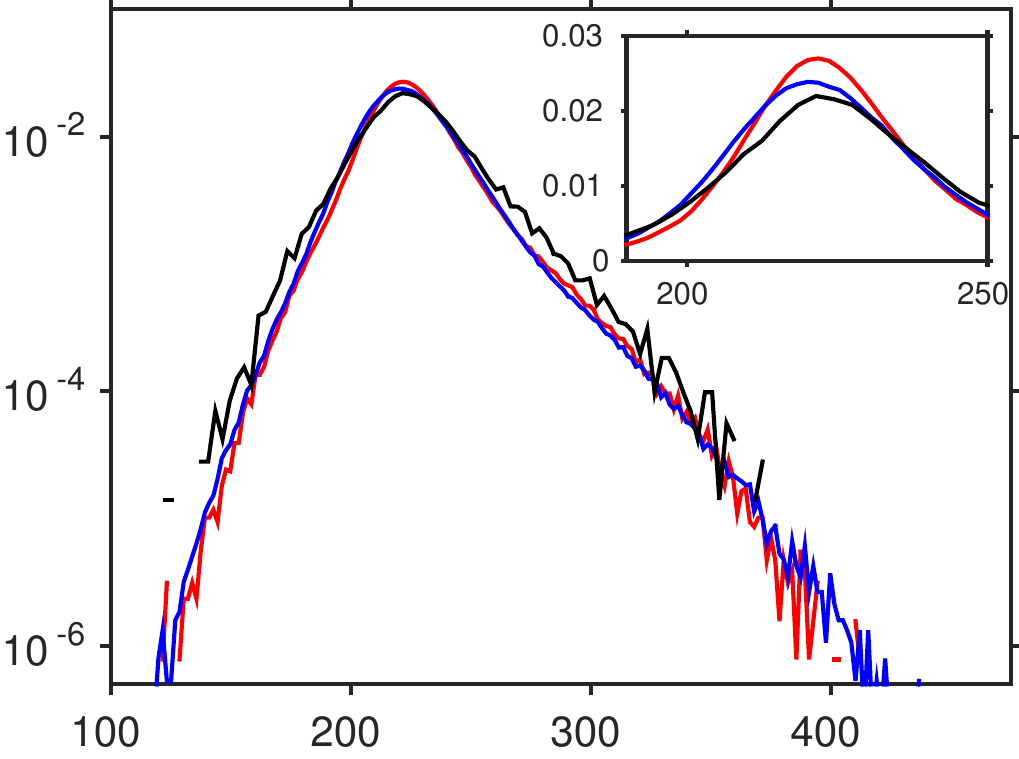}\\
      \centerline{$Re_p^{\mathcal{S}}$}
   \end{minipage}
%---
\caption{(a) P.d.f.\ of the angle $\alpha_p$ between the relative
  particle velocity vector and the vertical axis (in degrees). 
  The vertical lines indicate the location of the reference angles
  $\alpha_{ref}^i$ defined as $tan(\alpha_{ref}^i)=(2u_{rms}^{SP})^{1/2}/ \vert V_T+i \times u_{rms}^{SP} \vert$. The dashed lines corresponds to $\alpha_{ref}^0$ and the dotted lines to $\alpha_{ref}^{\pm1}$ (we have 
  $\alpha_{ref}^{-1}<\alpha_{ref}^{0}<\alpha_{ref}^{1}$). 
(b) P.d.f.\ of the Reynolds number of the particles based on the shell averaged relative velocities: $Re_p^{\mathcal{S}_{(i)}}=D\left|\mathbf{u}_{rel}^{\mathcal{S}_{(i)}} \right|/\nu$.
 Linestyle:  
$\color{blue}{\solidthick}$ G180-R140, 
$\color{red}{\solidthick}$ G178-R95, 
${\solidthick}$ G178.
The inset shows the same data in linear scale. }
\label{fig:settling_angle_Rep_shell}
\end{figure}

We introduce the instantaneous angle $\alpha_p$ between the relative particle velocity vector and the vertical axis, defined by the following relation
\begin{eqnarray}
   tan(\alpha_p) & = &
                       \frac{\sqrt{u_{rel,x}^2+u_{rel,y}^2}}{\left| u_{rel,z} \right|}
                       \,.
\end{eqnarray}
The p.d.f.\ of the settling angle is depicted on figure
\ref{fig:settling_angle_Rep_shell} and shows a shift of the peak of
probability towards larger angles with increasing turbulence
intensity, as well as a broadening of the distributions. Note that the
peak of probability for the case G178 is approximately located at the
angle which would be observed in the case of one isolated sphere
settling in a still fluid (i.e.\ $\alpha_p=5.2 ^\circ$).  
We introduce three reference angles 
$tan(\alpha_{ref}^i)=(2u_{rms}^{SP})^{1/2}/ \vert V_T+i \times u_{rms}^{SP} \vert$
 with $i=\{ -1, 0, 1 \}$, and observe that the peak is centered around the reference angle $\alpha_{ref}^0$. 
The spreading of the distribution seems also to be linked with the enlargement of the range $[\alpha_{ref}^{-1},\alpha_{ref}^{1} ]$ with the turbulence intensity, which implies an increase of the relative horizontal motion of the particles induced by turbulence.

Figure \ref{fig:settling_velocities}(b) shows the normalized p.d.f.\ of the
resulting Reynolds number (based on the sphere-averaged relative
velocity). Interestingly, the shape of the distribution does not highlight any 
significant influence of the forced turbulence suggesting that the
evolution of the local motion of the particle is dominated by the
wakes. 
This last point can also be emphasized by considering the p.d.f. of the
fluid velocity (not the relative velocity), represented on figure
\ref{fig:velocity_pdf_with_shell}. The distinction has been indeed
made between p.d.f.\ accounting for the total fluid domain, and the
domain occupied by the spheres $\mathcal{S}_{(i)}$. It shows a mild
increase of the probability for finding large vertical and horizontal
velocities in the vicinity of the particles. 
These points will be further discussed in
\S~\ref{sec-discussion-pref-sample}. 
We then slightly adapted the definition of the local velocity $\mathbf{u}^{\mathcal{S}_{(i)}}$ such as to account only for the flow upstream of the particles. To do so we first introduce an "incoming" flow velocity $\mathbf{U}_{in}^{(i)}$ taken as the negative particle velocity, viz.
\begin{eqnarray}
   \mathbf{U}_{in,1}^{(i)}(t)&=&-\mathbf{v}^{(i)}\;, \label{eq:u_in1}
\end{eqnarray}
or as the negative sphere-averaged relative velocity similar to the work of \citet{cisse:13}:
\begin{eqnarray}
   \mathbf{U}_{in,2}^{(i)}(t)&=&-\mathbf{u}_{rel}^{\mathcal{S}_{(i)}}(t) \;.
      \label{eq:u_in2}
\end{eqnarray}
Next we average over the set of points $\mathbf{x}$ located upstream of the particle, i.e. those for which the following condition holds:
\begin{eqnarray}
   (\mathbf{x}-\mathbf{x_p}^{(i)}) \cdot \mathbf{U}_{in,j}^{i} < 0 \text{  with }j=1,2 \;.
\end{eqnarray}
In the following we will denote the relative velocity obtained with the first and second definitions by $\mathbf{u}_{rel,f1}^{\mathcal{S}_{(i)}}$ and $\mathbf{u}_{rel,f2}^{\mathcal{S}_{(i)}}$ respectively, the corresponding fluid velocity sampled at the front of the sphere by $\mathbf{u}_{f1}^{\mathcal{S}_{(i)}}$ and $\mathbf{u}_{f2}^{\mathcal{S}_{(i)}}$ respectively.\\
Figure \ref{fig:pdf_velocity_front} gives the evolution of the p.d.f. of the fluid sampled either on the totality of the fluid domain, or on the totality of the particle-centered sphere $\mathcal{S}_{(i)}$ with a radius equal to three times the particle radius, or on the front of the spheres according to both definitions. As we can see the distinction made on the portion of the spheres to be sampled for the statistics has little influence. 
This distinction has a slightly larger affect on the resulting relative velocity between the particles and the incoming fluid (figure \ref{fig:settling_velocities}).
The absolute value of the settling velocity obtained with the two variants where the fluid velocity seen by the particles is computed on the upstream-facing, particle-centered hemisphere is systematically smaller by one to two percent. This result is presumably related to the typical structure of the wake flow in the vicinity of the particles which does not only feature a (low-velocity) downstream recirculation region, but also a (high-velocity) transverse displacement region. This point probably merits further investigation in the future. For the present purpose, however, the comparison of the three different definitions of a relative velocity confirm the observed mild trend of a decrease (in absolute value) with turbulence intensity of its mean value.

% ------------------------------------------------------------------------- %
\subsection{Clustering}
% ------------------------------------------------------------------------- %
We now turn to the tendency of the particles to form clusters.
We choose to base our characterization of the clustering on the Vorono\"{i} tesselation analysis proposed by \citet{Monchaux_PoF2010} and \citet{monchaux:12}. It consists in dividing the total volume into $N_p$ subvolumes $\mathcal{V}_{Vor}^{(i)}$ defined for each particle (i), such that every point contained in the volume $\mathcal{V}_{Vor}^{(i)}$ is closer to the ith particle's center than to any other particle.
The volumes obtained can then be interpreted as the inverse of a concentration
since dense regions containing more particles than the rest will be characterized by small Vorono\"{i} volumes
and conversely. This method has several advantages: it provides an objective diagnostic for clustering by comparing the p.d.f. of the Vorono\"{i} volumes normalized to its random equivalent.
A set or randomly distributed particles, with positions taken from random Poisson process will give a p.d.f. following a gamma distribution. This p.d.f. is universal with coefficients numerically estimated by \citet{ferenc:07}, but in the case of finite-size particle the p.d.f. looses its universality as it will be affected by the condition on particles to not overlap.
The coefficients are then function of the particle volume fraction and the size of the total domain.
The comparison between the p.d.f.'s of the cases considered here and the random equivalent represented on figure \ref{fig:voronoi_cells_volume_aspect_ratio} is a typical example for clustering diagnostic: we indeed observe that the p.d.f.'s differ for all simulated cases to the random equivalent, in favor of an increase of
probability to see small and large Vorono\"{i} volumes. The increase of the probability to have small Vorono\"{i}
volumes is indeed to be interpreted as the sign of local enhancement of particle concentration (i.e. clusters) while the increase of the probability to have large volumes is the sign of local decrease of this concentration
(i.e. voids).
The two volumes ($\mathcal{V}_{Vor}^{clus}$ and $\mathcal{V}_{Vor}^{void}$) corresponding to the crossing points between the p.d.f.s of the simulated cases and their random equivalent can then be taken as an objective criterion for determining whether particles belong to a cluster or a void region: every particle with a Vorono\"{i} volumes smaller than $\mathcal{V}_{Vor}^{clus}$ will belong to a cluster while every particle with a volume larger than $\mathcal{V}_{Vor}^{void}$ will belong to a void region \citep{Monchaux_PoF2010}.

We quantify the level of clustering with the evolution of the standard deviation of the Vorono\"{i} volumes. 
Its time evolution is represented on figure \ref{fig:stand_dev_vor_cells}(a) and
compared to the random configuration. A set of randomly distributed particles is
characterized by a standard deviation approximately 0.4 while clustering is be
characterized by an increase of this standard deviation.
Figure \ref{fig:stand_dev_vor_cells} recalls that the pure multiparticle settling case G178 is characterized by significant clustering \citep{uhlmann:14a}, and that the non-settling case G0-R120 with turbulence shows small but statistically significant clustering \citep{uhlmann:16a}.
Here we observe in the case G178-R95 that turbulence tends to rapidly disturb the clusters present in the initial condition without completely destroying them, the resulting standard deviation of the Vorono\"{i} volume being larger than its random equivalent with an increase of roughly 10$\%$.
Interestingly the influence of the turbulence intensity is non-monotonous (fig \ref{fig:stand_dev_vor_cells}(b)), as the level of clustering is larger for G180-R140 than G178-R95.
Both of them are indeed lower than in the pure multiparticle settling, meaning that the disturbance induced by the forced turbulence tends to decrease the level of clustering.
But an increase of the turbulence intensity from G178-R95 to G180-R140 does not induce a decrease of the level of clustering.
To our knowledge the work of \citet{fiabane:12} is the only one that details experimental evidence of clustering of finite-size for heavy particles in forced turbulence.
The Galileo numbers investigated were of the order of 10, and the configurations studied evidenced strong clustering.
However, the relative turbulence intensity in those experiments was much higher than here with $u_{rms}/V_T$ ranging from 0.7 to 4.5. 

Let us now move on to the characteristics of the clusters, which are reconstructed as follows: we first extract the Vorono\"{i} volumes below the threshold $\mathcal{V}_{Vor}^{clus}$ and reconstruct the cluster by finding all cells that are connected.
The simulations employ periodic boundary conditions in the three spatial directions and we included this periodicity in the reconstruction of the clusters.
We slightly adapted the choice of the threshold $\mathcal{V}_{Vor}^{clus}$ in
order to take a value common for all the cases.
We indeed observed that this threshold has an impact on the volume
$\mathcal{V}_{clus}/\langle \mathcal{V}_{Vor} \rangle$ featuring the
highest probability. We selected $\mathcal{V}_{Vor}^{clus}=
0.64\langle \mathcal{V}_{Vor} \rangle$, which is representative of the
crossing points obtained for the cases simulated here
(fig.\ \ref{fig:voronoi_cells_volume_aspect_ratio}). 
We obtained that the mean cluster volume $\langle
\mathcal{V}_{clus}\rangle$ approximately equals to 235$\eta^3$ for
G178-R95 and 3000$\eta^3$ for G180-140, corresponding to a linear
dimension of the clusters of  6.2$\eta$ and 14.6$\eta$ respectively,
which is of the same order as the previous observations of the
literature \citep{aliseda:02,Obligado_JoT2014}.
Figure \ref{fig:pdf_cluster_vol_number_part} depicts the p.d.f.s of the obtained cluster volumes scaled by the mean Vorono\"{i} volume, for the cases simulated here as well as a set of random distributed particles which possesses the same geometrical properties (particle radius and size of the domain) and solid volume fraction as cases G178-R95 and G180-R140.
It presents the same features as the p.d.f.'s observed with two-dimensional experimental visualizations by \citet{sumbekova:16} and in three-dimensional visualizations by \citet{uhlmann:16a}, with a peak of cluster volume followed by an power law decay over nearly two decades and an exponential decay for the largest clusters.

\citet{goto:08} proposed a model distribution of clusters and voids for two-dimensional simulations by assuming that they mimic the self-similarity of turbulent eddies.
By doing so they obtained an exponential decay of clusters area of -5/3, which
have been further observed experimentally by \citet{sumbekova:16}. An extension of this model to three-dimensional cluster detection gives a -16/9 exponential decay \citep{uhlmann:16a}, and provides a good trend of the decay for all the cases explored here.
Interestingly the set of random particle positions considered here for comparison provides the same trend 
with a peak located between approximately 0.4$\langle
\mathcal{V}_{Vor} \rangle$ and 0.8$\langle \mathcal{V}_{Vor} \rangle$, 
and an inertial range with exponential decay following the -16/9 slope. The difference to the randomness lies
in the structure of the smallest and largest clusters 
that are more probable in the simulations, 
and in the extent of the inertial range.
A possible scenario would be that turbulence tends to affect the
smallest accumulation zones by making them smaller and therefore
increasing the particle concentration.
On the other side of the volume distribution we observe that turbulence increases the probability to observe large clusters compared to the random reference. 
The graph also shows that gravity likewise tends to increase the
probability of finding large-volume clusters as compared to the random
data, i.e.\ all cases involving particle settling are above the
zero-gravity data for $\mathcal{V}_c/\langle \mathcal{V}_{Vor}
\rangle\geq200$. 
The observed differences between the settling case without turbulence (G178)
and those with the two different levels of background turbulence
(G178-R95, G180-R140) in terms of the cluster volume at the
large-scale end of the spectrum are probably not significant as
compared to the statistical uncertainty. 
However, we will see below that the clusters' geometrical aspect ratio
is clearly affected by turbulence, which suggests that the principal
action of turbulence is to modify the shape of gravity-induced
clusters. 

In order to get more insight into the geometrical properties of the
clusters, we considered their spatial extension. 
To do so, we looked for the smallest cuboid, with faces aligned with the
three global coordinate directions, that is able to contain each cluster, and we
define the directional extension of the cluster as the side length of the
corresponding cuboid.
We
then
analyzed the aspect ratio, defined here as the ratio between the
extension of the cluster in the vertical direction $L_c^z$ to its
extension in any of the horizontal directions $L_c^x$ or $L_c^y$.
The evolution of the mean aspect ratio of the clusters is shown in 
fig.~\ref{fig:aspect_ratio_clusters}. %(a) 
It shows that settling tends to elongate the largest clusters in the
vertical direction. For any significant cluster size, the aspect ratio
in general increases with the cluster volume. Here the largest values
are obtained for the pure settling case G178, while in the forced
turbulence cases G178-R95 and G180-R140 the elongation is
significantly smaller. Interestingly, the curves corresponding to the
two different turbulence intensities presently simulated practically
collapse. 
% ------------------------------------------------------------------------- %

\section{Discussion}
\label{sec:discussion}
% ######################################################################### %
We have seen in the previous sections that the relative intensity $I$ of the forced turbulence with respect to the original settling velocity does not necessarily bring a monotonous trend: although  it progressively decreases the falling speed of the particles, it decreases the level of clustering without displaying a clear progression of this reduction with $I$.
The visualizations based upon the $Q$ criterion highlighted a flow dominated at small scales either by worm-like vortical structures (in the absence of buoyancy) or by wakes (for $Ga \neq 0$).
The large scales are characterized by large vortices that are progressively disrupted by the wakes' contribution as turbulence intensity decreases.
Here we will 
discuss several aspects of the interaction mechanisms suggested in the
literature, focusing on the potential influence of the
various time scales, 
on the centrifugal mechanism, and on the potentially related preferential
sampling of the flow. 
A special attention will also be given regarding the tendency of gravity to increase particle accumulation compared to the situations featuring no buoyancy effect.
As the large eddies of the flow appear to be affected by the presence of the particles, in the settling configurations, we discuss to which extent those two-way coupling effects can have an impact on the aforementioned interaction mechanisms.
% ------------------------------------------------------------------------- %
\subsection{
Considerations on the time scales
}
% ------------------------------------------------------------------------- %
It is common to introduce a Stokes number based upon the Kolmogorov time scale to measure the ratio between the time-scale of the particles and the typical time scale of the smallest eddies of the flow, namely $St_\eta=\tau_p/\tau_\eta$, where the relaxation time scale of the particles $\tau_p=D^2\rho_p/(18\nu\rho_f)$ is taken from the Stokes drag.
Note that this definition of the particle time-scale is sometimes
adapted in the literature in order to account for added mass effects,
except if the density ratio is very large. 
In the case of point-particles it is commonly expected to see maximal preferential accumulation for Stokes number of the order of 1 \citep{bec:07,coleman:09}.
\citet{yoshimoto:07} have shown that clustering can still appear for larger Stokes number, meaning that not only the smallest eddies are playing a role in the clustering mechanism but larger ones can be involved. They introduced a resonance condition, namely that eddies of size $\ell$ and timescale $\tau(\ell)$ will centrifuge out a heavy particle with a relaxation time scale $\tau_p$ if the corresponding Stokes number $St_\ell=\tau_p/\tau(\ell)$ is contained between 0.1 and 2.
Preferential accumulation results then from the interaction with the range of eddies for which the resonance condition is fulfilled.

In the absence of gravity, we observe here a relatively small degree of clustering for G0-R120, which is in accordance with the resonance model since $St_\eta \approx 2.5$ and $St_e \approx 0.05$ for this case. The flow should then contain eddies that are capable of expelling particles.
The particle will barely feel the influence of any smaller eddies, and will be swept by the larger ones.
The order of magnitude of $St_\eta$ indicates however that the smallest scales verifying this resonance condition should be of the same order as the Kolomogorov length-scale, and consequently smaller than the particles themselves (we remind that this case features $D/\eta \approx 5.5$). The centrifugal scenario is hence not realistic at the small range, which might explain that the accumulation level achieved is actually low. 
\citet{uhlmann:16a} considered another configuration at $Ga=0$ (cf. their case ``D11'')  which features the same turbulence characteristics as the current G180-R140 case, with slightly larger particles ($D/\eta=11$). This case features Stokes numbers $St_\eta=10.7$ and $St_e=0.29$, and, consistently, presents smaller accumulation level than G0-R120. 
Judging from the consideration of these two reference cases without gravity, it would therefore be expected from a configuration presenting the same geometrical and turbulence characteristics as G180-R140 but no gravity effects to display a relatively small clustering level. This means that once gravity effects set in, we obtain a global enhancement of this particle accumulation, which can be potentially explained by different mechanisms that we will discuss next.

The first one is the attraction between particles, that is induced by their wakes and which is clearly at play in the pure settling case (G178).
It would be then expected to see the level of clustering monotonously decreasing with the turbulence intensity, contrarily to the trend observed.
Consequently this mechanism does not completely explain the enhancement of the accumulation.

Another candidate that can be proposed concerns the influence that
gravity has on the interaction time scales between fluid and
particles. In the original resonance model, the time-scale
$\tau(\ell)$ of an eddy $\ell$ does not correspond to the time
actually seen by a particle settling through this eddy,
which can be roughly estimated by
$\tau^{(S)}(\ell)=\ell/V_T$.
This concept can be connected to the so-called "crossing trajectory
effect" introduced by \citet{yudine:59}, which refers to the tendency
of heavy particles to continuously change their fluid neighborhood
while falling, resulting in a decorrelation of the fluid velocity
along particles trajectories \citep{wells:83,oesterle:07}.  
Crossing trajectory effects appear to decrease particle dispersion and
to have a negligible influence for turbulence intensity larger than
unity \citep{csanady:63,wells:83}. 
It can be first verified that the settling time scale does not exceed its characteristic time scale with the condition $\tau(\ell) > \tau^{(S)}(\ell)$. This is indeed the case, since for the large eddies we have $\tau_e/\tau_e^{(S)}=1/I$ (with $I<1$), and for the Kolmogorov length scale $\tau_\eta/\tau_\eta^{(S)}\sim\sqrt{Re_\lambda}/I$.
We introduce the surrogate Stokes number $St_\ell^{(S)}=\tau_p/\tau_\ell^{(S)}$, which takes into account the reduced interaction time due to the mean relative velocities.
It shows that turbulence intensity tends to decrease the Stokes number based on the large scales according to $St_e^{(S)}=St_e/I$,  and that both Reynolds number and relative turbulence intensity affect the Stokes number of the small scales with $St_\eta^{(S)}=St_\eta\sqrt{Re_\lambda}/(15^{1/4}I)$.
Gravity will consequently have the effect of increasing the value of the
apparent Stokes number representative of the largest eddies, which would tend to
promote the occurrence of preferential accumulation.
Gravity will also have the effect of filtering out the influence of the smallest eddies of the flow upon the particle motion by increasing $St_\eta^{(S)}$. \\

In order to have a better impression on how this can affect the motion of the particle, we go back to the size $\ell$ of the eddy that would realize $\tau_p/\tau(\ell)=\alpha$. This eddy should verify
\begin{eqnarray}
   \frac{\ell}{D}&=&\left( \frac{D}{\eta}\right)^2
                    \left( \frac{\rho_p}{18\alpha\rho_f}\right)^{{3/2}}
\end{eqnarray}
This can be used to estimate the size of the smallest eddy realizing the resonance condition ($\alpha=2$), which in the present case turns out to yield non realistic values ($\ell/D \approx 0.4$ for G178-R95 and $\ell/D \approx 0.6$ for G180-R140). This means that this resonance can only be verified at larger scales, leading to a narrower range of resonance. Looking now at the size of the eddy $\ell$ that would verify the settling-aware condition on the time-scale ratio, viz. $\tau_p/\tau^{(S)}(\ell)=\alpha$  leads to the following relation for the eddy size:
\begin{eqnarray}
  \frac{\ell}{D}&=&\frac{\rho_p}{18\alpha\rho_f}Re_T
\end{eqnarray} 
with $Re_T=DV_T/\nu$. In this context, the smallest eddies verifying the resonance condition ($\alpha=2$) would be of the order of $10\eta$.
This means that the centrifuge mechanism leading to particle accumulation that these flow structures have upon the particle motion is potentially active.
%
% ------------------------------------------------------------------------- %
\subsection{Preferential sampling}
\label{sec-discussion-pref-sample}
\begin{figure}
   \begin{minipage}{2ex}
      \rotatebox{90}{\centerline{p.d.f.}}
   \end{minipage}
   \begin{minipage}{0.45\linewidth}
      \centerline{(a)}
      \includegraphics[width=0.95\linewidth]
         {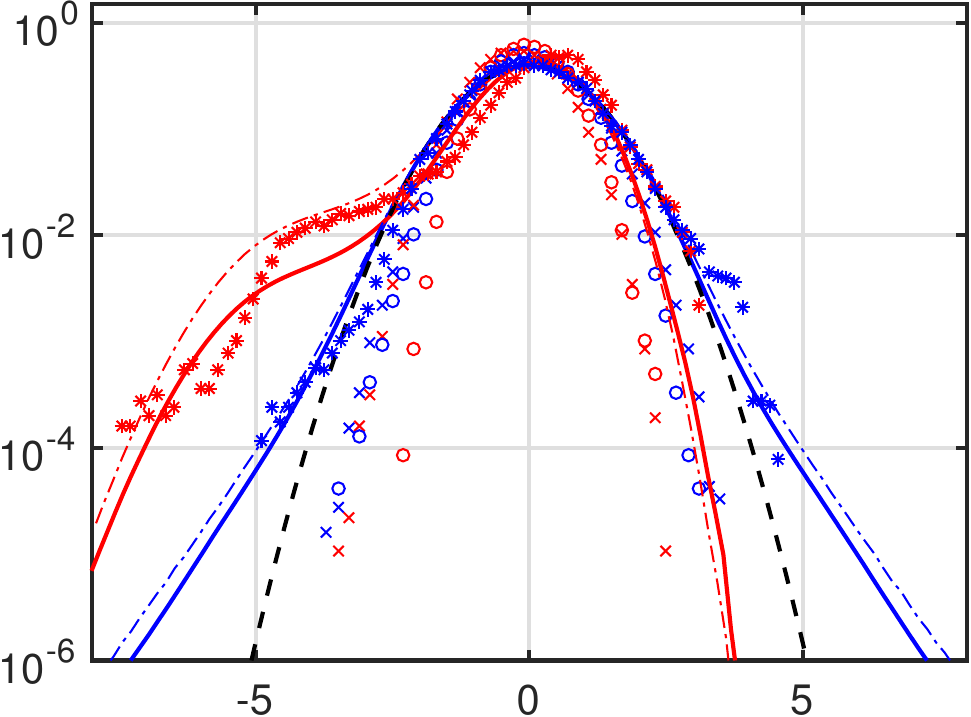}\\
      \centerline{$u^{'}/\sigma(u^{'})$}
   \end{minipage}
   \begin{minipage}{2ex}
      \rotatebox{90}{\centerline{p.d.f.}}
   \end{minipage}
   \begin{minipage}{0.45\linewidth}
      \centerline{(b)}
      \includegraphics[width=0.95\linewidth]
         {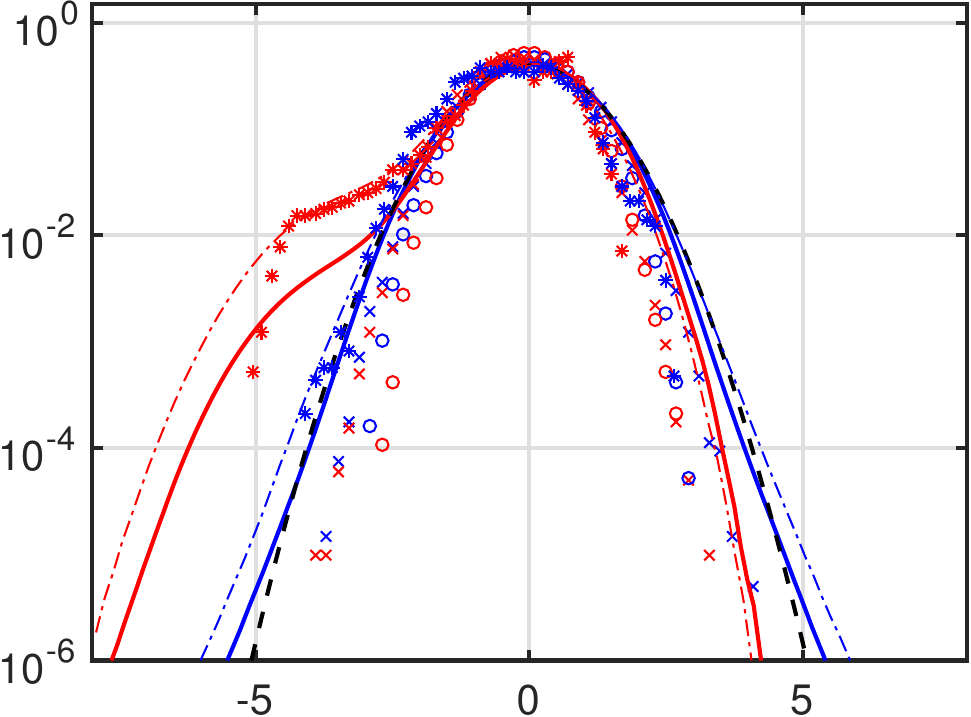}\\
      \centerline{$u^{'}/\sigma(u^{'})$}
   \end{minipage}
%---
\caption{P.d.f.\ of the fluid velocity computed on different regions
  of the flow for the case (a) G178-R95 and (b) G180-R140.
  Continuous lines represent the full domain occupied by the fluid
  $\Omega_f$, dashed-dotted lines
  refer to the flow regions sampled by the spheres
  $\mathcal{S}_{(i)}$ used for the computation of the local relative
  velocity.
  The stars 
  correspond to sampling on the particle-centered spheres of those
  particles which are members of a cluster.
  Crosses refer to the velocities obtained when sampling over the
  particle-centered spheres, and then averaging;
  the open circles are data from samples on the surface of
  randomly-placed spheres, and then averaging. 
  The dashed line indicates the Gaussian distribution. 
  Colorstyle: 
 $\color{red}{\solidthick}$ vertical component,
 $\color{blue}{\solidthick}$ horizontal component. }
\label{fig:velocity_pdf_with_shell}
\end{figure}

\begin{figure}[h]
   \begin{minipage}{2ex}
      \rotatebox{90}{\centerline{p.d.f.}}
   \end{minipage}
   \begin{minipage}{0.45\linewidth}
      \centerline{(a)}
      \includegraphics[width=\linewidth]
         {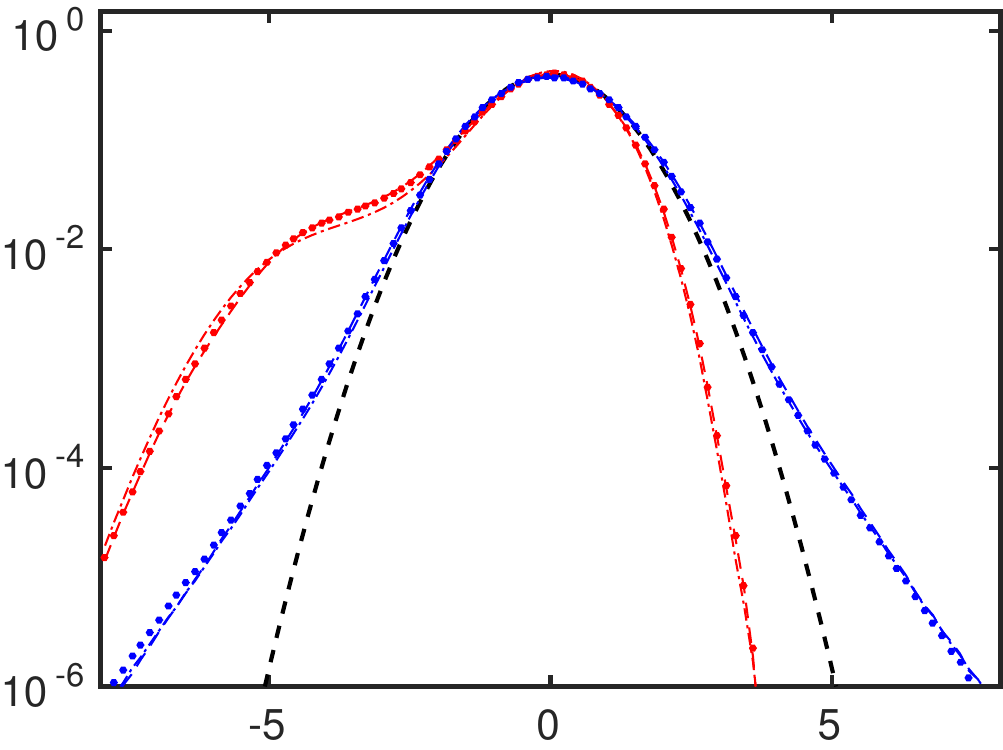}
      \centerline{$u^{'}/\sigma(u^{'})$}
   \end{minipage}
   % ---
   \begin{minipage}{2ex}
      \mbox{}\\
   \end{minipage}
   % ---
   \begin{minipage}{2ex}
      \rotatebox{90}{\centerline{p.d.f.}}
   \end{minipage}
   \begin{minipage}{0.45\linewidth}
      \centerline{(b)}
       \includegraphics[width=\linewidth]
         {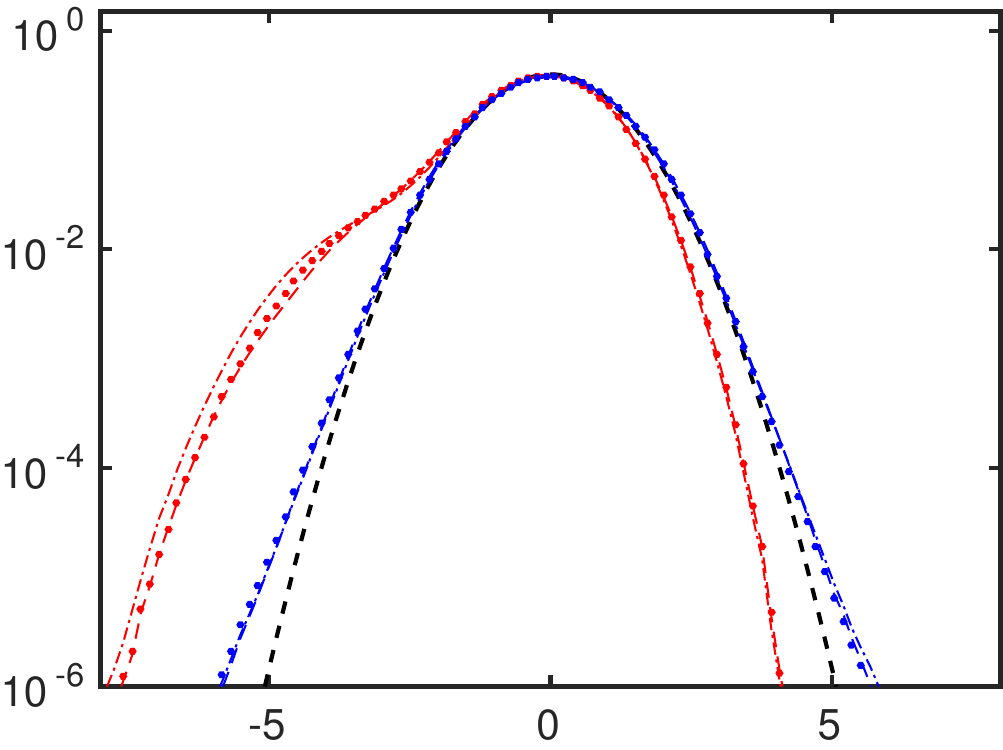}
      \centerline{$u^{'}/\sigma(u^{'})$}
   \end{minipage}
   % ---
   %
   \caption{P.d.f.\ of the fluid velocity computed on different regions
  of the flow for the case (a) G178-R95 and (b) G180-R140.
  Dashed-dotted lines
  refer to the flow regions sampled by the spheres
  $\mathcal{S}_{(i)}$ used for the computation of the local relative
  velocity, dashed lines to the velocities sampled at the front of the spheres according to the first definition and the dotted lines to the front of the spheres according to the second definition.
  The black dashed line remind the Gaussian distribution.
 Colorstyle: 
 $\color{red}{\solidthick}$ vertical component,
 $\color{blue}{\solidthick}$ horizontal component.}
   \label{fig:pdf_velocity_front}
\end{figure}

\begin{figure}
   \begin{minipage}{2ex}
      \rotatebox{90}{\centerline{$\sigma(\mathcal{V}_{Vor}/\langle \mathcal{V}_{Vor}\rangle)$}}
   \end{minipage}
   \begin{minipage}{0.45\linewidth}
      \centerline{(a)}
      \includegraphics[width=0.95\linewidth]
         {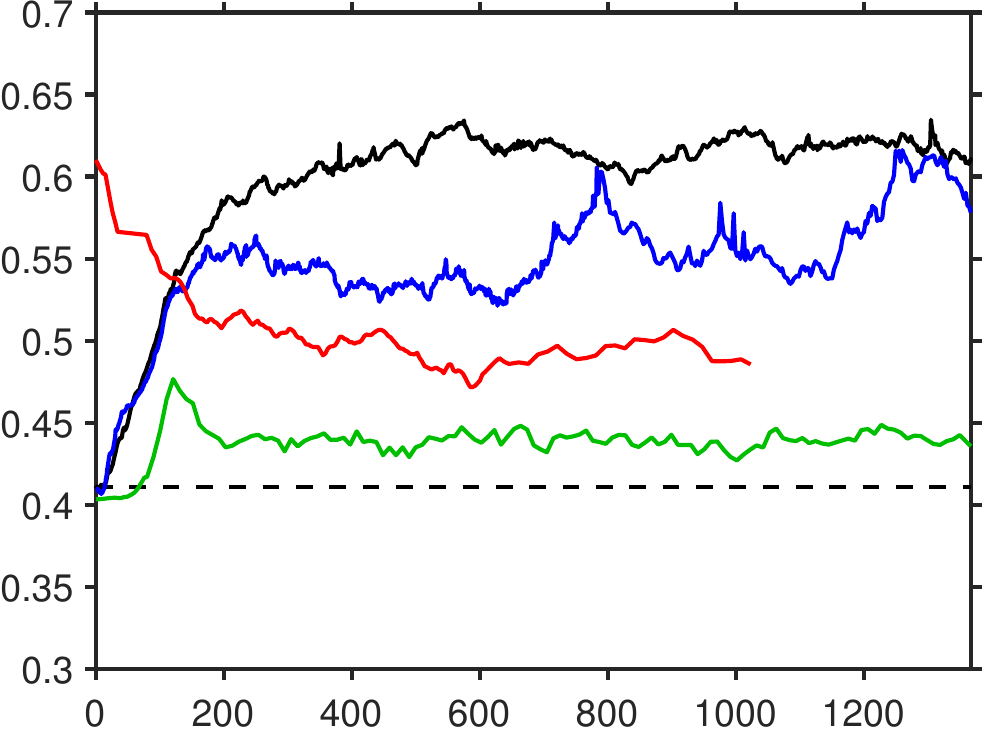}\\
      \centerline{$t/\tau_g$ , $t/\tau_D$}
   \end{minipage}
   \begin{minipage}{2ex}
      \rotatebox{90}{\centerline{$\sigma/\sigma_{Rand}$}}
   \end{minipage}
   \begin{minipage}{0.45\linewidth}
      \centerline{(b)}
      \includegraphics[width=0.95\linewidth]
         {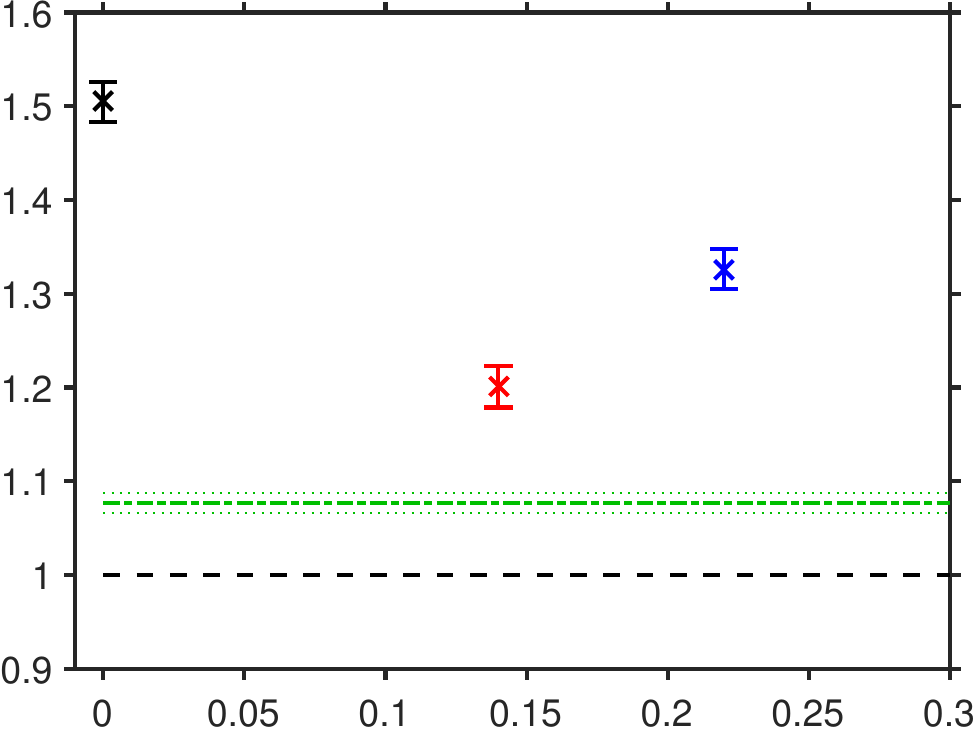}\\
      \centerline{$u_{rms}^{SP}/V_T$}
   \end{minipage}
%---
\caption{(a) Time evolution of the standard deviation of the volume of
  the Vorono\"i cells. For the cases G180-R140, G178-R95 and G178 time
  is scaled by $\tau_g$, for the case G0-R120 as gravity is set to
  zero we used the time scale $\tilde{\tau}_D=D/u_{rms}^{SP}$ for the
  scaling.
  (b) Time-mean of the standard deviation in (a) scaled by its
  equivalent for a corresponding set of randomly distributed
  particles, plotted as a function of the relative turbulence
  intensity. The crosses indicates the cases with settling, the
  green dash-dotted line the case G0-R114, and the dashed line the
  random reference. Linestyle:    
  $\color{red}{\solidthick}$ G178-R95, 
  $\color{blue}{\solidthick}$ G180-R140, 
  ${\solidthick}$ G178, 
  $\color{green}{\solidthick}$ G0-R120, 
  $\dashed$ randomly distributed particles. 
} 
\label{fig:stand_dev_vor_cells}
\end{figure}

\begin{figure}
   \centering
   \begin{minipage}{2ex}
      \rotatebox{90}{\centerline{p.d.f.}}
   \end{minipage}
   \begin{minipage}{0.45\linewidth}
      \centerline{(a)}
      \includegraphics[width=0.95\linewidth]
         {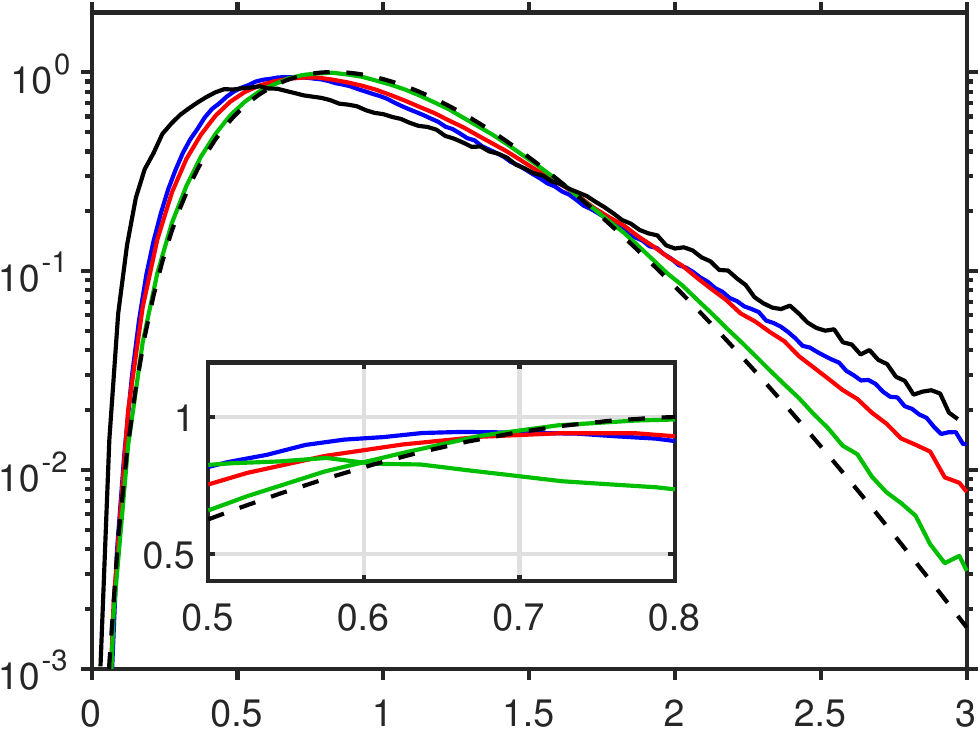}\\
      \centerline{$\mathcal{V}_{Vor}/\langle \mathcal{V}_{Vor}\rangle$}
   \end{minipage}
   \caption{(a) P.d.f.\ of the volume of the Vorono\"i cells.
     The inset
     shows a close-up of the same data around the lower cross-over
     points. 
     Linestyle:    
     $\color{red}{\solidthick}$ G178-R95, 
     $\color{blue}{\solidthick}$ G180-R140, 
     ${\solidthick}$ G178, 
     $\color{green}{\solidthick}$ G0-R120, 
     $\dashed$ randomly distributed finite-size particles.
   }
   \label{fig:voronoi_cells_volume_aspect_ratio}
 \end{figure}

\begin{figure}
   \begin{minipage}{2ex}
      \rotatebox{90}{\centerline{p.d.f.}}
   \end{minipage}
   \begin{minipage}{0.45\linewidth}
      \centerline{(a)}
      \includegraphics[width=0.95\linewidth]
         {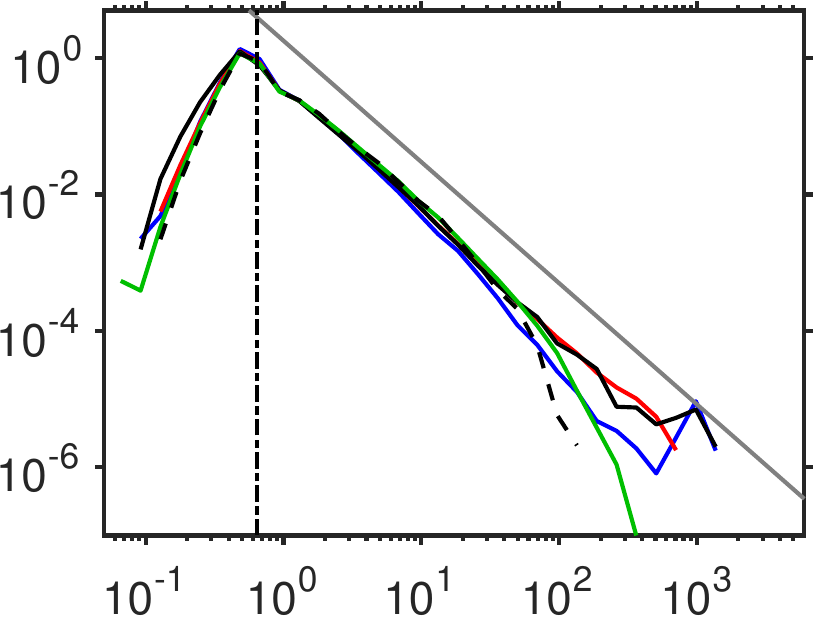}\\
      \centerline{$\mathcal{V}_c/\langle \mathcal{V}_{Vor} \rangle$}
   \end{minipage}
   \begin{minipage}{2ex}
      \rotatebox{90}{\centerline{p.d.f.}}
   \end{minipage}
   \begin{minipage}{0.45\linewidth}
      \centerline{(b)}
      \includegraphics[width=0.95\linewidth]
         {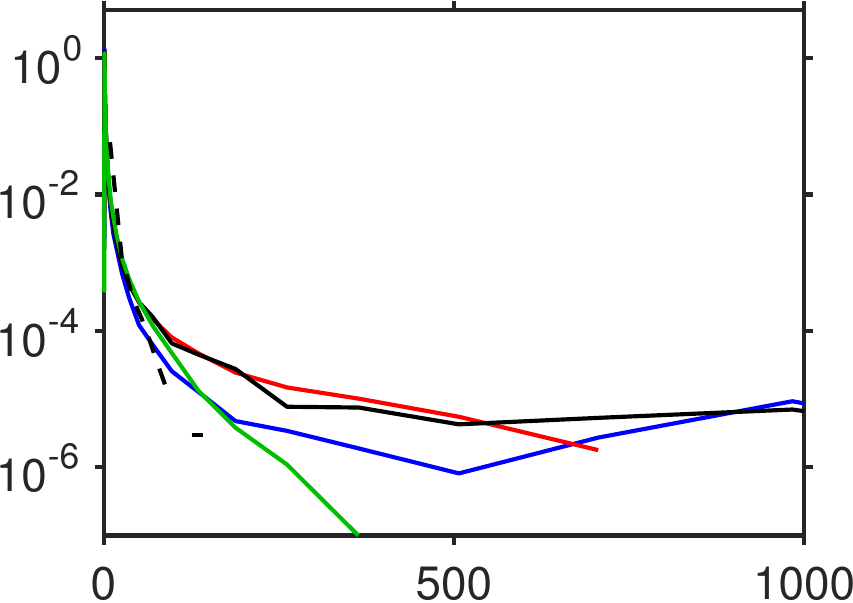}\\
      \centerline{$\mathcal{V}_c/\langle \mathcal{V}_{Vor} \rangle$}
   \end{minipage}
%---
\caption{P.d.f.\ of the cluster volume $\mathcal{V}_c$ in logarithmic
  scale (a) and in linear scale (b).  
 Linestyle:   
$\color{red}{\solidthick}$ G178-R95, 
$\color{blue}{\solidthick}$ G180-R140, 
${\solidthick}$ G178, 
$\color{green}{\solidthick}$ G0-R120,
$\dashed$ randomly distributed particles.
In (a) the vertical dash-dotted line indicates the clustering threshold
volume, $\mathcal{V}_{Vor}^{clus}$; 
the dark grey line is proportional to $\mathcal{V}_c^{-16/9}$. 
}
\label{fig:pdf_cluster_vol_number_part}
\end{figure}

\begin{figure}
  \centering
   \begin{minipage}{2ex}
      \rotatebox{90}{\centerline{$\mathcal{A}_{c}$}}
   \end{minipage}
   \begin{minipage}{0.45\linewidth}
      %\centerline{(a)}
      \includegraphics[width=0.95\linewidth]
         {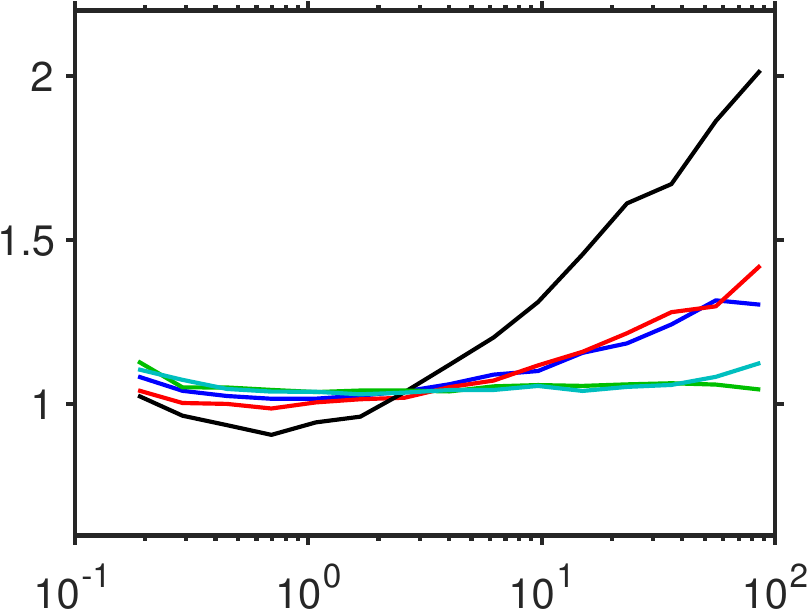}\\
      \centerline{$\mathcal{V}_c/\langle \mathcal{V}_{Vor} \rangle$}
   \end{minipage}
   \caption{
     Mean geometrical aspect ratio $\mathcal{A}_{c}=L_c^z/L_c^{x,y}$
     of the clusters plotted as a function of the cluster volume. 
$\color{red}{\solidthick}$ G178-R95, 
 $\color{blue}{\solidthick}$ G180-R140, 
${\solidthick}$ G178, 
$\color{green}{\solidthick}$ G0-R120, 
$\color{cyan}{\solidthick}$ randomly distributed particles.}
\label{fig:aspect_ratio_clusters}
\end{figure}

\begin{figure}
   \begin{minipage}{2ex}
      \rotatebox{90}{\centerline{p.d.f.}}
   \end{minipage}
   \begin{minipage}{0.45\linewidth}
      \centerline{(a)}
      \includegraphics[width=0.95\linewidth]
         {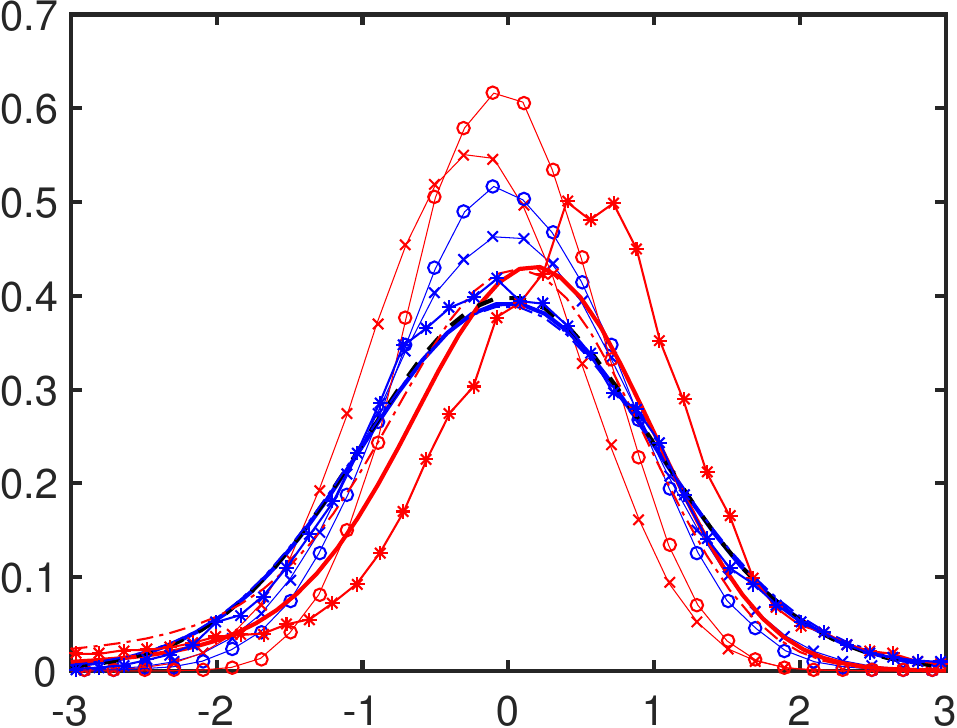}\\
      \centerline{$u^{'}/\sigma(u^{'})$}
   \end{minipage}
   \begin{minipage}{2ex}
      \rotatebox{90}{\centerline{p.d.f.}}
   \end{minipage}
   \begin{minipage}{0.45\linewidth}
      \centerline{(b)}
      \includegraphics[width=0.95\linewidth]
         {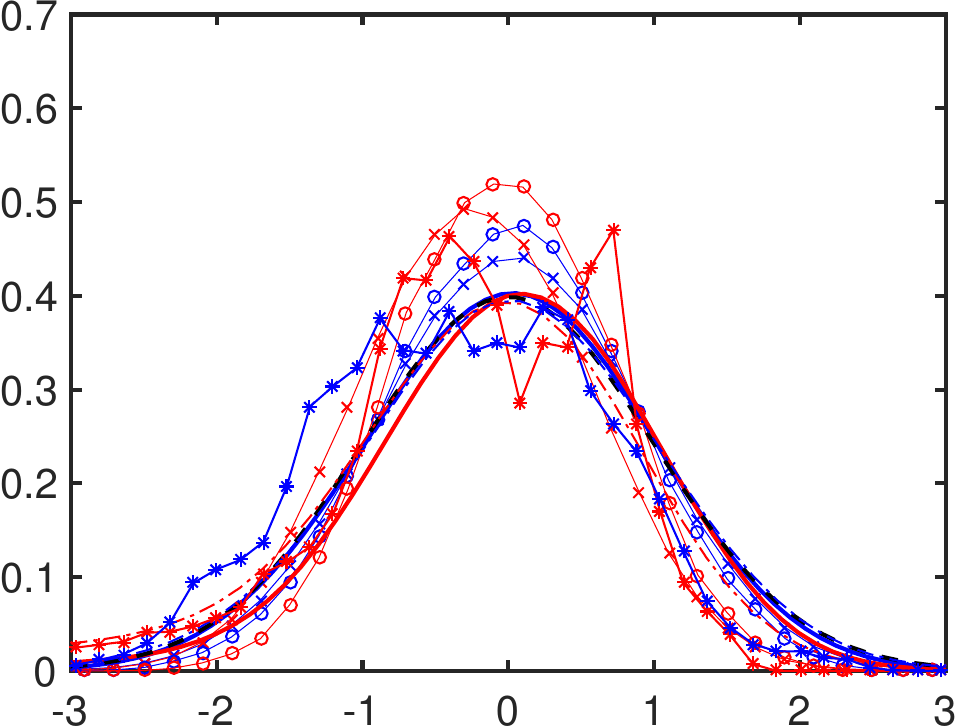}\\
      \centerline{$u^{'}/\sigma(u^{'})$}
   \end{minipage}
%---
\caption{P.d.f.\ of the fluid velocity (as in
  fig.~\ref{fig:velocity_pdf_with_shell}), but here plotted in linear
  scale
  for case (a) G178-R95 and (b) G180-R140.
  Continuous lines represent the full domain occupied by the fluid
  $\Omega_f$, dash-dotted lines
  refer to the flow regions sampled by the spheres
  $\mathcal{S}_{(i)}$ used for the computation of the local relative
  velocity.
  The stars 
  correspond to sampling on the particle-centered spheres of those
  particles which are members of a cluster.
  Crosses refer to the velocities obtained when sampling over the
  particle-centered spheres, and then averaging;
  the open circles are data from samples on the surface of
  randomly-placed spheres, and then averaging. 
  The dashed line indicates the Gaussian distribution. 
  Colorstyle: 
 $\color{red}{\solidthick}$ vertical component,
 $\color{blue}{\solidthick}$ horizontal component.
}
\label{fig:velocity_pdf_with_shell_LinScale}
\end{figure}

% ------------------------------------------------------------------------- %
We now
turn to 
the structure of the velocity field. According to the preferential sweeping scenario of  \citet{wang:93}, the particles preferentially sample the downward-flowing regions, resulting in increased settling speed.
Contrarily, in the present case we have seen that no clear
modification of the settling speed is observed
(section~\ref{sec:General_Stats}), with a small decrease of the
velocity obtained for G180-R140 and a small increase for G178-R95. 
We have seen that non-linear drag effects could explain the slight speed reduction for the case G180-R140, but that by itself it would induce a slight reduction in the other one as well. 
We have therefore investigated whether particles sample specific regions of the flow to test if there is a potential competition between fast tracking and non-linear drag effects.
We considered for this the probability density of the fluid velocity field, as represented on figure \ref{fig:velocity_pdf_with_shell}.
We separated the horizontal and vertical components and observed a
similar trend as what was obtained at lower Reynolds number
\citep{chouippe:15a} or in vertical turbulent channel flow \citep{Uhlmann_PoF2008}, namely that the fluid velocity fluctuations lose their Gaussian behaviour for amplitudes larger than approximately  three times their standard deviations, and that a negative skewness arises, due to the formation of wakes.  Note that a similar trend was also observed for G178 in the case of purely particle induced turbulence \citep{doychev:14}.
Comparing the velocity p.d.f.\ for both G178-R95 and G180-R140 shows also that, as expected, the system with the lowest turbulence intensity will deviate more from the Gaussian distribution than the other one that keeps a nearly Gaussian evolution of the horizontal velocity for fluctuation amplitudes up to five times the standard deviation.
We then explore the fluid velocity field in the vicinity of the particles, by looking at the field on the sphere surface $\mathcal{S}^{(i)}$ used for the computation of the local relative velocities in section \ref{sec:General_Stats}. 
Note that the difference in the current analysis is that we look at the very velocity seen by the particles, not at the relative velocity with respect to the particle motion.
For more clarity we represented the p.d.f.'s of the velocity fluctuations in linear scale on figure \ref{fig:velocity_pdf_with_shell_LinScale}, and compared the results with a set of randomly distributed particles. It shows that the particles tend to sample downward-flowing regions, and an estimation of the mean vertical fluid velocity seen by the particle yields
$\langle u_z^{\mathcal{S}_{(i)}}\rangle_p - \langle u_z\rangle_{\Omega_f} \approx -0.072u_g \;(-0.079u_g)$ for G178-R95 (G180-R140).
This point is however difficult to separate from the particles'
influence on a meso-scale, since large downward flow events are
generated by the wakes of multiple particles in proximity. 
We have also tested whether particles belonging to a cluster sample
the flow differently than the entire set of particles. We observe that
particles in clusters seem to avoid regions with the largest velocity
amplitude
(fig. \ref{fig:velocity_pdf_with_shell}), 
with a slight shift towards positive $u_z$ for G178-M95
(cf. fig. \ref{fig:velocity_pdf_with_shell_LinScale}).
In this case, particles falling through $u_z > 0$ regions would decelerate, which could be in favour of preferential accumulation but not of an augmentation of the falling speeds. 
In case G180-R140, the particles in clusters seem to avoid the regions of low velocity amplitude, which can be a sign that particles are expelled outside of the vortices.
We explored the evolution of $Q$ in the vicinity of the particles
(i.e.\ estimated on the particle-centered spheres $\mathcal{S}_{(i)}$)
and the corresponding p.d.f.'s show an increase of the probability to
sample regions of large strain or large vorticity
(fig. \ref{fig:qhunt_pdf_filtered}). 
This last result indicates that several mechanisms involved are here
in competition, with no clear distinction on the
dominant 
one. 
%
%
% ------------------------------------------------------------------------- %
\subsection{Influence of two-way coupling}
% ------------------------------------------------------------------------- %
The precedent considerations tend to propose a scenario to explain how gravity can promote the formation of clusters by shifting the apparent Stokes numbers towards larger values, and filtering by this the influence of the small scale structures in favour of the larger eddies. But this model does not account for any transformation of the eddies induced by the settling.
The surrogate Stokes number representative of the largest scales yields for instance $St_e^{(S)}=1.1 $ and $0.75$ respectively for G178-R95 and G180-R140, which does not necessarily justify the augmentation of $\sigma(\mathcal{V}_{Vor})$ observed from G178-R95 to G180-R140. 

However, we have seen that a decrease of the relative turbulence
intensity can also affect the coherence of the background turbulence,
with a partial destruction of the large scale by the wakes
(fig.~\ref{fig:visus_coherent_structures_wakes2}).   
The aforementioned Stokes numbers refer to the single phase large eddy
time scale, but we have seen that the particles tend to increase the
dissipation, with the effect of decreasing this large eddy time-scale
(assuming that the relation
$\tau_e^{TP}=u_{rms}^{TP}/\varepsilon^{TP}$ still holds). As a result,
the corresponding time-scales yield $\tau_e^{TP}/\tau_e^{SP} \approx
0.10 \;(0.18)$ for G178-R95 (G180-R140), which can partially explain
the difference observed in the amount of clustering.
This destruction can also be the signature of the formation of
large-scale columnar flow structures of the pure settling
configurations \citep{uhlmann:14a}, which will further affect the
preferential sampling in favor of particle-induced downward-flowing
regions.   
This means that increasing the influence of the settling has two effects in competition: on the one hand it filters the influence of the small scales and enhances the impact of the large scales, on the other hand it disrupts the structure of the large scales, reducing by this their effect on the particle motion. As those effects are difficult to decouple, it would be of interest in the future to systematically vary the turbulence intensity.
%

%--------------------------------------------------------------------%
\section{Conclusion}
% --------------------------------------------------------------------%
We have simulated the motion of a dilute suspension of fully-resolved,
finite-size particles under the combined action of gravity and
homogeneous-isotropically forced turbulence in triply-periodic
domains.
The solid-to-fluid density ratio was fixed at 1.5 and the Galileo
number at approximately 180 which corresponds to a parameter point for
which strong wake-induced clustering has been reported in the absence
of turbulence \citep{uhlmann:14a}. 
Two simulations have been performed with Taylor-scale Reynolds numbers
95, 140 and particle diameters corresponding to 7, 8.5 Kolmogorov
lengths, respectively. 
In the absence of gravity,
these parameters
fall into the range investigated
by 
\citet{uhlmann:16a},
who have reported slight turbulence-induced particle clustering which
has been statistically linked to the properties of the fluid
acceleration field.
The present work therefore brings the two effects of turbulence and
gravity together at parameter points intersecting as much as possible
with these earlier studies. 
The two present simulations differ in their value for the relative
turbulence intensity (0.14 and 0.22), which describes the relative
importance of these two effects, arguably the principal parameter
here.

We find that the average settling velocity (defined as an apparent
slip velocity, i.e.\ averaging separately over each phase and in time,
then subtracting both values) is only slightly different from the one 
for an isolated particle settling in ambient fluid, with an increase
by approximately 1\% at the lower turbulence intensity and a decrease
by 0.6\% for the higher turbulence intensity. 
At the same time the level of particle clustering, quantified with the
aid of Vorono\"i tessellation analysis, is presently found to be
intermediate between the strong clustering of the ambient settling
case and the weak clustering of the gravity-free case.
This result is not surprising, since forced background turbulence can
be expected to interfere with the formation of wake-induced clusters
on the different scales of the turbulent spectrum. 
However, it turns out that the present simulation with the larger
value of relative turbulence intensity leads to the larger level of
clustering. 
This means that the tendency to cluster does not monotonously decay
as a function of turbulence intensity.
As forced background turbulence reduces the level of wake-induced
clustering, it simultaneously enables clustering due to preferential 
concentration.
Unfortunately, the limited data which we have available (essentially
four data points, cf.\ figure~\ref{fig:stand_dev_vor_cells}$b$) does not
allow us to further map out the competition between these two opposing
effects. 

We have analyzed the velocity seen by the particles through sampling
and averaging the fluid velocity field on a sphere with radius equal to
three particle radii, centered on each particle.
It is observed that on average the particles in the present
simulations are exposed to more negative vertical fluid velocity than
the simple space average. 
We believe that this result is related to the collective effect of
particles having a relatively high probability of being located in the
wake region of one or more other particles.
On the other hand, the vertical component of the relative velocity
computed with the fluid velocity seen by each particle is on average
significantly smaller (in amplitude) than the apparent slip velocity,
and, therefore, also smaller than the single-particle settling speed
(by 5-7\%).
A straightforward estimate of the non-linear drag effect (based upon
standard quasi-steady drag and the assumption of identically
distributed Gaussian fluid velocity fluctuations, as e.g.\ shown by
\citet{homann:13}) predicts a reduction of the settling speed of the
order of one percent for the present cases. 

Based upon the Vorono\"i tessellation data we have performed a cluster
identification analysis (connecting neighboring Vorono\"i cells which
have a cell volume below a threshold value). We find that the
probability distribution of the cluster volumes follows a power-law
consistent with the model proposed by \citet{goto:06} in two space
dimensions, as already reported in \citet{uhlmann:16a} in the absence
of gravity. We note that the same power-law is also exhibited by a
random particle distribution, the difference between the various
data-sets lying in the tails of the distribution. Most significantly,
strong clustering corresponds to much larger probabilities for
observing very large clusters. We also observe that forced turbulence
has a strong effect upon the geometrical aspect ratio of the 
clusters: the ratio between the cluster extent in the vertical and
the horizontal direction is strongly decreased in both present
simulations (as compared to the case without turbulence) for
practically all relevant cluster sizes.

Finally we have discussed different mechanisms through which the
coherent flow structures are commonly believed to affect the particle
motion. The discussion has focused on vortical structures
educed via the second invariant of the velocity gradient tensor,
and our arguments are based upon visualizations and considerations of
the various time scales of the system.
We have identified the influence of gravity upon the interaction time
scales between a given eddy and a settling particle (at the origin of
the crossing trajectory effect) as one potential reason for the
different clustering characteristics in our two simulations.
The essence is that, as shown by \citet{yoshimoto:07} for the simplest
point-particle model, particles can be expected to interact with a
range of flow scales for which a kind of resonance condition is
fulfilled, based upon a comparison of time scales (i.e.\ Stokes
numbers). Now taking into account the mean relative velocity (due to
settling) leads to a correction of the relevant Stokes number,
proportional to the inverse of the relative turbulence intensity. For
cases with turbulence intensity below unity, this has the effect of
reducing the expected relevance of the small scales for the motion of
particles which are larger than the Kolmogorov scale. On the other
hand it brings the largest flow scales closer towards resonance with
the particles by increasing their associated Stokes number.
Therefore, under this scenario gravity would further decrease the
effect of the smallest flow scales upon particles which are larger
than the Kolmogorov scale, and it would promote the effect 
(beyond pure sweeping) due to the largest scales. This gravity-induced
shift of the resonant range of scales might be at the origin of the
increase in clustering with turbulence intensity.
Other scenarios, however, cannot be excluded as an explanation, such
as the influence of the settling particles upon the carrier phase. 

As an outlook of the present study we recommend on the one hand to
extend the parameter space 
by performing additional systematic studies varying independently the relative turbulence intensity, the particle diameter and the Galileo number.
On the other hand, we believe that it will be highly
useful to generate time-resolved series of instantaneous data in order
to allow for a detailed analysis of the dynamics of the interaction
between particles and coherent structures.

%
%--------------------------------------------------------------------%
\section*{Acknowledgements}
%\begin{acknowledgements}
  % --------------------------------------------------------------------%
  This work was supported by the German Research Foundation (DFG)
  under project UH~242/1-2.  
  The simulations were partially performed at LRZ M\"unchen (under
  grant pr83la) and at SCC Karlsruhe (project DNSPARTHIT).      
  The computer resources, technical expertise and assistance 
  provided by these centers are thankfully acknowledged. 
  % --------------------------------------------------------------------%
%\end{acknowledgements}
% --------------------------------------------------------------------%

% % --------------------------------------------------------------------

\bibliography{references}

@article{bec:07,
  title = {Heavy Particle Concentration in Turbulence at Dissipative and Inertial Scales},
  author = {Bec, J. and Biferale, L. and Cencini, M. and Lanotte, A. and Musacchio, S. and Toschi, F.},
  journal = {Phys. Rev. Lett.},
  volume = {98},
  issue = {8},
  pages = {084502},
  numpages = {4},
  year = {2007},
  month = {Feb},
  publisher = {American Physical Society},
  doi = {10.1103/PhysRevLett.98.084502},
}

@article{csanady:63,
  doi = {10.1175/1520-0469(1963)020<0201:tdohpi>2.0.co;2},
  year  = {1963},
  month = {may},
  publisher = {American Meteorological Society},
  volume = {20},
  number = {3},
  pages = {201--208},
  author = {Csanady, G. T.},
  title = {Turbulent Diffusion of Heavy Particles in the Atmosphere},
  journal = {Journal of the Atmospheric Sciences}
}

@article{Eswaran_CF1988,
author = {Eswaran, V. and Pope, S.B.},
booktitle = {Computers {\&} Fluids},
doi = {10.1016/0045-7930(88)90013-8},
issn = {00457930},
pages = {257--278},
title = {{An examination of forcing in direct numerical simulations of turbulence}},
volume = {16},
year = {1988}
}

@Article{ferenc:07,
  author = 	 {J.-S. Ferenc and Z. Neda},
  title = 	 {On the size distribution of {P}oisson {V}oronoi cells},
  journal = 	 {Physica A},
  pages =        {518-526}, 
  year = 	 2007,
  volume = 	 385}

@Article{glowinski:99,
  author =   {R. Glowinski and T.-W. Pan and T.I. Hesla and D.D. Joseph},
  title =    {A distributed {L}agrange multiplier/fictitious domain method for particulate flows},
  journal =    {Int. J. Multiphase Flow},
  year =   1999,
  volume =   25,
  pages =  {755-794}
}

@article{Hwang_JFM2006,
author = {Hwang, Wo. and Eaton, J. K.},
journal = {J. Fluid Mech.},
doi = {10.1017/S0022112006001431},
isbn = {0022112006},
issn = {0022-1120},
pages = {361},
title = {{Homogeneous and isotropic turbulence modulation by small heavy (${St \sim 50}$) particles}},
volume = {564},
year = {2006}
}

@InProceedings{hunt:88,
  author =   {J.C.R. Hunt and A.A. Wray and P. Moin},
  title =    {Eddies, streams, and convergence zones in turbulent flows},
  booktitle =    {Proceedings of the Summer Programm},
  pages =    {193-208},
  year =   1988,
  publisher = {(Center for Turbulence Research, Stanford)}}

@article{Jenny_JFM2004,
author = {Jenny, M. and Du{\v{s}}ek, J. and Bouchet, G.},
doi = {10.1017/S0022112004009164},
issn = {0022-1120},
journal = {J. Fluid Mech.},
pages = {201--239},
title = {{Instabilities and transition of a sphere falling or ascending freely in a Newtonian fluid}},
volume = {508},
year = {2004}
}

@article{Kidanemariam_NJP2013,
author = {Kidanemariam, A.G. and Chan-Braun, C. and Doychev, T. and Uhlmann, M.},
doi = {10.1088/1367-2630/15/2/025031},
eprint = {1301.5771},
issn = {13672630},
journal = {New J. Phys.},
title = {{Direct numerical simulation of horizontal open channel flow with finite-size, heavy particles at low solid volume fraction}},
volume = {15},
year = {2013}
}

@article{Monchaux_PoF2010,
author = {Monchaux, R. and Bourgoin, M. and Cartellier, a.},
doi = {10.1063/1.3489987},
isbn = {1070-6631},
issn = {10706631},
journal = {Phys. Fluids},
number = {10},
pages = {103304},
title = {{Preferential concentration of heavy particles: A Vorono\"i analysis}},
volume = {22},
year = {2010}
}

@article{Obligado_JoT2014,
author = {Obligado, M. and Teitelbaum, T. and Cartellier, A. and Mininni, P. and
 Bourgoin, M.},
title = {Preferential concentration of heavy particles in turbulence},
journal = {Journal of Turbulence},
volume = {15},
number = {5},
pages = {293-310},
year  = {2014},
publisher = {Taylor & Francis},
doi = {10.1080/14685248.2014.897710},
}

@inproceedings{oesterle:07,
  doi = {10.2495/mpf070371},
  year  = {2007},
  month = {may},
  publisher = {{WIT} Press},
  author = {Oesterl\'{e}, B.},
  title = {A note on crossing-trajectory effects in gas-particle turbulent flows},
  booktitle = {Computational Methods in Multiphase Flow {IV}}
}

@article{Riboux_JFM2010,
   title={Experimental characterization of the agitation generated by bubbles rising at high Reynolds number},
   volume={643}, 
   DOI={10.1017/S0022112009992084}, 
   journal={Journal of Fluid Mechanics}, 
   publisher={Cambridge University Press}, 
   author={Riboux, G. and Risso, F.C and Legendre, D.}, 
   year={2010}, 
   pages={509-539}}

@article{Schiller_1935,
  title={A drag coefficient correlation},
  author={L. Schiller and A. Naumann},
  journal={VDI Zeitung},
  volume={77},
  number={318},
  pages={51},
  year={1935}
}

@article{Tanaka_PRL2008,
author = {Tanaka, T. and Eaton, J. K.},
doi = {10.1103/PhysRevLett.101.114502},
isbn = {0031-9007 (Print)$\backslash$r0031-9007 (Linking)},
issn = {00319007},
journal = {Phys. Rev. Lett.},
number = {September},
pages = {1--4},
pmid = {18851286},
title = {{Classification of turbulence modification by dispersed spheres using a novel dimensionless number}},
volume = {101},
year = {2008}
}

@article{Uhlmann_JCP2005,
author = {Uhlmann, M.},
doi = {10.1016/j.jcp.2005.03.017},
issn = {00219991},
journal = {Journal of Computational Physics},
pages = {448--476},
pmid = {230736700004},
title = {{An immersed boundary method with direct forcing for the simulation of particulate flows}},
volume = {209},
year = {2005}
}

@article{Uhlmann_PoF2008,
author = {Uhlmann, Markus},
doi = {10.1063/1.2912459},
journal = {Phys. Fluids},
number = {May 2013},
title = {{Interface-resolved direct numerical simulation of vertical particulate channel flow in the turbulent regime}},
volume = {20},
year = {2008}
}

@article{Uhlmann_IJMF2014,
author = {Uhlmann, M. and Du{\v{s}}ek, J.},
doi = {10.1016/j.ijmultiphaseflow.2013.10.010},
issn = {03019322},
journal = {International Journal of Multiphase Flow},
pages = {221--243},
publisher = {Elsevier Ltd},
title = {{The motion of a single heavy sphere in ambient fluid: A benchmark for interface-resolved particulate flow simulations with significant relative velocities}},
volume = {59},
year = {2014}
}

@article{wells:83,
title={The effects of crossing trajectories on the dispersion of particles in a turbulent flow},
volume={136},
DOI={10.1017/S0022112083002049},
journal={Journal of Fluid Mechanics}, 
publisher={Cambridge University Press}, 
author={Wells, M. R. and Stock, D. E.}, 
year={1983}, 
pages={31-62}}

@incollection{yudine:59,
title = "Physical Considerations on Heavy-Particle Diffusion",
editor = "H.E. Landsberg and J. Van Mieghem",
series = "Advances in Geophysics",
publisher = "Elsevier",
volume = "6",
pages = "185 - 191",
year = "1959",
issn = "0065-2687",
author = "Yudine, M.I."
}

@article{uhlmann:14a,
  author =        {M. Uhlmann and T. Doychev},
  journal =       {J.\ Fluid Mech.},
  pages =         {310-348},
  title =         {Sedimentation of a dilute suspension of rigid spheres
                   at intermediate {G}alileo numbers: the effect of
                   clustering upon the particle motion},
  volume =        {752},
  year =          {2014},
  doi =           {10.1017/jfm.2014.330},
}

@article{uhlmann:16a,
  author =        {M.\ Uhlmann and A.\ Chouippe},
  journal =       {J.\ Fluid Mech.},
  pages =         {991--1023},
  title =         {Clustering and preferential concentration of
                   finite-size particles in forced homogeneous-isotropic
                   turbulence},
  volume =        {812},
  year =          {2017},
  doi =           {10.1017/jfm.2016.826},
}

@article{shaw:03,
  author =        {R.A. Shaw},
  journal =       {Annu. Rev. Fluid Mech.},
  number =        {1},
  pages =         {183-227},
  title =         {Particle-turbulence interactions in atmospheric
                   clouds},
  volume =        {35},
  year =          {2003},
  doi =           {10.1146/annurev.fluid.35.101101.161125},
}

@article{grabowski:13,
  author =        {W.W. Grabowski and L.-P. Wang},
  journal =       {Annu. Rev. Fluid Mech.},
  number =        {1},
  pages =         {293-324},
  title =         {Growth of Cloud Droplets in a Turbulent Environment},
  volume =        {45},
  year =          {2013},
  doi =           {10.1146/annurev-fluid-011212-140750},
}

@article{squires:91,
  author =        {K.D. Squires and J.K. Eaton},
  journal =       {Phys. Fluids A},
  number =        {5},
  pages =         {1169-1178},
  title =         {Preferential concentration of particles by
                   turbulence},
  volume =        {3},
  year =          {1991},
}

@article{maxey:87,
  author =        {M.R. Maxey},
  journal =       {J. Fluid Mech.},
  pages =         {441-465},
  title =         {The gravitational settling of aerosol particles in
                   homogeneous turbulence and random flow fields},
  volume =        {174},
  year =          {1987},
}

@article{hogan:01,
  author =        {R.C. Hogan and J.N. Cuzzi},
  journal =       {Phys. Fluids},
  number =        {10},
  pages =         {2938-2945},
  title =         {Stokes and {R}eynolds number dependence of
                   preferential particle concentration in simulated
                   three-dimensional turbulence},
  volume =        {13},
  year =          {2001},
  doi =           {http://dx.doi.org/10.1063/1.1399292},
}

@article{balachandar:10,
  author =        {S. Balachandar and J.K. Eaton},
  journal =       {Ann. Rev. Fluid Mech.},
  pages =         {111-133},
  title =         {Turbulent dispersed multiphase flow},
  volume =        {42},
  year =          {2010},
}

@article{monchaux:12,
  author =        {R. Monchaux and M. Bourgoin and A. Cartellier},
  journal =       {Int. J. Multiphase Flow},
  pages =         {1-18},
  title =         {Analyzing preferential concentration and clustering
                   of inertial particles in turbulence},
  volume =        {40},
  year =          {2012},
}

@article{yoshimoto:07,
  author =        {H. Yoshimoto and S. Goto},
  journal =       {J. Fluid Mech.},
  pages =         {275-286},
  title =         {Self-similar clustering of inertial particles in
                   homogeneous turbulence},
  volume =        {577},
  year =          {2007},
}

@article{sumbekova:16,
  author =        {S. Sumbekova and A. Cartellier and A. Aliseda and
                   M. Bourgoin},
  journal =       {Phys. Rev. Fluids},
  number =        {2},
  pages =         {024302},
  title =         {Preferential concentration of inertial
                   sub-{K}olmogorov particles. {T}he roles of mass
                   loading of particles, {S}tokes and {R}eynolds
                   numbers},
  volume =        {2},
  year =          {2016},
  doi =           {10.1103/PhysRevFluids.2.024302},
}

@article{goto:08,
  author =        {S. Goto and J.C. Vassilicos},
  journal =       {Phys. Rev. Lett.},
  number =        {5},
  pages =         {054503},
  title =         {Sweep-Stick Mechanism of Heavy Particle Clustering in
                   Fluid Turbulence},
  volume =        {100},
  year =          {2008},
}

@article{coleman:09,
  author =        {S.W. Coleman and J.C. Vassilicos},
  journal =       {Phys. Fluids},
  number =        {11},
  pages =         {113301},
  title =         {A unified sweep-stick mechanism to explain particle
                   clustering in two-and three-dimensional homogeneous,
                   isotropic turbulence},
  volume =        {21},
  year =          {2009},
}

@article{zaichik:03,
  author =        {L.I. Zaichik and V.M. Alipchenkov},
  journal =       {Phys. Fluids},
  number =        {6},
  pages =         {1776--1787},
  title =         {Pair dispersion and preferential concentration of
                   particles in isotropic turbulence},
  volume =        {15},
  year =          {2003},
}

@article{chun:05,
  author =        {J. Chun and D.L. Koch and S.L. Rani and A. Ahluwalia and
                   L.R. Collins},
  journal =       {J. Fluid Mech.},
  pages =         {219--251},
  title =         {Clustering of aerosol particles in isotropic
                   turbulence},
  volume =        {536},
  year =          {2005},
}

@article{zaichik:07,
  author =        {L.I. Zaichik and V.M. Alipchenkov},
  journal =       {Phys. Fluids},
  number =        {11},
  pages =         {113308},
  title =         {Refinement of the probability density function model
                   for preferential concentration of aerosol particles
                   in isotropic turbulence},
  volume =        {19},
  year =          {2007},
}

@article{bragg:15a,
  author =        {A.D. Bragg and P.J. Ireland and L.R. Collins},
  journal =       {J. Fluid Mech.},
  pages =         {327-343},
  title =         {On the relationship between the non-local clustering
                   mechanism and preferential concentration},
  volume =        {780},
  year =          {2015},
  doi =           {10.1017/jfm.2015.474},
}

@article{gustavsson:16,
  author =        {K. Gustavsson and B. Mehlig},
  journal =       {Adv. Phys.},
  number =        {1},
  pages =         {1-57},
  title =         {Statistical models for spatial patterns of heavy
                   particles in turbulence},
  volume =        {65},
  year =          {2016},
  doi =           {10.1080/00018732.2016.1164490},
}

@article{fortes:87,
  author =        {A.F. Fortes and D.D. Joseph and T.S. Lundgren},
  journal =       {J. Fluid Mech.},
  pages =         {467-483},
  title =         {Nonlinear mechanics of fluidization of beds of
                   spherical particles},
  volume =        {177},
  year =          {1987},
}

@article{wu:98,
  author =        {J. Wu and R. Manasseh},
  journal =       {Int. J. Multiphase Flow},
  pages =         {1343-1358},
  title =         {Dynamics of dual-particles settling under gravity},
  volume =        {24},
  year =          {1998},
}

@article{kajishima:02,
  author =        {T. Kajishima and S. Takiguchi},
  journal =       {Int. J. Heat Fluid Flow},
  pages =         {639-646},
  title =         {Interaction between particle clusters and
                   particle-induced turbulence},
  volume =        {23},
  year =          {2002},
}

@article{kajishima:04b,
  author =        {T. Kajishima},
  journal =       {Int. J. Heat Fluid Flow},
  number =        {5},
  pages =         {721-728},
  title =         {Influence of particle rotation on the interaction
                   between particle clusters and particle-induced
                   turbulence},
  volume =        {25},
  year =          {2004},
}

@phdthesis{doychev:14,
  author =        {T. Doychev},
  school =        {Karlsruhe Institute of Technology},
  title =         {The dynamics of finite-size settling particles},
  year =          {2014},
  doi =           {10.5445/KSP/1000044723},
}

@article{lance:91,
  author =        {M. Lance and J. Bataille},
  journal =       {J. Fluid Mech.},
  pages =         {95-118},
  title =         {Turbulence in the liquid phase of a uniform bubbly
                   air-water flow},
  volume =        {222},
  year =          {1991},
}

@article{risso:11,
  author =        {F. Risso},
  journal =       {Phys. Fluids},
  number =        {1},
  pages =         {011701},
  publisher =     {AIP},
  title =         {Theoretical model for $k^{-3}$ spectra in dispersed
                   multiphase flows},
  volume =        {23},
  year =          {2011},
  doi =           {10.1063/1.3530438},
  eid =           {011701},
}

@article{wang:93,
  author =        {L.-P. Wang and M.R. Maxey},
  journal =       {J. Fluid Mech.},
  pages =         {27-68},
  title =         {Settling velocity and concentration distribution of
                   heavy particles in homogeneous isotropic turbulence},
  volume =        {256},
  year =          {1993},
}

@article{bec:14,
  author =        {J. Bec and H. Homann and S.S. Ray},
  journal =       {Phys. Rev. Lett.},
  month =         {May},
  pages =         {184501},
  title =         {Gravity-Driven Enhancement of Heavy Particle
                   Clustering in Turbulent Flow},
  volume =        {112},
  year =          {2014},
  doi =           {10.1103/PhysRevLett.112.184501},
}

@article{good:14,
  author =        {G.H. Good and P.J. Ireland and G.P. Bewley and
                   E. Bodenschatz and L.R. Collins and Z. Warhaft},
  journal =       {J. Fluid Mech.},
  pages =         {R3},
  title =         {Settling regimes of inertial particles in isotropic
                   turbulence},
  volume =        {759},
  year =          {2014},
  doi =           {10.1017/jfm.2014.602},
}

@article{ireland:16b,
  author =        {P.J. Ireland and A.D. Bragg and L.R. Collins},
  journal =       {J. Fluid Mech.},
  pages =         {659-711},
  title =         {The effect of {R}eynolds number on inertial particle
                   dynamics in isotropic turbulence. {P}art 2.
                   {S}imulations with gravitational effects},
  volume =        {796},
  year =          {2016},
  doi =           {10.1017/jfm.2016.227},
}

@article{nielsen:93,
  author =        {P. Nielsen},
  journal =       {J. Sed. Res.},
  number =        {5},
  pages =         {835-838},
  title =         {Turbulence effects on the settling of suspended
                   particles},
  volume =        {63},
  year =          {1993},
  doi =           {10.1306/D4267C1C-2B26-11D7-8648000102C1865D},
}

@article{aliseda:02,
  author =        {A. Aliseda and A. Cartellier and F. Hainaux and
                   J.C. Lasheras},
  journal =       {J. Fluid Mech.},
  pages =         {77-105},
  title =         {Effect of preferential concentration on the settling
                   velocity of heavy particles in homogeneous isotropic
                   turbulence},
  volume =        {468},
  year =          {2002},
  doi =           {10.1017/S0022112002001593},
}

@article{monchaux:17,
  author =        {R. Monchaux and A. Dejoan},
  journal =       {Phys. Rev. Fluids},
  month =         {Oct},
  pages =         {104302},
  title =         {Settling velocity and preferential concentration of
                   heavy particles under two-way coupling effects in
                   homogeneous turbulence},
  volume =        {2},
  year =          {2017},
  doi =           {10.1103/PhysRevFluids.2.104302},
}

@article{yang:03,
  author =        {T.S. Yang and S.S. Shy},
  journal =       {Phys. Fluids},
  number =        {4},
  pages =         {868-880},
  title =         {The settling velocity of heavy particles in an
                   aqueous near-isotropic turbulence},
  volume =        {15},
  year =          {2003},
  doi =           {10.1063/1.1557526},
}

@article{tunstall:68,
  author =        {E.B. Tunstall and G. Houghton},
  journal =       {Chem. Eng. Sci.},
  number =        {9},
  pages =         {1067--1081},
  title =         {Retardation of falling spheres by hydrodynamic
                   oscillations},
  volume =        {23},
  year =          {1968},
}

@article{homann:13,
  author =        {H. Homann and J. Bec and R. Grauer},
  journal =       {J. Fluid Mech.},
  month =         {4},
  pages =         {155--179},
  title =         {Effect of turbulent fluctuations on the drag and lift
                   forces on a towed sphere and its boundary layer},
  volume =        {721},
  year =          {2013},
  doi =           {10.1017/jfm.2013.66},
  issn =          {1469-7645},
}

@article{chouippe:15a,
  author =        {A. Chouippe and M. Uhlmann},
  journal =       {Phys.\ Fluids},
  number =        {12},
  pages =         {123301},
  title =         {Forcing homogeneous turbulence in {D}{N}{S} of
                   particulate flow with interface resolution and
                   gravity},
  volume =        {27},
  year =          {2015},
  doi =           {10.1063/1.4936274},
}

@article{fornari:16a,
  author =        {W. Fornari and F. Picano and L. Brandt},
  journal =       {J. Fluid Mech.},
  pages =         {640--669},
  title =         {Sedimentation of finite-size spheres in quiescent and
                   turbulent environments},
  volume =        {788},
  year =          {2016},
  doi =           {10.1017/jfm.2015.698},
}

@article{fornari:16b,
  author =        {W. Fornari and F. Picano and G. Sardina and
                   L. Brandt},
  journal =       {J. Fluid Mech.},
  pages =         {153-167},
  title =         {Reduced particle settling speed in turbulence},
  volume =        {808},
  year =          {2016},
  doi =           {10.1017/jfm.2016.648},
}

@article{fiabane:12,
  author =        {L. Fiabane and R. Zimmermann and R. Volk and
                   J.-F. Pinton and M. Bourgoin},
  journal =       {Phys. Rev. E},
  number =        {035301(R)},
  title =         {Clustering of finite-size particles in turbulence},
  volume =        {86},
  year =          {2012},
}

@article{goto:06,
  author =        {S. Goto and J.C. Vassilicos},
  journal =       {Phys. Fluids},
  number =        {11},
  pages =         {115103},
  title =         {Self-similar clustering of inertial particles and
                   zero-acceleration points in fully developed
                   two-dimensional turbulence},
  volume =        {18},
  year =          {2006},
}

@article{buerger:12a,
  title={Vortices within vortices: hierarchical nature of vortex tubes in turbulence},
  author={B{\"u}rger, Kai and Treib, Marc and Westermann, R{\"u}diger and Werner, Suzanne and Lalescu, Cristian C and Szalay, Alexander and Meneveau, Charles and Eyink, Gregory L},
  journal={arXiv preprint arXiv:1210.3325},
  year={2012}
}

@Article{ferrante:03,
  author = 	 {A. Ferrante and S. Elghobashi},
  title = 	 {On the physical mechanisms of two-way coupling in particle-laden isotropic turbulence},
  journal = 	 {Phy. Fluids},
  year = 	 2003,
  volume =	 15,
  number =	 2,
  pages =	 {315-329}
}

@article{cisse:13,
  doi = {10.1017/jfm.2013.490},
  year  = {2013},
  month = {oct},
  publisher = {Cambridge University Press ({CUP})},
  volume = {735},
  author = {M. Cisse and H. Homann and J. Bec},
  title = {Slipping motion of large neutrally buoyant particles in turbulence},
  journal = {Journal of Fluid Mechanics}
}

@article{huisman:16,
  title = {Columnar structure formation of a dilute suspension of settling spherical particles in a quiescent fluid},
  author = {Huisman, S. G. and Barois, T. and Bourgoin, M. and Chouippe, A. and Doychev, T. and Huck, P. and Bello Morales, C. E. and Uhlmann, M. and Volk, R.},
  journal = {Phys. Rev. Fluids},
  volume = {1},
  issue = {7},
  pages = {074204},
  numpages = {11},
  year = {2016},
  month = {Nov},
  publisher = {American Physical Society},
  doi = {10.1103/PhysRevFluids.1.074204},
}

@article{rosa:16,
title = "Settling velocity of small inertial particles in homogeneous isotropic turbulence from high-resolution {D}{N}{S}",
journal = "Int. J. Multiphase Flow",
volume = "83",
pages = "217-231",
year = "2016",
doi = "10.1016/j.ijmultiphaseflow.2016.04.005",
author = "B. Rosa and H. Parishani and O. Ayala and L.-P. Wang"
}
%% % --------------------------------------------------------------------
%
\clearpage
\clearpage
 % ---------------------------------------------------------------------------- %
% ---------------------------------------------------------------------------- %

 % ---------------------------------------------------------------------------- %
% ---------------------------------------------------------------------------- %

 % ---------------------------------------------------------------------------- %
% ---------------------------------------------------------------------------- %

 % ---------------------------------------------------------------------------- %
% ---------------------------------------------------------------------------- %

 % ---------------------------------------------------------------------------- %
% ---------------------------------------------------------------------------- %
% ---------------------------------------------------------------------------- %
% ---------------------------------------------------------------------------- %

%
 % ---------------------------------------------------------------------------- %
% ---------------------------------------------------------------------------- %

 % ---------------------------------------------------------------------------- %
% ---------------------------------------------------------------------------- %

 % ---------------------------------------------------------------------------- %
% ---------------------------------------------------------------------------- %

 % ------------------------------------------------------------------------- %
% ------------------------------------------------------------------------- %
% ------------------------------------------------------------------------- %

%
 % ---------------------------------------------------------------------------- %
% ---------------------------------------------------------------------------- %

 % ---------------------------------------------------------------------------- %
% ---------------------------------------------------------------------------- %

%
 % ---------------------------------------------------------------------------- %
% ---------------------------------------------------------------------------- %

 % ---------------------------------------------------------------------------- %
% ---------------------------------------------------------------------------- %

%
 % ---------------------------------------------------------------------------- %
% ---------------------------------------------------------------------------- %

% ************************************************************************* %
\end{document}